%% file: main.tex
\documentclass[journal=jacsat,manuscript=article]{achemso}

\usepackage{chemformula} 
\usepackage[T1]{fontenc} 
\usepackage{import}
\usepackage{gensymb}
\usepackage{array}
\usepackage{tabularray}
\usepackage{multirow}
\usepackage{hyperref}
\mciteErrorOnUnknownfalse

\newcommand{\bra}[1]{\langle #1 |}
\newcommand{\ket}[1]{| #1 \rangle}
\newcommand{\braket}[2]{\langle #1 | #2 \rangle}

\author{Rae A. Corrigan Grove}
\affiliation[T1]{Theoretical Division, Los Alamos National Laboratory, Los Alamos, 87545, NM, USA}
\email{rcgrove@lanl.gov}

\author{Kevin G. Kleiner}
\affiliation[UIIC]{Department of Physics, University of Illinois Urbana-Champaign, Urbana-Champaign, 61820, IL, USA}

\author{Joshua Finkelstein}
\affiliation[T1]{Theoretical Division, Los Alamos National Laboratory, Los Alamos, 87545, NM, USA}

\author{Ivana Matanovic}
\affiliation[T1]{Theoretical Division, Los Alamos National Laboratory, Los Alamos, 87545, NM, USA}

\author{Michael E. Wall}
\affiliation[CAID]{Computing and Artificial Intelligence Division, Los Alamos National Laboratory, Los Alamos, 87545, NM, USA}

\author{Travis E. Jones}
\affiliation[T1]{Theoretical Division, Los Alamos National Laboratory, Los Alamos, 87545, NM, USA}

\author{Anders M. N. Niklasson}
\affiliation[T1]{Theoretical Division, Los Alamos National Laboratory, Los Alamos, 87545, NM, USA}

\author{Christian F. A. Negre}
\affiliation[T1]{Theoretical Division, Los Alamos National Laboratory, Los Alamos, 87545, NM, USA}
\email{cnegre@lanl.gov}

\title{Modeling Reactions on the Solid-Liquid Interface with Next Generation Extended Lagrangian Quantum-Based Molecular Dynamics}

\begin{document}

\begin{tocentry}

Some journals require a graphical entry for the Table of Contents.
This should be laid out ``print ready'' so that the sizing of the
text is correct.

Inside the \texttt{tocentry} environment, the font used is Helvetica
8\,pt, as required by \emph{Journal of the American Chemical
Society}.

The surrounding frame is 9\,cm by 3.5\,cm, which is the maximum
permitted for  \emph{Journal of the American Chemical Society}
graphical table of content entries. The box will not resize if the
content is too big: instead it will overflow the edge of the box.

This box and the associated title will always be printed on a
separate page at the end of the document.

\end{tocentry}

\begin{abstract}
  We present a series of simulations of the oxygen reduction reaction (ORR) using a novel framework for atomistic simulations of surface catalysis under electrochemical bias. The framework makes use of quantum-mechanical extended Lagrangian Born-Oppenheimer molecular dynamics (XL-BOMD) simulations, which provide the speed and accuracy required for explicit atomistic treatment of both electrode and electrolyte. Simulations of solvated O$_2$ near nitrogen-doped graphene (NG) were performed to gain insight into the ORR, and different mechanisms were observed, depending on the applied bias. Under higher bias the ORR occurred by an outer-sphere mechanism, without adsorption of O$_2$ to NG. In this mechanism, electron transfer between the catalyst and the O$_2$ was mediated by the solvent. Under lower bias the ORR occurred by an inner-sphere mechanism involving adsorption of O$_2$ to NG, leading to direct electron transfer. Our extensive, all-atom quantum-mechanical molecular dynamics simulations also show clear differences between the kinetics of the ORR on this ideally polarizable electrode and commonly used kinetic theories, leading to new insights regarding mechanistic changes with varied overpotentials.
  Combining quantum accuracy with explicit solvation and electrostatic potential bias, XL-BOMD opens a route to predictive, atomistic insight into electrocatalytic processes, as demonstrated with the ORR.
\end{abstract}


\section{Introduction}\label{sec1}

As global energy demands rise, the development of cost-effective and efficient electrocatalysts is a central challenge for clean energy technologies \cite{Morozan11, Matanovic18}. In fuel cells, the oxygen reduction reaction (ORR) at the cathode is the rate-limiting step. While platinum group metals (PGMs) remain the benchmark catalysts, their scarcity and expense motivate the search for alternatives \cite{Markovic01, Song08}. Among these, nitrogen-doped graphene (NG) has emerged as a promising candidate due to its activity, natural abundance, and suitability as a model platform for studying ORR mechanisms \cite{Reyimjan06, Yang19, Ma2022-xo}. Despite extensive experimental and theoretical work, however, the fundamental nature of charge transfer at NG surfaces -- particularly the role of the electrolyte -- remains unresolved \cite{Okamoto2009-tg, Boukhvalov2012-yw, Man2020-py, Ganyecz2021-ne, Ma2022-xo, Low24}.

Both inner- and outer-sphere pathways have been proposed to describe ORR on PGM-free catalysts \cite{Ramaswamy01,Low24}. In the inner-sphere mechanism, O$_2$ chemisorbs to the catalyst surface prior to electron transfer, whereas in the outer-sphere mechanism, charge transfer proceeds through the solvent without direct adsorption. The balance between these routes depends on the solvation and applied potential, and the complexity is challenging to capture computationally. Density functional theory (DFT), often employed with the computational hydrogen electrode (CHE), has provided valuable insights into intermediate energetics and adsorption processes \cite{Norskov2004-qu, Yu2011-ua, Zhang2011-cw}, however such methods inherently assume proton-coupled electron transfer and cannot describe non-adsorptive charge transfer. Conventional quantum-based molecular dynamics (QMD) can in principle address electrode–electrolyte interactions \cite{Kronberg21, Zhao22, Tian25}, but the formal cubic scaling of Kohn-Sham DFT places severe restrictions on the practically accessible length and time-scales. In general, the need for repeated self-consistent field iterations makes simulations of reactive, solvated interfaces using DFT expensive and prone to convergence difficulties as system sizes increase.\cite{Rabuck99, Kudin02, Liao22, Grob23}

Quantum-mechanical extended Lagrangian Born-Oppenheimer molecular dynamics (XL-BOMD) offers a possible solution to these challenges. By treating the electronic degrees of freedom as auxiliary dynamical variables, XL-BOMD enables efficient and stable propagation of both nuclear and electronic motion, avoiding the bottlenecks of traditional QMD simulations, e.g.\ tightly converged, iterative self-consistent field optimizations prior to each force evaluation \cite{Niklasson08, Niklasson17, Niklasson21}. This framework retains ground-state accuracy while allowing explicit inclusion of both the electrode and electrolyte, thus making possible energy-conserving simulations of electrocatalytic processes that were previously inaccessible.

In XL-BOMD, the electronic degrees of freedom are propagated as dynamical variables in the spirit of extended Lagrangian Car-Parrinello molecular dynamics \cite{CarParrinello85, Hutter_Book}. However, the extended Lagrangian framework is different and has another set of equations of motion that more closely follow the exact Born-Oppenheimer potential energy surface. This is particularly true in its most recent shadow potential formulation \cite{Niklasson08, Niklasson21}. Moreover, XL-BOMD is not limited by a finite gap between the highest occupied molecular orbital and the lowest unoccupied molecular orbital (known as the HOMO-LUMO gap) and can be used for vanishing energy gap systems, which is important for capturing chemical reactions. Additionally, in contrast to Car-Parrinello molecular dynamics, no tuning of fictitious electron mass parameters is necessary. The ability of XL-BOMD to maintain long term energy stability in microcanonical \textit{NVE} simulations is also important when a thermostat is added for canonical \textit{NVT} simulations \cite{Martinez2015-vm} as this is necessary for accurately sampling a canonical distribution and yielding correct equilibrium thermodynamics. 


In this work, we employ the most recent shadow potential formulation of XL-BOMD to simulate the ORR on NG in aqueous solution, and introduce a novel method to apply an electrochemical potential bias to the NG sheet, thereby changing the observed ORR mechanism. This generalizable method applies a controlled bias to the electrode, directly modulating the driving force of the reaction under study. We demonstrate stable, picosecond timescale simulations of charge transfer across the NG–water interface and identify mechanistic signatures of both inner- and outer-sphere reduction. Which mechanism we see depends on the bias applied to the NG sheet: simulations with a higher applied bias exhibit outer-sphere electron transfer mediated by the solvent; and simulations with a lower bias show inner-sphere pathways involving direct O$_2$ adsorption. 
Additionally, we see distinct differences in the mechanistic changes observed in simulations and those described by Butler-Volmer and Marcus-Hush-Chidsey kinetic theories. Our observations of the shift between inner- and outer-sphere mechanisms across simulations with varied applied biases suggest that the atomistic picture of reactions provided by advanced simulation methods is necessary to form a clear and complete description of electrode kinetics. 
By explicitly representing electrode and electrolyte, we are able to model the ORR under varied conditions and our biasing approach establishes a new route for probing potential-dependent reactivity in atomistic simulations of this important electrochemical system.

\section{Methods} 
    
\subsection{Density Functional Tight Binding Theory}
\label{DFTB}

DFT has been a primary workhorse for computational chemistry for the last three decades, as it is a relatively inexpensive technique for analyzing the energetics of simple molecules in vacuum and small crystal units \cite{HohenbergKohn64, ROJones89, RParr89, Haunschild2019-rj, Clark2021-py}. However, a time-resolved dynamic picture of chemical reactions including several hundreds to thousands of atoms has remained out of reach for direct DFT-based QMD simulations. In such QMD simulations, forces are evaluated for each new configuration, and these calculations have to be repeated over thousands of time steps. For this reason, first-principles DFT is often too expensive for many dynamics applications. 
Simplifying approximations are, therefore, needed to reduce the computational cost. The self-consistent charge density functional tight-binding theory (SCC-DFTB) provides a well-balanced approximation between speed and accuracy \cite{Elstner98, Frauenheim00, Aradi07, Hourahine20}. SCC-DFTB is based on a second-order approximation of the Kohn-Sham energy functional in DFT with respect to fluctuations around a set of overlapping neutral atomic electron densities. 
The SCC-DFTB approximation of the Kohn-Sham Hamiltonian \cite{KohnSham65} can be written as:

\begin{equation}  \label{autocon-DFTB}
    H_{i\alpha,j\beta} = \bra{\phi_{i \alpha}}\hat{H}_0 \ket{\phi_{j \beta}} + \frac{1}{2} (S_{i\alpha, j\beta} V^{\rm C}_{j} + V^{\rm C}_{i} S_{i\alpha, j \beta}),
\end{equation}
where
\begin{equation}
    V_i^{\rm C} = \sum_k \gamma_{ik} \Delta q_k.
\end{equation}
is the Coulomb potential including onsite, $i = k$, terms. 
Here $\phi_{i \alpha}$ and $\phi_{j \beta}$ are atomic orbitals $\alpha$ and $\beta$ centered at atoms $i$ and $j$, respectively, $\hat{H}_0$ is the non-SCC Hamiltonian operator \cite{Elstner98}, and $S$ is the overlap matrix. Using a composite atom+orbital indexing $\mu = \{i \alpha\}$ and $\nu = \{j \beta\}$, the elements of $S_{\mu \nu}$ are given by the inner product $\braket{\phi_{\nu}}{\phi_{\mu}}$. The last term of Equation (\ref{autocon-DFTB}) contains the dependence on the fluctuation of the charge.  
$\gamma_{ik}$ is a function that scales as the inverse distance between atoms, $1/||{R}_i - {R}_k||$, in the long-range limit, leading to a pure Coulombic potential produced by charges $\Delta q_k$. In the opposite case, when both orbitals are close to each other, $\gamma_{ik}$ becomes $\gamma_{ii} = U_{i}$, where $U$ is the Hubbard parameter that accounts for the onsite self-interaction of electrons and is related to the chemical hardness of the atomic species. 
Charge fluctuations $\Delta q_i = -(q_i - q_i^0)$ are obtained from the Mulliken analysis such that
\begin{equation}
    \Delta q_i = - \Big(\sum_{\nu \in i} \frac{1}{2}(S \rho + \rho S)_{\nu \nu}  - z_i \Big)
\end{equation}
where $\rho$ is the density matrix and $z_{i}$ is the electron population of the isolated species, $i$. The summation runs through every orbital, $\nu$, that belongs to atom $i$.  The density matrix is computed as $\rho = C f(\epsilon) C^{T}$, where $C$ is the matrix of the eigenvectors of $H$; $\epsilon$ is the diagonal matrix containing the eigenvalues of $H$; and $f$ is the Fermi distribution function, defined as $f(x) = 1/(1 + \exp(-\beta (x - \mu_e))$, where $\beta$ is the inverse electronic temperature and $\mu_e$ is the chemical potential for the electrons. 

The SCC-DFTB method is appropriate for cases where there is a large charge redistribution \cite{Elstner98, Frauenheim00} and is faster than standard Kohn-Sham DFT calculations by up to three orders of magnitude, often without significant loss of accuracy \cite{Cui00, Elstner01, Krishnapriyan17}. 
SCC-DFTB is implemented in the ${\rm DFTB}+$ \cite{Aradi07} and LATTE \cite{latte} software packages. In LATTE, the Hamiltonian elements are expressed as a decaying exponential of a polynomial function of the interatomic distances. The coefficients for the polynomial function depend on the nature of the orbitals and the species they belong to. 

A special characteristic of the LATTE Hamiltonian is the transferability. 
The new \textit{lanl}31 parametrization has proven to be highly transferable and has been used in studies involving different types of systems such as proteins \cite{Negre23}, high-explosives \cite{Perriot2018-cg, Cawkwell2019}, and molecular crystals \cite{Singh2024-ms}. We performed some specific tests to assess the suitability of these parameters for simulations of oxygen reduction. The adsorption energies of oxygen and hydrogen species are similar between DFTB and DFT (Table \ref{table: adsEnergies}), and the optimized molecular conformation of water using DFTB is very similar to that found using DFT (Table \ref{table: vac-h2o}). Moreover, the average bond length of O$_2$ solvated in water matches the previously reported experimental value \cite{Radjenovic19} (Figure \ref{fig:o2h2o-o2bondlength}). The radial distribution functions (RDF) computed from DFTB simulations of a water box are qualitatively similar to experiment, but with quantitative differences, as has been previously observed \cite{Goyal14}.  

The code used for performing the simulations in this work, graph-partitioned molecular dynamics with a kernel (GPMDK) \cite{Negre23,gpmdk}, is DFTB-based and uses the LATTE Hamiltonian method \cite{latte}. It shows good scaling with system size (Table \ref{table: waterboxes}, Figure \ref{fig:waterboxTimings}) and conserves energy over the 20 picosecond maximum simulation time, both for non-reactive test systems and for the reactive ORR systems, tested below (Figure \ref{fig:nve}).

\subsection{Extended Lagrangian Born-Oppenheimer MD}
\label{xlbomd}
The extended-Lagrangian Born-Oppenheimer MD method provides a smooth evolution of the charge which appears as a dynamical variable in the equations of motion \cite{Niklasson21}. For non-reactive systems, XL-BOMD also allows us to use a relatively long integration time step, as in regular Born-Oppenheimer MD, while maintaining system stability. This is of particular importance when transitioning to studies of reactive chemical systems, which, in general, require shorter integration time steps compared to the regular non-reactive MD simulations often used to sample statistical averages. A shorter integration time step is necessary to capture and analyze rapid local changes in the molecular motion and electronic structure as the reactions occur. The fast computational speed of XL-BOMD makes it feasible to employ these shorter time steps for the reactive simulations of the ORR, performed in this work.

We will briefly present the XL-BOMD method as explained elsewhere \cite{Niklasson17, Niklasson21}. In regular Born-Oppenheimer molecular dynamics, the potential energy surface, ${U}_{\rm BO}(\textbf{R})$ with ${\bf R} = \{R_i\}_{i=1}^N$ being the nuclear coordinates for each atom $i$, is determined through an iterative self-consistent optimization of a non-linear energy function of the electron density or charge distribution. 
In XL-BOMD, $U_{\rm BO}({\bf R})$ is replaced by a \textit{shadow} -- or approximated -- potential energy surface, ${\cal U}_{\rm BO}(\textbf{R},n)$, which is based on the relaxed ground state of an approximate shadow energy functional. The shadow energy functional is based on a linearization of the regular energy density functional around an approximate ground state density, $n(\textbf{r})$. Thanks to this linearization, the optimized density of the $n$-dependent ground state, $\varrho[n](\textbf{r})$, of the linearized energy functional is given directly through a single diagonalization, avoiding expensive SCC iterations. This reduces the cost and avoids convergence errors. The density in XL-BOMD, $n(\textbf{r})$, is then included as a dynamic field variable that evolves through a harmonic oscillator that is centered around the exact density of the ground state, or the best approximation available of the exact ground state, that is, $\varrho[n](\textbf{r})$. In this way, ${\cal U}_{\rm BO}(\textbf{R},n)$ closely follows the exact surface of the potential energy with an error that scales quadratically as ${\cal O}(||\varrho[n](\textbf{r})-n(\textbf{r})||^2)$ \cite{Niklasson17}. The accuracy of the shadow potential can be further improved by a first-level update \cite{ANiklasson23} which guarantees energy conservation for systems which are highly unstable when using SCC.

The extended Lagrangian functional, $L(\mathbf{R}, \dot{\mathbf{R}}, n, \dot{n})$, includes the nuclear kinetic energy, the shadow potential energy surface, the motion of the electron density, and a harmonic oscillator for the electron density centered around $\varrho[n](\textbf{r})$. The harmonic oscillator also includes a metric tensor, $T = K^T K$, given through an integral kernel, $K$, which is defined as the inverse Jacobian of the residual function, $\varrho[n](\textbf{r})-n(\textbf{r})$. This kernel ensures that the density variable $n({\bf r})$ oscillates closely around the exact, fully optimized Born-Oppenheimer ground state density, $\rho_{\rm min}({\bf r})$.

In our XL-BOMD finite-dimensional DFTB-based formulation \cite{Aradi15, Niklasson17}, the continuous charge density, $n(\textbf{r})$, is represented by a set of net Mulliken charges, ${\bf n} = \{n_i\}_{i=1}^N$, where $N$ is the total number of atomic sites, $i$, in the system. The equations of motion for this DFTB-based XL-BOMD scheme govern the evolution of both the coarse-grained atomic charge distribution, ${\bf n}$, and the corresponding nuclear configuration, ${\bf R}$. These equations, derived in a classical adiabatic limit assuming $\textbf{n}({\bf r})$ evolves on a fast time scale\cite{Niklasson14, Aradi15, Niklasson17} relative to the nuclear degrees of freedom, are 
\begin{eqnarray}
    M_i \ddot{R}_i & =& -\left. \frac{\partial {\cal U}_{\rm BO}(\mathbf{R}, \textbf{q}[\textbf{n}])}{\partial R_i}\right\vert_{\textbf{n}}, \notag \\
    \ddot{\textbf{n}}& = &-\omega^2 K(\textbf{q}[\textbf{n}] - \textbf{n}).
    \label{EOM_DFTB} 
\end{eqnarray}
\normalsize
\noindent 
Here, $M_i$ denotes the nuclear mass of atom $i$ and $\textbf{q}[{\bf n}] = \{q_i[{\bf n}]\}_{i=1}^N$ is a vector containing approximate ground state occupations for each atom that are given from the optimization of the linearized energy expression, that is, the coarse-grained Mulliken net charge version of $\varrho[n]({\bf r})$ above. The nuclear equations of motion are integrated with the standard Verlet scheme, and the electronic equation of motion is integrated with a modified Verlet scheme that includes a weak dissipative term \cite{Niklasson09, Zheng11, Niklasson17} that removes the accumulations of numerical noise and helps keep the electronic and nuclear degrees of freedom synchronized. At the first QMD time step, the extended dynamical variables for the electronic degrees of freedom, ${\bf n}(t_0)$, are set to the fully converged regular Born-Oppenheimer ground state densities. The deviation of charge from the exact ground state, which can be estimated by the residual function, scales with the size of the integration time step, $\delta t$, as $||\textbf{q}[\textbf{n}]-\textbf{n}|| \propto \delta t^2$ \cite{Niklasson17}.

In the DFTB-based formalism for XL-BOMD, the kernel $K$ is represented by an $N\times N$ matrix

\begin{equation}
    K = \left(\left \{ \frac{\partial \textbf{q}[\textbf{n}]}{\partial {n_i}}\right \}_{i=1}^N - I \right)^{-1} = J^{-1},
    \label{Kernel_Matrix} 
\end{equation}
\noindent
where $\{ \partial \textbf{q}[\textbf{n}]/\partial {n_i}\}_{i=1}^N$ is a set of $N$ column vectors each having length $N$ containing occupation response derivatives per atom; $I$ is the identity matrix; and $J$ is the Jacobian matrix of the residual function. The kernel $K$ appears as a preconditioner that acts on the residual function of Equation (\ref{EOM_DFTB}) and keeps ${\bf n}$ oscillating close to the exact fully-optimized density of the ground state. The kernel helps to keep the integration stable for simulations involving highly reactive chemical systems. Recalculating the full inverse Jacobian matrix in Equation (\ref{Kernel_Matrix}) at each time step is computationally too expensive to be practical. We therefore simplify this matrix computation by using a preconditioned rank-$m$ ($m<N$)  approximation of $K$ \cite{Niklasson17, ANiklasson20}. As a preconditioner, we use the exact kernel, $K_0$, from the first time step. The rank-$m$ update can be calculated from quantum perturbation theory \cite{Niklasson15, ANiklasson20} and the rank can be adjusted adaptively. Typically, $m \leq 4$ is used during simulation based on the accuracy of a preconditioned low-rank resolution of identity acting on the residual error, $\textbf{q}[\textbf{n}] - \textbf{n}$. If this error is still too large following a rank-4 update, the preconditioner, $K_0$, is recalculated. In general, this adds little extra cost to an average time step, but significantly improves the stability of the resulting molecular trajectories. 

\subsection{Applying an Electrochemical Potential Bias} \label{section:bias}

To modify the electrochemical potential of the electrode, we develop an approach to alter the onsite energies of the orbitals of these atoms. This method either stabilizes or destabilizes the electrons in the selected atoms (which, in our case, constitutes the NG sheet) depending on the bias applied. This approach is used in subsequent simulations to introduce an electric potential difference between the cathode and the solution. It is distinct from constant potential methods \cite{Lozovoi01, Taylor06, Yeh11, Bonnet12, Hormann19, Melander19, Schwarz2020}, and the ensemble used for all simulations in this work is \textit{NVT}.

Formally, the shifted Hamiltonian is given by
\begin{equation}
H_{i\alpha,j\beta} = H_{i\alpha,j\beta} + \frac{1}{2} (S_{i\alpha, j\beta} V_{j} + V_{i} S_{i\alpha, j \beta}),
\end{equation}
where $V_i$ are the shifts to the onsite orbital energies for each atom $i$. It should be noted that the chemical potential will be displaced by a constant $\phi$ in the event that all $V_i$ are equal to $\phi$. 
This implementation allows us to set a potential difference between the electrode and the solution, effectively raising or lowering the electrochemical potential of the electrons in the NG sheet. Using this method, we are able to adjust the adjust the driving force of the reaction (i.e., adjust the overpotential) and investigate mechanisms of ORR catalysis under diverse conditions. Examples of simulations with applied bias are presented in the "Simulations With Applied Bias" section.

\subsection{MD Simulation System Set Up} \label{system_setup}
\begin{figure}
    \centering
    \includegraphics[width=0.5\columnwidth]{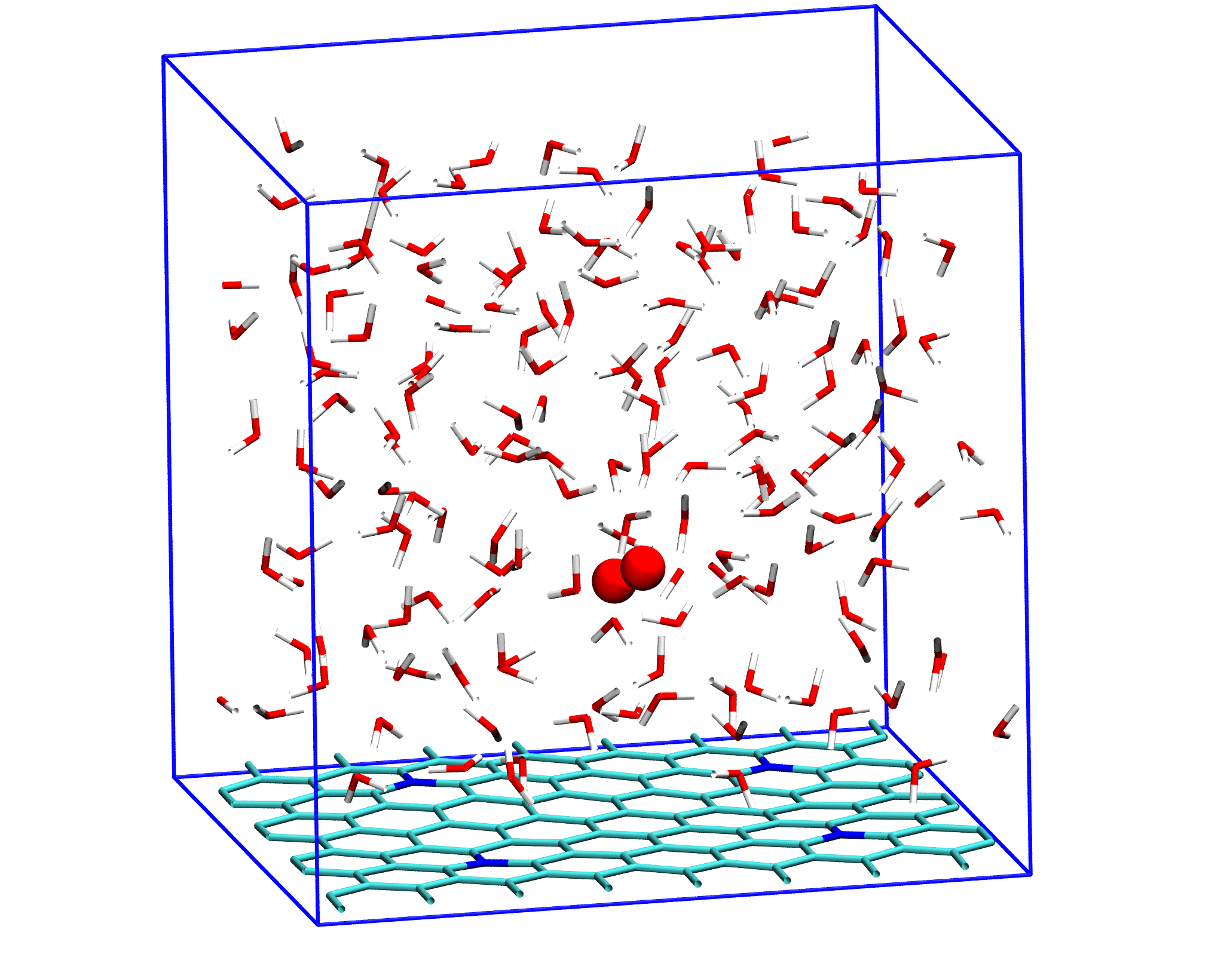}
    \caption{Full simulation system containing NG, O$_2$, and water. Water molecules and the NG sheet are shown using a \textit{dynamic bonds} representation with a cut-off of 1.8 \AA, O$_2$ is shown as spheres \cite{Humphrey96}. C, H, N, and O atoms are displayed in teal, white, blue, and red, respectively. The periodic bounding box (19.686 x 17.048 x 20.000 \AA$^3$) is shown in blue.}
    \label{fig:mainsystem}
\end{figure}

The main system we used consisted of a 19.686 x 17.048 x 20.000 \AA$^3$ box containing O$_2$, 164 water molecules (solvent), and the NG sheet. The NG sheet was set up along one face of the water box. A total of four nitrogen atoms were included in the NG sheet, each fully bound to carbon atoms (graphitic nitrogen doping). This system (622 atoms in total with 1840 electrons) is shown in Figure \ref{fig:mainsystem}. Variants containing molecular oxygen in different environments were also constructed for comparison purposes. These variant "control systems" are as follows: O$_2$ with NG in vacuum, O$_2$ in water, and O$_2$ with pure graphene in water. The starting coordinates for all three control systems were taken as pieces of the main system.

Simulations were run for up to 20 ps (a set 10 ps for simulations without applied bias and variable lengths up to 20 ps until reactions were observed for simulations with applied bias) using a 0.1 fs time step and periodic boundary conditions. A Langevin thermostat was used to maintain a constant temperature of 300 K throughout the simulations \cite{Sivak} with a friction coefficient of $\gamma = 0.01$ fs$^{-1}$. This thermostat belongs to a special class of stochastic Langevin integrators which correctly describe free diffusion and the statistical mechanics of linear systems \cite{NGronbechJensen20, JFinkelstein21}. Alternative methods in this class may be preferable for larger values of $\gamma \Delta t$, such as GJF \cite{NGJensen13}, and may be explored in future work. A $\beta$ value of 40 eV$^{-1}$ (corresponding to room temperature) was used to ``smear'' the Fermi distribution and improve the stability of the QMD simulations. Simulating with the shadow XL-BOMD method makes it possible to use a physically reasonable electronic temperature, which improves simulation accuracy compared to the much higher electronic temperatures which are often used to control SCF convergence. Rank-\textit{N} updates to the kernel were performed with a rank of 4. Simulations were run using Nvidia A100 GPUs as accelerators and each MD time step took $\sim$0.15 seconds of wall-clock time.

\section{Results and Discussion}\label{results}

We simulated the ORR under varied applied potential bias conditions using the XL-BOMD simulation method with a low-rank updated kernel approximation. The goal of these simulations was to 
 verify the robustness and reliability of XL-BOMD as a valid simulation tool for reactive systems and to 
shed light on the details of the ORR on NG as a function of applied electrochemical bias. 

\subsection{Preconditioning XL-BOMD}

\begin{figure}
    \centering
    \includegraphics[width=0.7\columnwidth]{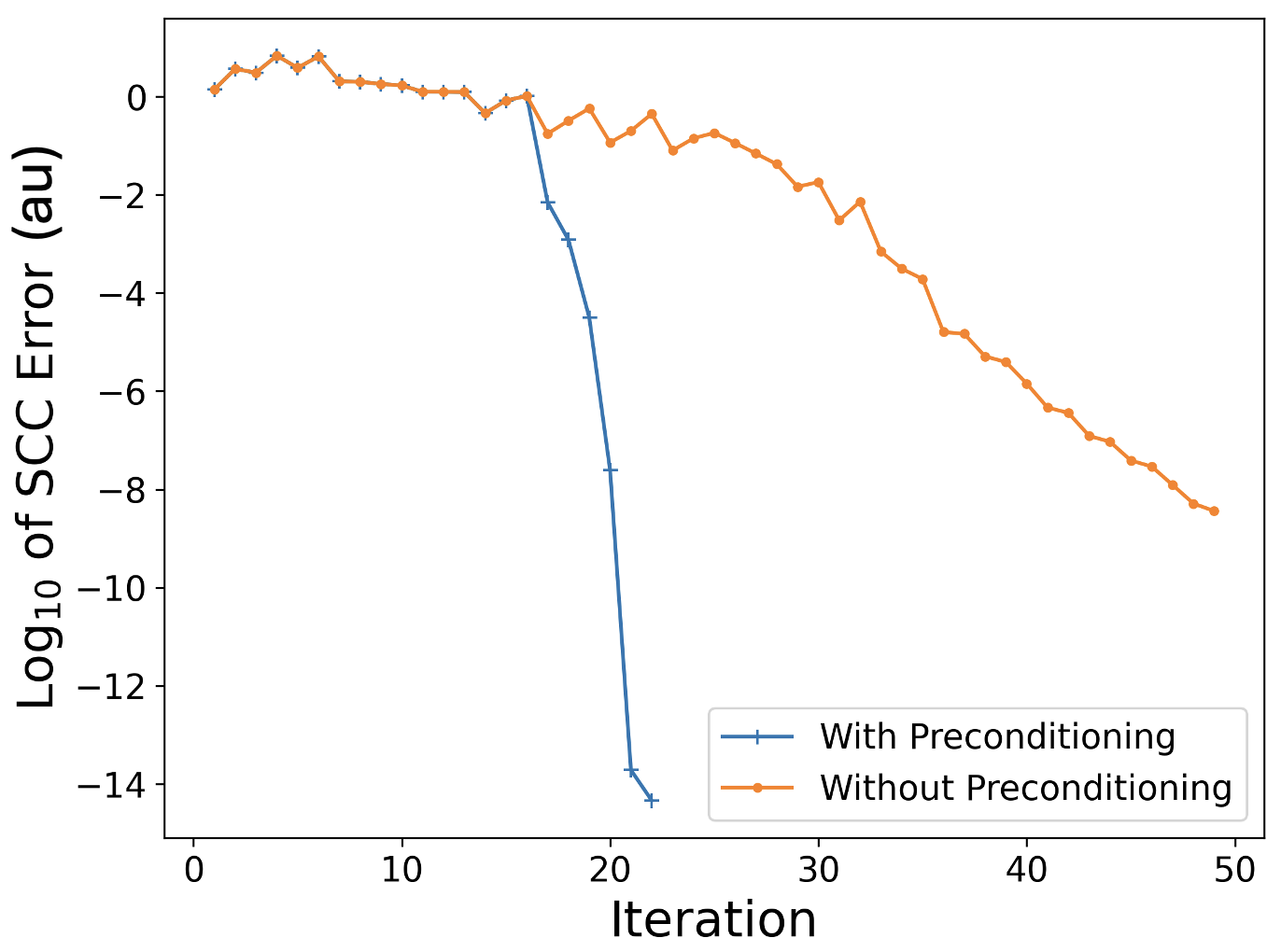}
    \caption{Logarithm of the SCC Error (in arbitrary units) as a function of the SCC iteration number. The results of using and not using preconditioning are shown in blue with pluses and orange with dots, respectively. This figure shows the fast convergence attained when preconditioning (using a pre-calculated kernel) is used, which typically follows a second order power law ($\log_{10} (\mathrm{Error}_{i}) \simeq 2 \log_{10}(\mathrm{Error}_{i-1}) $) beyond a sufficiently small (10$^{-3}$) SCC error. This verification has been done for the case of NG with water and molecular oxygen.}
    \label{fig:scciterations}
\end{figure}

We used preconditioning with XL-BOMD to help improve the stability of the QMD simulations. The preconditioned low-rank kernel approximation in XL-BOMD can also be used as a SCC-accelerator \cite{ANiklasson20}. Figure \ref{fig:scciterations} shows the SCC iterations required for the initial convergence of the main system with and without preconditioning. This result demonstrates how preconditioning can greatly accelerate the SCC convergence required in the first time step to initialize the charges of the XL-BOMD simulation. After the first time step, no SCC optimization is performed.

\subsection{Oxygen Species Mulliken Charge Analysis}

As mentioned in the "Density Functional Tight Binding Theory" section, we need to compute the Mulliken charges at every simulation time step to determine the electronic structure. 
Although for most purposes, Mulliken charges are just auxiliary variables, they can be used to gauge the overall charge of an atomic species. They also can be used to characterize species with different total charges and oxidation states, both for species adsorbed on an electrode and in solution.

To characterize oxygen-related species of interest in this study, we performed preliminary MD simulations and computed the average Mulliken charges. Each simulation was run for 1 ps using the same conditions described in the "MD Simulation System Set Up" section and each species was simulated in a water box without an NG sheet. Average Mulliken charge values along with standard deviations and oxidation states for the various oxygen species are shown in Table \ref{table: Mulliken}. 
\begin{table}
\caption{Mulliken charges in \textit{e} units (mean and standard deviation) within the DFTB level of theory and oxidation states for various aqueous oxygen species. Mulliken charges were computed based on 1 ps MD simulations of each species in an 19.686 x 17.048 x 20.000 \AA$^3$ water box.
For the peroxide anion (HOO$^-$), the oxygen of interest is bold and underlined. Species are ordered based on mean charge.}
\begin{tabular}{ c c c c }
 \hline
 Species &Mean&Standard Deviation&Oxidation State\\
 \hline
 OH$^-$ &-1.01&  0.02 & -2\\
 R-O$^{-}$ & -0.86 & 0.02 & -2\\
 HO\underline{\textbf{O}}$^-$  & -0.65 & 0.03 & -1\\
 H$_2$O & -0.65 & 0.01 & -2\\
 HO$\cdot$ & -0.59 & 0.03 & -1\\
 H\underline{\textbf{O}}O$^-$  & -0.46 & 0.05 & -1\\
 H$_3$O$^+$ & -0.44 & 0.01 & -2\\
 H$_2$O$_2$ & -0.32 & 0.01 & -1\\
 O$_2$ & 0.00 & 0.01 & 0\\
 \hline
\end{tabular}
\label{table: Mulliken}
\end{table}
The spread of average values ranges from OH$^-$ with an average Mulliken charge (in \textit{e} units) of -1.01 to O$_2$ with an average Mulliken charge of 0.00. 
Although there is not a strict correlation between the oxidation state and the average Mulliken charge, we can still use the latter as a means of identifying aqueous oxygen-related species within the context of a complex MD system. 
The Mulliken charges of these species can be expressed by combining selections of the following components with the tested oxygen: -H, -e$^-$, -H$^{+}$, -OH, -O$^{-}$, -O, -R, $\cdot$ (representing a radical). Solving a system of linear equations based on Table \ref{table: Mulliken} yields the following partial charges for these components: -0.33, -0.67, 0.23, 0.02, -0.13, 0.00, -0.19, and -0.26, respectively (Supporting Information, Table \ref{table: contributions}). 

\subsection{Simulations Without Applied Bias} \label{mdsims}
MD simulations for the main and three control systems (Methods) were run without applied bias (or said equivalently, with an applied bias of 0.0 eV) for 10 ps and results have been analyzed based on both geometry and Mulliken charges. 

\subsubsection{Control Systems}\label{controlSims}
 No oxygen reduction was observed during the simulation timescale for any of the three control systems. The system containing O$_2$ with pure graphene in water showed an initial slow electron injection, but no evidence of oxygen cleavage - the results for all three control simulations are shown in the Supporting Information (Figure \ref{fig:controlcomp}).

\subsubsection{Main System}
\begin{figure}[ht]
    \centering\includegraphics[width=0.7\columnwidth]{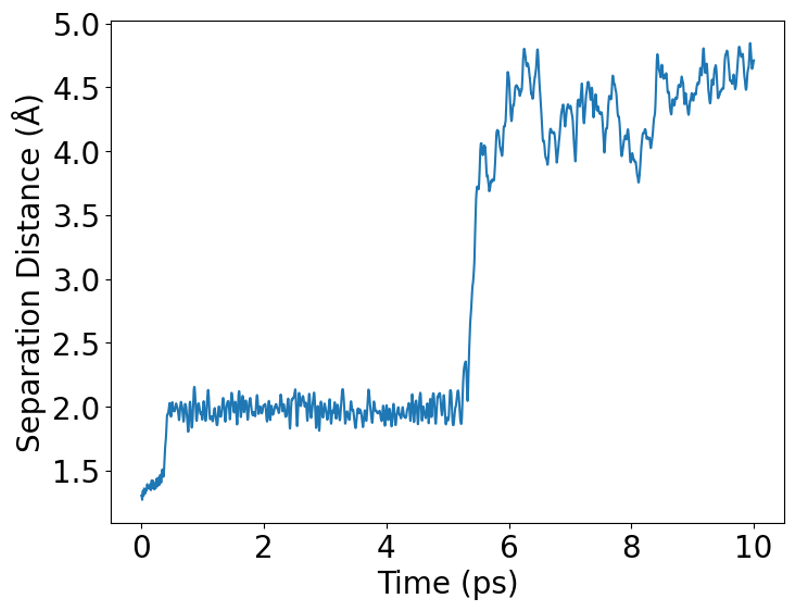}
    \caption{\small Distance between oxygen atoms in molecular oxygen as a function of time for a 10 ps QMD simulation. At around 6 ps, a sharp increase in the O-O separation distance is clearly observed, demonstrating cleavage of the O-O bond.}
    \label{O_2}
\end{figure}

\begin{figure}[ht]
    \centering
    \includegraphics[width=0.7\columnwidth]{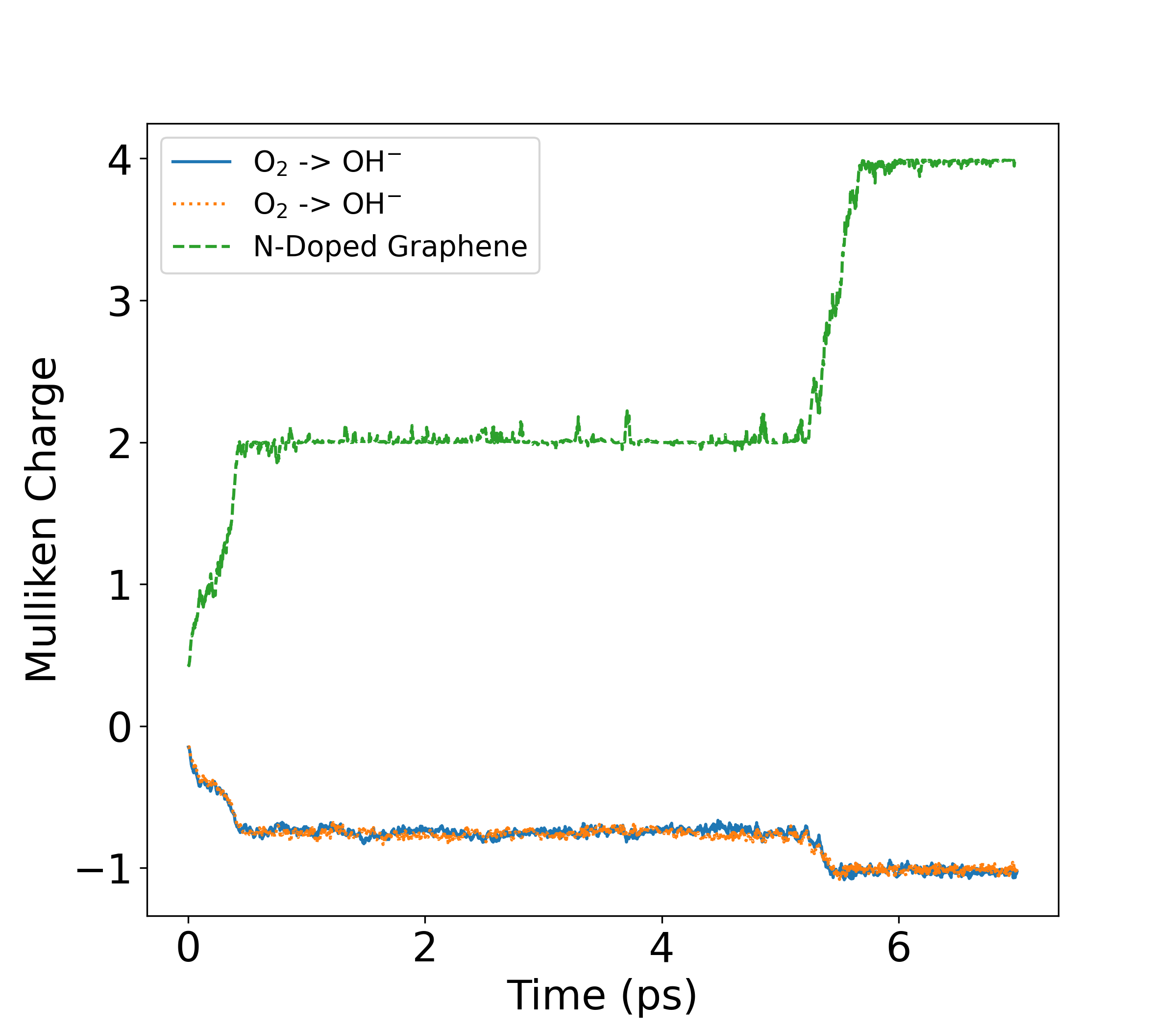}
    \caption{Mulliken charges in \textit{e} units for the two oxygen atoms that are initially part of the oxygen molecule (solid blue and dotted orange lines) and the sum of Mulliken charges for all NG sheet atoms (dashed green line) are shown.  
    }
    \label{fig:o2ncpop}
\end{figure}

For the main system (containing NG, O$_2$, and water), oxygen reduction was observed within 6 picoseconds. 
Figure \ref{O_2} shows a plot of the distance between the atoms of molecular oxygen versus simulation time. Around 6 ps, a sharp increase in the O-O separation distance is clearly observed, demonstrating cleavage of the O-O bond.
Figure \ref{fig:o2ncpop} shows the Mulliken charges of both atoms of molecular oxygen as they undergo reduction (solid blue line and dotted orange line). We can infer that both of these oxygen atoms get transformed into OH$^-$ after $\sim$ 6 ps of simulation as a result of the reaction since both oxygen atoms have Mulliken charges which are close to the average charge shown in Table \ref{table: Mulliken} for an OH$^{-}$ oxygen.
The sum of the Mulliken charges of all atoms in the NG sheet is shown as a green dashed line on the same plot. Two important points are noticeable here: 1) A total of four electrons are transferred to the solution from the NG sheet, and, 2) The electron transfer proceeds in two steps. The final step can be clearly observed as a two-electron transfer beginning at around 6 ps. The electrons that are not transferred directly to the molecular oxygen are transferred to the solvent, which acts as a buffer of electrons during the reduction steps. These reduction steps can be written  as follows: 
\begin{equation} \label{step1}
    {\rm 2 e^- + O_2 \rightarrow O_2^{2-} }
\end{equation}
\begin{equation} \label{step2}
    {\rm 2 e^- + O_2^{2-} + 2H_2O \rightarrow 4OH^- }
\end{equation}
\begin{center}
\rule{0.5\textwidth}{0.4pt}
\end{center}
\begin{equation} \label{fullrxn}
    {\rm 4 e^- + O_2 + 2H_2O \rightarrow 4OH^- }
\end{equation}
The integer charges assigned in reactions \ref{step1}-\ref{fullrxn} were determined by comparing the Mulliken charges of the species with the data presented in Table \ref{table: Mulliken}. It is crucial to note that, according to the Mulliken charge analysis, the transfer of four electrons ultimately culminates in the formation of four OH$^-$ anions, which occurs via water-mediated (outer-sphere) electron transfer in this case. Even though the oxygen atoms in water molecules cannot be further reduced (i.e., to a lower oxidation state), from the Mulliken perspective, the partial charges of these oxygen atoms must go down by approximately -0.36 for them to be transformed into OH$^-$ anions. This means that, when analyzed from the DFTB and Mulliken charge viewpoint, an oxygen atom in OH$^-$ appears to be more reduced than an oxygen atom in water since it has a more negative partial charge. Therefore, although the Mulliken charge analysis is useful as a tool to identify species by referencing Table \ref{table: Mulliken}, it does not formally indicate the oxidation state. 

\begin{figure}[ht]
    \centering        
    \includegraphics[width=0.7\columnwidth]{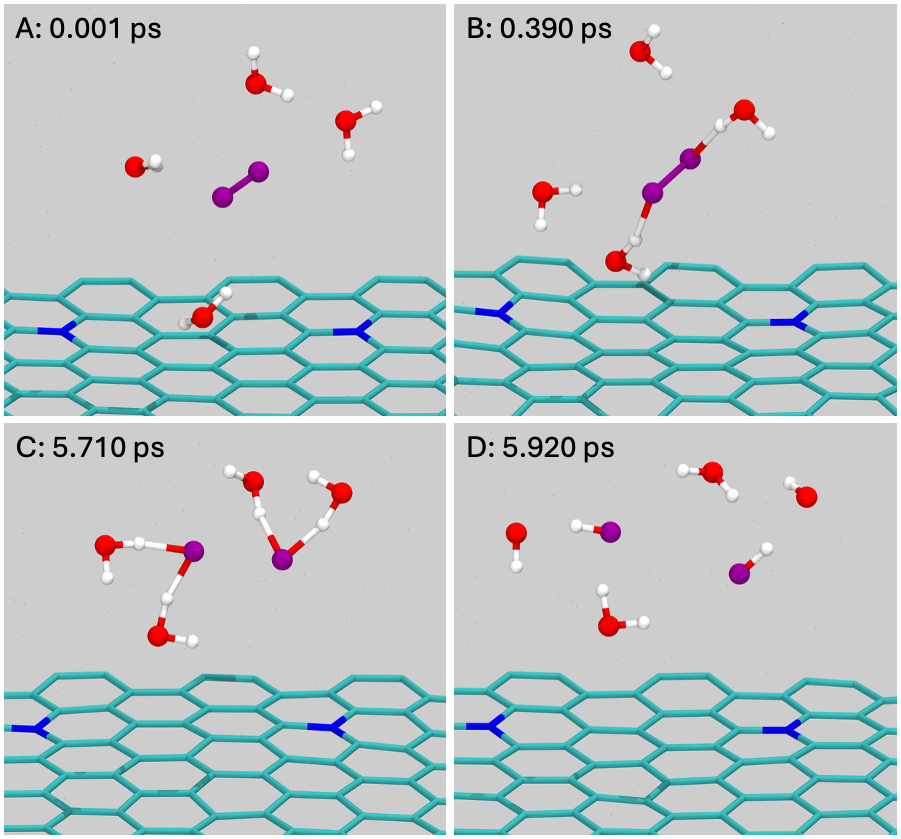}
    \caption{\small System configuration at simulation times of A) 0.001 ps, B) 0.390 ps, C) 5.710 ps D) 5.920 ps. C, N, H, and O atoms are depicted in teal, blue, white, and red colors respectively. The oxygen molecule is depicted in purple.
    For clarity, only 4 of the 164 water molecules are shown. Dynamic bonds shown are $\leq$ 1.8 \AA \hspace{1mm} in length, except for the O$_2$ bond, which is $\leq$ 2.0 \AA \hspace{1mm} in length.} 
    \label{fig:10psframes}        
\end{figure}

Figure \ref{fig:10psframes} shows key frames of the simulation of the ORR with NG via an outer-sphere mechanism. The oxygen molecule is shown as purple spheres for distinctiveness and the four water molecules that most directly participate in the reaction are shown as red and white spheres (oxygen and hydrogen atoms, respectively). The NG sheet is shown as dynamic bonds with carbon and nitrogen atoms depicted in teal and blue, respectively, and additional waters are not shown for clarity. Frame A shows the initial configuration of the system, with the molecular oxygen atoms tightly bound (1.26 \AA\, separation distance). Frame B shows what appears to be the formation of a solvated peroxo species with an O-O separation distance of 1.70 \AA. This time point corresponds to the injection of the first two electrons from the NG sheet (Figure \ref{fig:o2ncpop}).
Comparison of oxygen Mulliken charges at this time point with the average Mulliken charge values in Table \ref{table: Mulliken} suggests that this species may be more similar to O$_2^{2-}$ than HOO$^-$ or H$_2$O$_2$ since both oxygen atoms have Mulliken charges $\sim$0.6-0.7. 
Because of the relatively long bond distance of $\simeq$2.0 \AA$\,$ in this section of the simulation, this species could also be interpreted as two weakly bonded atomic oxygen radical anions (O$^{\cdot-}$ ) coordinated by water molecules. A superoxide anion can be ruled out at this point in the simulation since the Mulliken charge on any of the oxygen atoms of a superoxide anion would be -0.36 - the average between the charges of the O and O$^-$ in that anion, which can be computed using the component contributions from Table \ref{table: contributions}. 
Frame C shows a further separation of the O-O bond of molecular O$_2$ (separation distance of 2.52 \AA) and formation of a 2O$^{\cdot-}$ + 2H$_2$O transition state. This point closely corresponds to the beginning of the second injection of electrons from the NG sheet as well (Figure \ref{fig:o2ncpop}). Frame D shows the formation of the OH$^-$ anions from O$_2$ and two water molecules. 
At the end of the simulation, O$_2$ was fully reduced to four OH$^-$ anions and a total of 4 electrons were transferred from the NG sheet to the solution. This electron transfer occurs under conditions of fully converged, self-consistent electronic structure. 

It should be noted that we have also observed an alternative mechanism proceeding via an HOO$^{-}$ species instead of a peroxo species. The most important configurations for this alternative mechanism are detailed in the Supporting Information (Figure \ref{fig:overlay-hydroperoxo}). In essence, reactions \ref{step1} - \ref{fullrxn} are reformulated as follows:

\begin{equation} \label{step1-hoo}
    {\rm 2 e^- + O_2 + H_2O \rightarrow HOO^{-} + OH^{-} }
\end{equation}
\begin{equation} \label{step2-hoo}
    {\rm 2 e^- + HOO^{-} + H_2O \rightarrow 3OH^- }
\end{equation}
\begin{center}
\rule{0.5\textwidth}{0.4pt}
\end{center}
\begin{equation} \label{fullrxn-hoo}
    {\rm 4 e^- + O_2 + 2H_2O \rightarrow 4OH^- }
\end{equation}

\begin{figure}
\centering
\includegraphics[width=4cm]{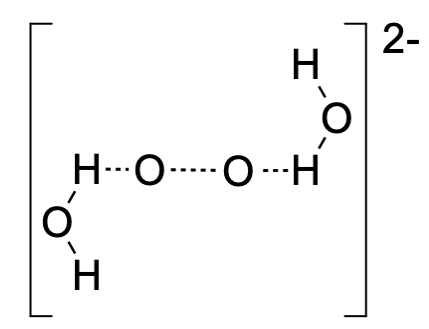}
\caption{Skeletal representation of the transient state in the observed ORR reactions.}
\label{fig:watero2inter}
\end{figure}

While the parameters that control reaction mechanisms are not fully understood, these simulations do depend on the initial randomized velocities established by the 300 K Langevin thermostat. One could say that the transient state formed by O$_2$ and two water molecules could be more generally depicted with the representation in Figure \ref{fig:watero2inter}, leading to O$_2^{2-}$ + 2H$_2$O, 2O$^{\cdot-}$ + 2H$_2$O, HOO$^{-}$ + OH$^{-}$ + H$_2$O, or H$_2$O$_2$ + 2OH$^{-}$ combinations that are each plausible depending on the distances of the ``dotted'' bonds and the Mulliken charges on the oxygen atoms. It is worth noting that transient species formation in the current simulations as part of the observed outer-sphere mechanism might be due to limitations of the simulation conditions, where there is no steady state condition, there are no electron reservoirs, and no constant bias is applied. Since our method is currently implemented with the tight binding approximation (Methods), all of the reactions we observe are consistent with that level of theory; our present studies provide a foundation for a deeper investigation of mechanisms using a variety of theories. 

\subsubsection{Extended Systems}
To investigate the ORR using a larger NG sheet with multiple O$_2$ molecules in solution, two extended systems were created by duplicating the main system in the $x$- or $x$- and $y$-directions. These simulations also showed evidence of the ORR and electron transfer from the NG sheet through water via an outer-sphere mechanism in simulations at zero applied bias and +1.0 eV applied bias (Figures \ref{fig:oosepdouble}-\ref{fig:oosep2x2plus1}). A further description of these extended system simulations is available in the Supporting Information.

\subsubsection{Reactivity Indicators}
Both the HOMO-LUMO energy gap and the residual error function are important elements of the electronic structure that can be monitored and used to provide more information on the processes occurring in the simulated system. In this context, the HOMO is the highest occupied molecular orbital and the LUMO is the lowest unoccupied molecular orbital. The residual error function is the norm of the difference between the exact ground state charge vector ($\mathbf{\varrho} [\mathbf{n}(\mathbf{r})]$) and the approximate charge vector ($\mathbf{n}(\mathbf{r})$) and is used as an auxiliary variable in the shadow energy function, as described in the "Extended Lagrangian Born-Oppenheimer MD" section. It is important to note that these residual values are unique to the extended Lagrangian framework. Both of these aforementioned elements can be used to identify reactions.

\begin{figure}[ht]
    \centering
    \includegraphics[width=\columnwidth]{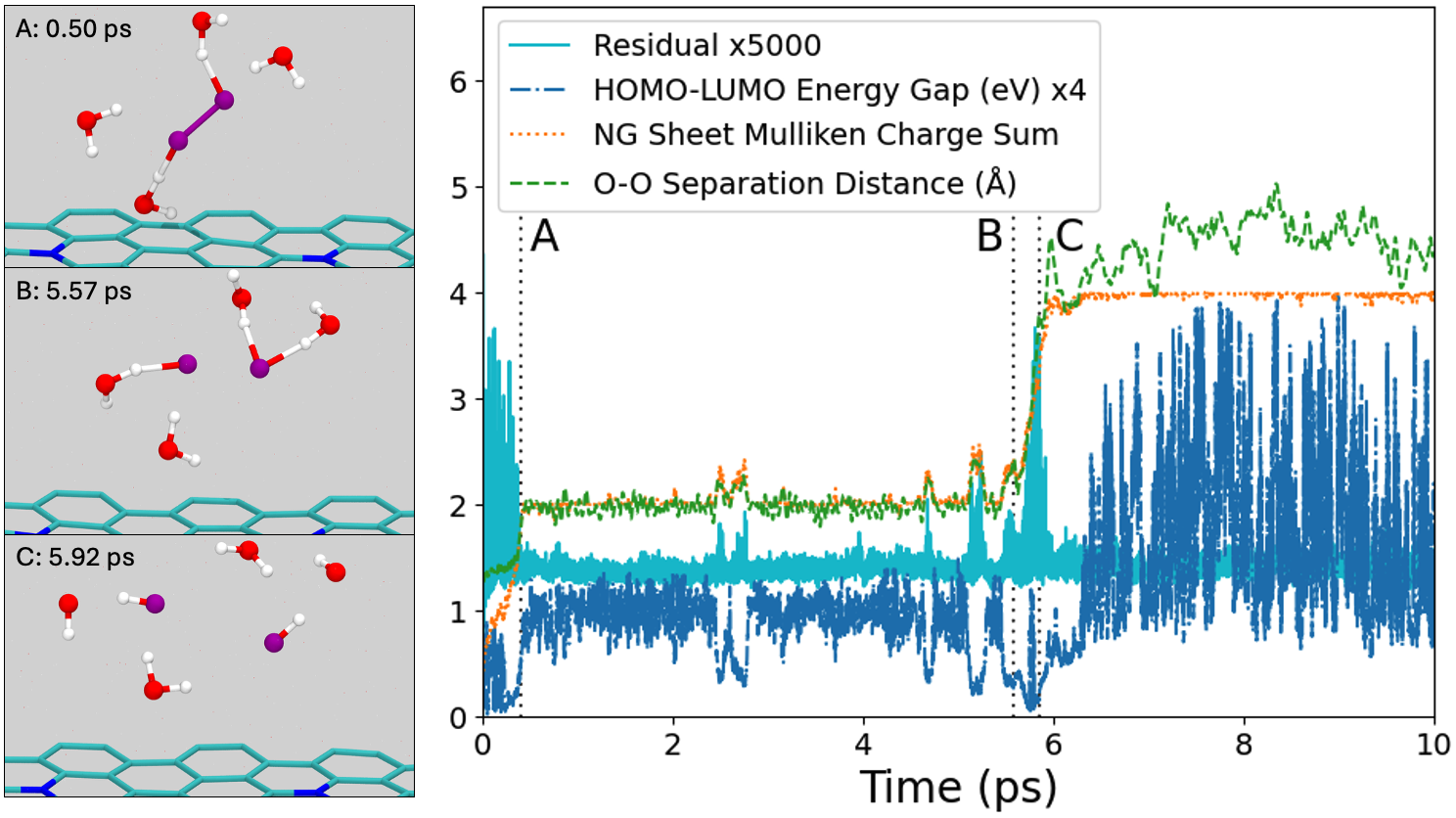}
    \caption{Overlay of four plots: norm of the residuals ($||q[n] - n||$) multiplied by 5000 to show features (solid teal), HOMO-LUMO energy gap multiplied by 4 to show features (dot-dashed blue), separation distance between the two oxygen atoms that originally form molecular oxygen (dashed green), and the sum of the NG sheet atom Mulliken charges in \textit{e} units (dotted orange). Panels A, B, and C to the left correspond to gray dotted vertical lines A, B, and C on the plot.}
    \label{fig:overlay}
\end{figure}

Figure \ref{fig:overlay} shows the HOMO-LUMO energy gap, the norm of the residuals, the sum of Mulliken charges for all atoms in the NG sheet, and the O-O separation distance over simulation time. It is known that when reactions occur, the HOMO-LUMO gap tends to close (i.e., has a value near zero) \cite{Bredas14}. The gap drops near zero at two clear points during the simulation - 0.40 ps and 5.85 ps. These time points coincide with the two waves of electrons that are injected into the solution from the NG sheet. It is important to note that the O-O bond does not fully cleave until the second pair of electrons is injected, as previously analyzed in Figure \ref{O_2}. 

The norm of the residuals within the context of XL-BOMD can also be used to identify reactions, since this norm tends to increase when reactions occur \cite{Niklasson21}. Although not as clear as the HOMO-LUMO gap closures, we can see that the norm of the residual function is large near points A and C in Figure \ref{fig:overlay}. These time points show critical configurations of the reactive species, corresponding to the points where the HOMO-LUMO gap goes to zero. The same behavior of the HOMO-LUMO gap and the norm of the residuals is observed for when the reaction forms HOO$^{-}$ as an intermediate (Figure \ref{fig:overlay-hydroperoxo}).
\subsubsection{Electron Hole Delocalization}
\begin{figure}
    \centering
    \includegraphics[width=0.7\columnwidth]{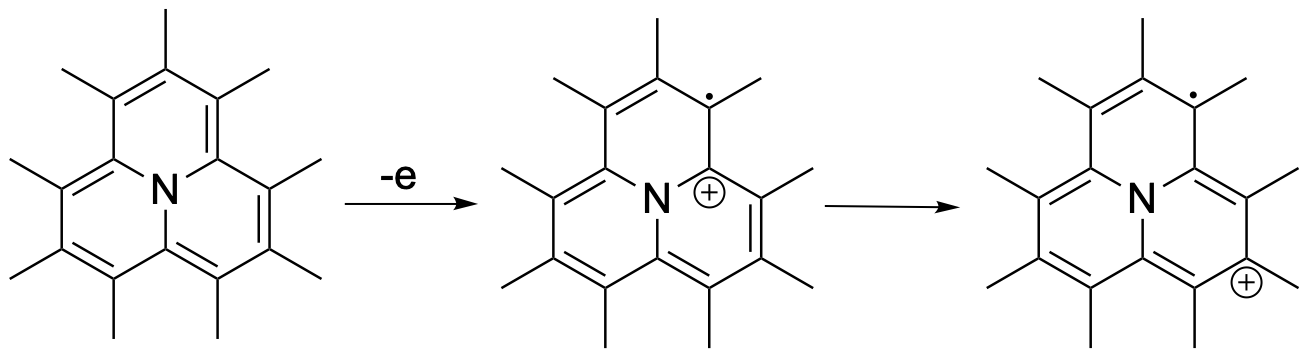}
    \caption{Possible resonance structures to explain the localization of the holes (positive charges) on the carbon atoms surrounding nitrogen doping atoms in the NG sheet following oxidation. Mulliken charge averages of nitrogen atoms and neighboring carbon atoms are available in Table \ref{table:mus}.}
    \label{fig:resonance}
\end{figure}

As molecular oxygen is reduced, the NG catalyst undergoes an oxidation, on the basis of the evidence discussed above. This oxidation (NG $\rightarrow$ NG$^{4+}$ + 4e$^-$) leads to four electron holes (net positive charges) in the NG sheet. 
The resulting electron holes are mostly localized over the carbon atoms directly bonded to the nitrogen atoms (Table \ref{table:mus}).
As a possible explanation for this, we have proposed the resonance structures shown in Figure \ref{fig:resonance}. In these structures, we see that the holes are positioned such that they could be de-localized across the entire NG sheet. Additional possible resonance structures are shown in Figure \ref{fig:mesomeric}.

\subsubsection{Density of States}
\begin{figure}
    \centering
    \includegraphics[width=0.7\columnwidth]{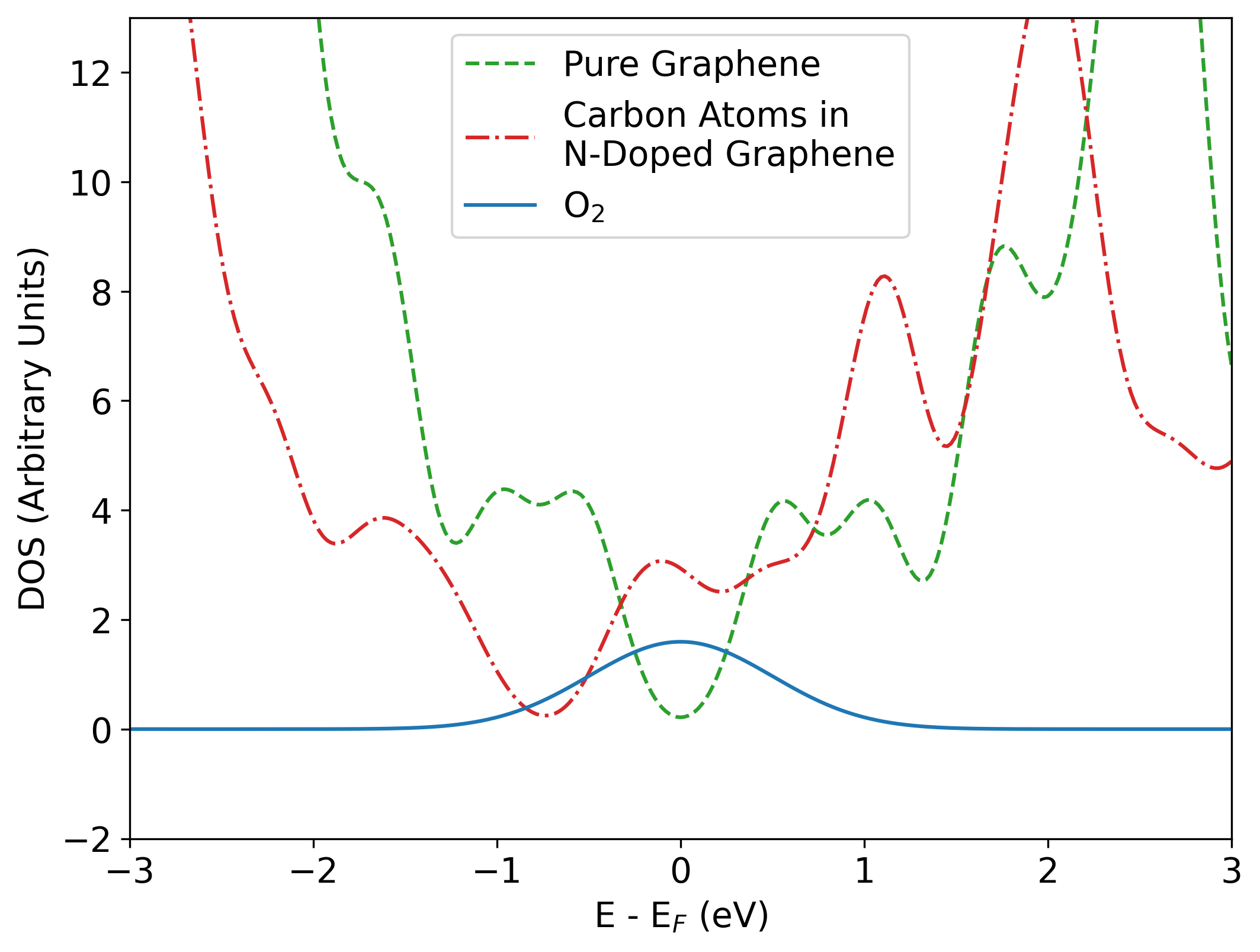}
    \caption{Total DOS for O$_2$ (solid blue) and pure graphene (dashed green) are shown along with the DOS corresponding to the carbon atoms bound to nitrogen dopant atoms in NG (dot-dashed red). Plots are shifted such that the $\mu_e$ of pure graphene is zero.
    }
    \label{fig:dos}
\end{figure}

 From our computational studies, we have shown that under the present simulation conditions, NG is a more active catalyst than pure graphene (see the "Control Systems" section). One possible explanation for this comes from the density of states (DOS) around the chemical potential of the electrons $\mu_e$ (Fermi level) for both materials.
 In Figure \ref{fig:dos}, we show the DOS of O$_2$ (solid blue), carbon atoms near nitrogen doping atoms in NG (dot-dashed red), and pure graphene (dashed green). 

Both graphene and NG have states with electrons that can be donated to O$_2$. However, in comparison to pure graphene, we can see that the effect of nitrogen doping is twofold - shifting $\mu_e$ to increase the reduction potential and increasing the density of states around $\mu_e$, thus providing a source of electrons. The change in the value of $\mu_e$ is visible in Figure \ref{fig:dos} as the difference between the minimum points in the red and green lines, while the increase in DOS is visible as the widening of the parabolic section around the minimum point of the NG plot compared to the same section of the pure graphene plot. 

%
\subsubsection{Outer-Sphere Mechanism}
An important point to note in these simulations without applied bias is that the oxygen molecule does not need to be chemisorbed onto the NG sheet to undergo reduction. Reduction occurs via an outer-sphere mechanism mediated by the solvent molecules. Compelling evidence indicates that molecular oxygen adsorption may not be essential for the ORR to take place on N-doped graphene. \cite{Mosyak96, Negre11, Choi14}.

Our initial XL-BOMD simulations without applied bias capture essential features of the oxygen reduction reaction at the NG–water interface. During O$_2$ cleavage, the catalytic surface underwent formal oxidation, indicating direct charge transfer between the solvated oxygen and the electrode. This redox process proceeded through an outer-sphere pathway, with molecular oxygen and the NG sheet acting as the oxidizing and reducing agents, respectively. While the lack of a steady-state current means these conditions do not represent a complete electrochemical reaction, the simulations nevertheless show clear atomistic dynamics of electron transfer at a solvated catalytic interface, which is beyond the reach of conventional approaches. This capability lays the foundation for the applied bias simulations that follow, where reaction conditions are systematically tuned to probe distinct mechanistic regimes and better understand reaction kinetics.

\subsection{Simulations With Applied Bias}\label{sec:appliedBias}

\begin{figure}
    \centering
    \includegraphics[width=\columnwidth]{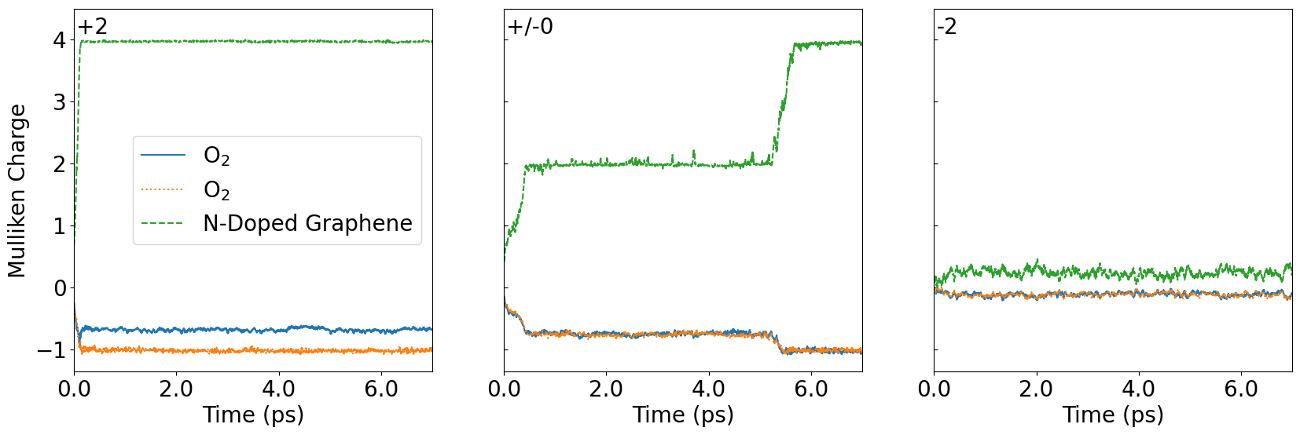}
    \caption{Initial 7 ps of simulation time for solvated O$_2$ on NG with and without applied $\mu_e$ shift. Applying a $\mu_e$ shift of +2.0 eV to the NG sheet (Left Panel) accelerates the outer-sphere reaction relative to not applying a shift (Center Panel); outer-sphere ORR occurs at the point in each simulation where the sum of Mulliken charges for the NG sheet reaches $\sim$4, signaling the loss of 4 electrons from the sheet. Applying a $\mu_e$ shift of -2.0 eV (Right Panel) eliminates the outer-sphere mechanism across the full 10 ps simulation time (Figure \ref{fig:biasMinus2}). All Mulliken charge values are presented in \textit{e} units.}
    \label{fig:bias-plusMin2}
\end{figure}

We investigated the effect of applying an electrochemical bias (Methods) on simulations of oxygen reduction; both the dynamics and the kinetics of reduction changed depending on the applied bias. Figure \ref{fig:bias-plusMin2} illustrates the Mulliken charge dynamics over a period of 7 ps of simulation applying three different $\mu_e$ shifts (+2.0, 0.0, and -2.0 eV) to the atoms of the NG sheet, demonstrating the influence of substantially altering the electrochemical potential of the cathode within the simulation. For all simulations with applied bias, the electrode bias is set at the beginning of the simulation and remains fixed. It is important to note that, although the potential applied to the NG sheet is held constant throughout the simulations, the ensemble remains canonical (\textit{NVT}, constant charge), not grand-canonical (\textit{$\mu$VT}, constant potential) since the chemical potential, $\mu$, of the system is not fixed, but adjusted in each time step to achieve charge neutrality (or, hypothetically, any fixed net charge).

Next, we sought conditions where reduction occurs by an inner-sphere mechanism: (1) We applied a negative shift of $\mu_e$(NG), hence, lowering the bias for electron injection from the NG sheet towards the oxygen molecule. In future work, we will explore adjusting the potential of the oxygen molecule following adsorption to the NG sheet, but for the present simulations, only the $\mu_e(\rm NG)$ is shifted, not the $\mu_e(\rm O_{ads})$. (2) We used a modified version of the main test system containing NG, O$_2$, and water with the O$_2$ molecule located closer to the NG surface. (3) We employed a steering method \cite{CorriganGrove25} to attract the oxygen molecule to the NG sheet and increase the likelihood of O$_2$ adsorption during simulation time. 

\begin{figure*}
    \centering
    \includegraphics[width=\textwidth]{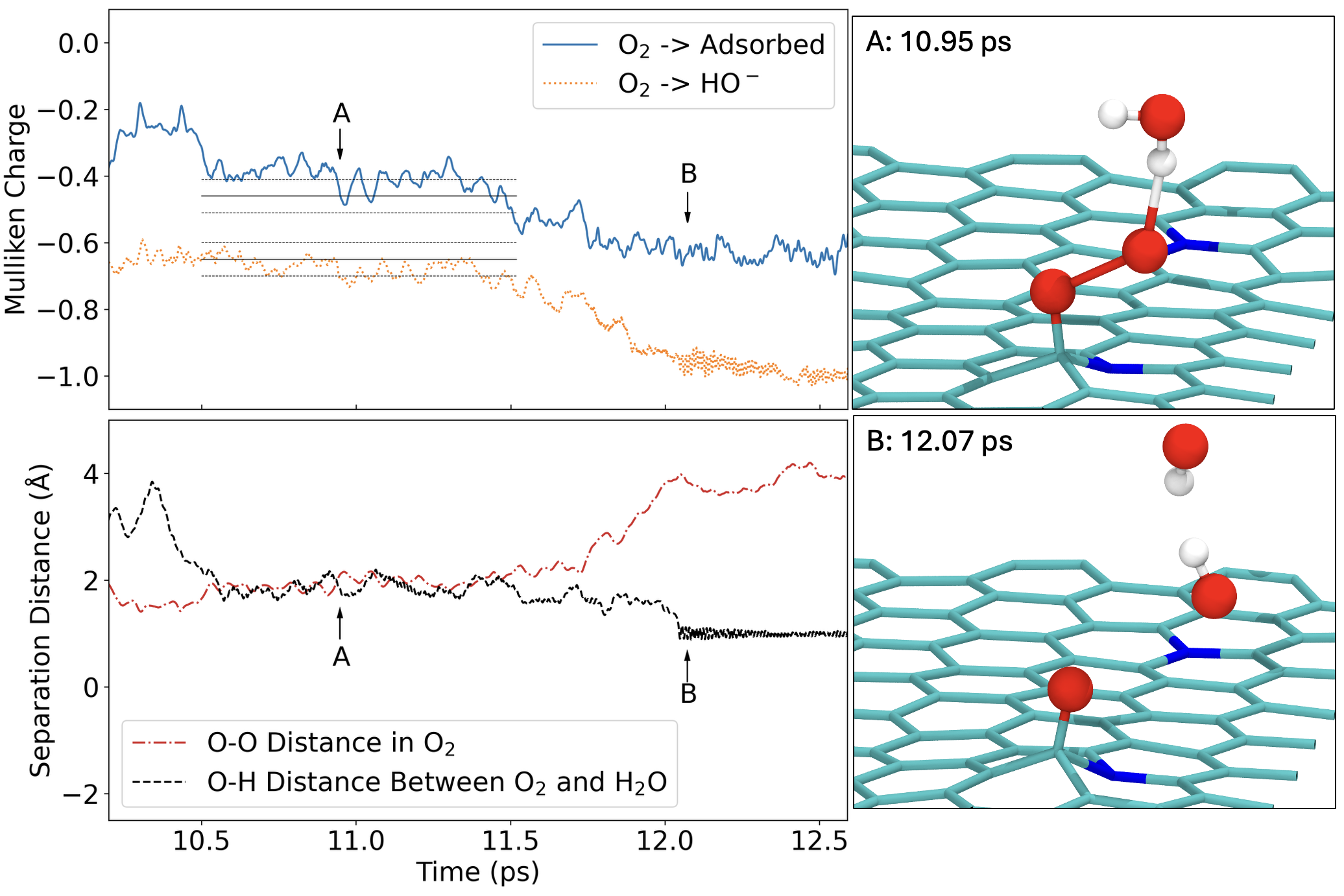}
    \caption{Adsorbed oxygen species observed in trajectory analysis and charge dynamics analysis for a simulation with an applied bias of -0.9 eV. Left, Top: Mulliken charges in \textit{e} units for both O$_2$ oxygen atoms for a 2.4 ps section of simulation time. Solid horizontal lines show the average Mulliken charges for H\underline{O}O$^-$ (-0.46) and HO\underline{O}$^-$ (-0.65) from Table \ref{table: Mulliken} while dotted horizontal lines show the standard deviations, also given in Table \ref{table: Mulliken}. Points A and B, labeled with arrows on the Mulliken charge plot, correspond to the A and B snapshots in the right panel VMD images. Left, Bottom: Separation distances between the two oxygen atoms that initially form O$_2$ (dot-dashed red) and between the non-adsorbed O$_2$ oxygen and the water hydrogen (dashed black).  Points A and B, labeled with arrows on the separation distance plot, correspond to the A and B snapshots in the right panel VMD images. Right: VMD representation of the adsorbed O$_2$ intermediate at 10.95 ps (A) and 12.07 ps (B). At point (A), a peroxo intermediate forms, which is reduced to form 2 OH$^-$ anions and a single adsorbed oxygen at point (B). Dynamic bonds are shown at $\leq$ 2.1 Angstroms. Other water molecules (not interacting with O$_2$) are not shown for clarity.}
    \label{fig:adsorbed}
\end{figure*}

We scanned $\mu_e$ shifts that reduced the driving force of the reaction and found that with lower applied biases (corresponding to lowered overpotentials), we no longer observe an outer-sphere reaction. Instead, we are able to observe an inner-sphere reaction according to the following steps:
\begin{equation} \label{step1-is}
    {\rm 2e^- + O_2 \rightarrow (O_{2}^{2-})_{ads} }
\end{equation}
\begin{equation} \label{step2-is}
    {\rm 2e^- + (O_{2}^{2-})_{ads} + H_2O \rightarrow O^{2-}_{ads} + 2OH^- }
\end{equation}
\begin{center}
\rule{0.5\textwidth}{0.4pt}
\end{center}
\begin{equation} \label{fullrxn-is}
    {\rm 4e^- + O_2 + H_2O \rightarrow O^{2-}_{ads} + 2OH^-}
\end{equation}
where O$^{2-}_{\rm ads}$ stays adsorbed to the NG surface. 
Alternatively, the adsorbed oxygen might undergo a less complete reduction, forming an O$^{\cdot-}_{\rm ads}$ species (-O$\cdot$) according to the following steps:
\begin{equation} \label{step1-is2}
    {\rm 2e^- + O_2 \rightarrow (O_{2}^{2-})_{ads} }
\end{equation}
\begin{equation} \label{step2-is2}
    {\rm  e^- + (O_{2}^{2-})_{ads} + H_2O \rightarrow O^{\cdot-}_{\rm ads} + 2OH^- }
\end{equation}
\begin{center}
\rule{0.5\textwidth}{0.4pt}
\end{center}
\begin{equation} \label{fullrxn-is2}
    {\rm 3e^- + O_2 + H_2O \rightarrow O^{\cdot-}_{\rm ads} + 2OH^-}
\end{equation}
In both cases, the adsorbed O$_2$ forms a transient species during reduction where the partial charges on the two oxygen atoms resemble the oxygen partial charges of HOO$^-$ anion (H\underline{\textbf{O}}O$^-$ = -0.46 and HO\underline{\textbf{O}}$^-$ = -0.65, Table \ref{table: Mulliken}). Figure \ref{fig:adsorbed} shows the adsorbed oxygen interacting with a nearby water to form the transient bound peroxo species prior to reduction. Additionally, we show Mulliken charges across simulation time for both oxygen atoms of the molecular oxygen and highlight the portion of the simulation during which the observed Mulliken charges for O$_2$ fall within one standard deviation of the average Mulliken charge for the HOO$^-$ anion, as given in Table \ref{table: Mulliken}. We also show the separation distance between the two oxygen atoms of molecular oxygen as well as the separation distance between the protonating water hydrogen and the non-adsorbed oxygen from O$_2$ across the same portion of simulation time. The Mulliken charge analysis of the remaining adsorbed oxygen atom in the $\mathrm{O}^{2-}_{\rm ads}$ or O$^{\cdot-}_{\rm ads}$ species shows an average value of about $\simeq$ -0.63 (Figure \ref{fig:adsorbed}). Both adsorbed oxygen species have a similar Mulliken charge when simulated in isolation (Figure \ref{fig:compOads}), so we are not able to tell them apart using Mulliken charges - this issue distinguishing between species has also come up in metal-oxygen interactions \cite{Montemore17}. Hence, we have to consider both possibilities. Moreover, it is possible that the NG sheet in the isolated case has not lost electrons as it would in the reductive case. This could explain the difference in Mulliken charge between the adsorbed oxygen in ORR simulations (-0.63) and in isolated simulations (-0.86). Our results are consistent with other reports that adsorbed O$^{\cdot -}$ and O$_2$ species can be stabilized by the water solvation shell, questioning the suitability of implicit solvent models for modeling the ORR on N-doped graphene.\cite{Vegge18} 

\begin{figure}
    \centering
    \includegraphics[width=0.7\columnwidth]{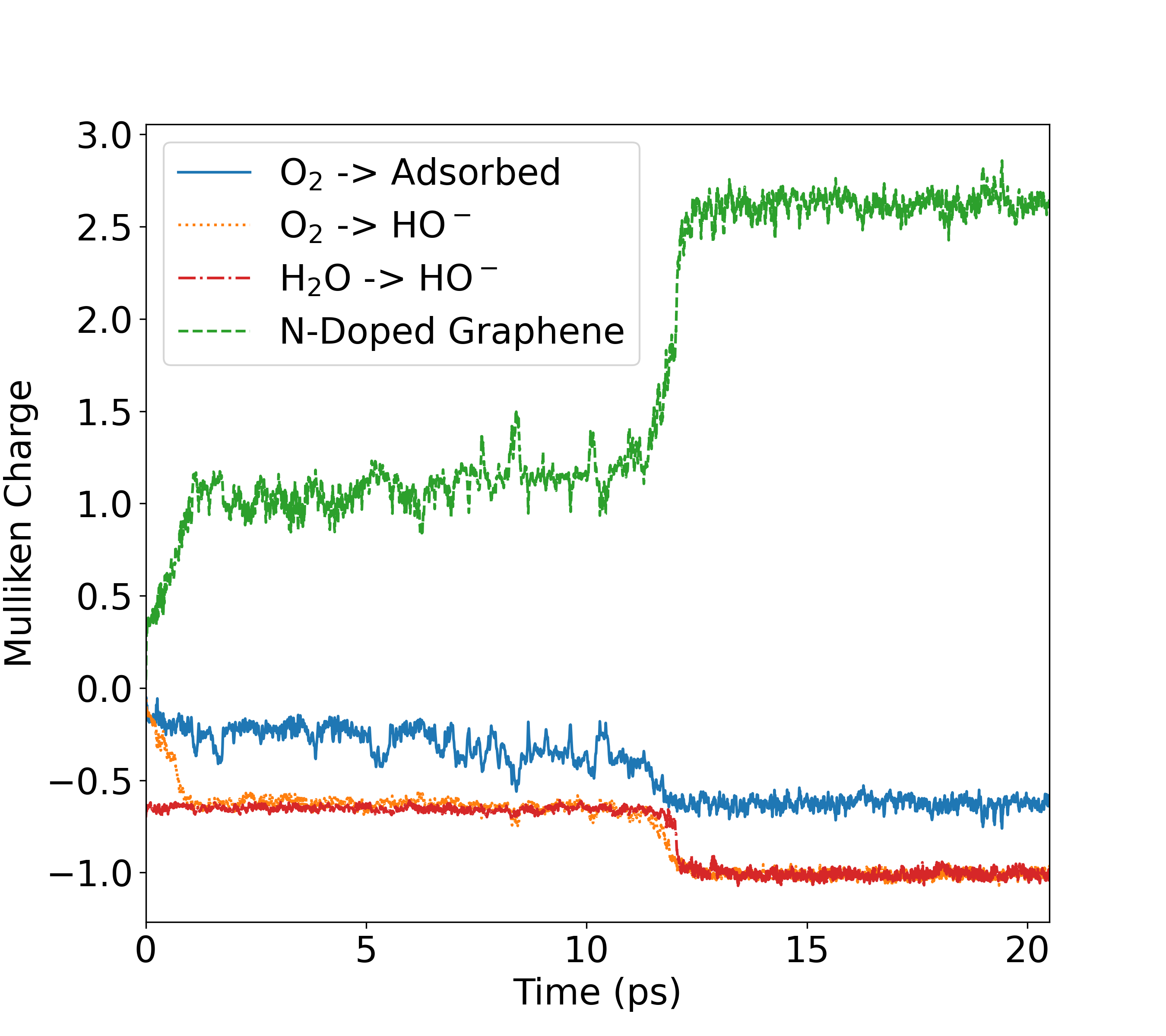}
    \caption{Mulliken charges in \textit{e} units for the two oxygen atoms that are initially part of the oxygen molecule, the oxygen atom of the reacting water molecule (dot-dashed red line), and the sum of Mulliken charges for all NG sheet atoms (dashed green line) are shown for a simulation with an applied bias -0.9 eV. The Mulliken charge of the oxygen atom from molecular oxygen that gets adsorbed is shown as a solid blue line and the Mulliken charge of the oxygen atom from molecular oxygen that becomes an OH$^-$ anion is shown as a dotted orange line.}
    \label{fig:mullMin09}
\end{figure}

\begin{figure}
    \centering
    \includegraphics[width=0.7\columnwidth]{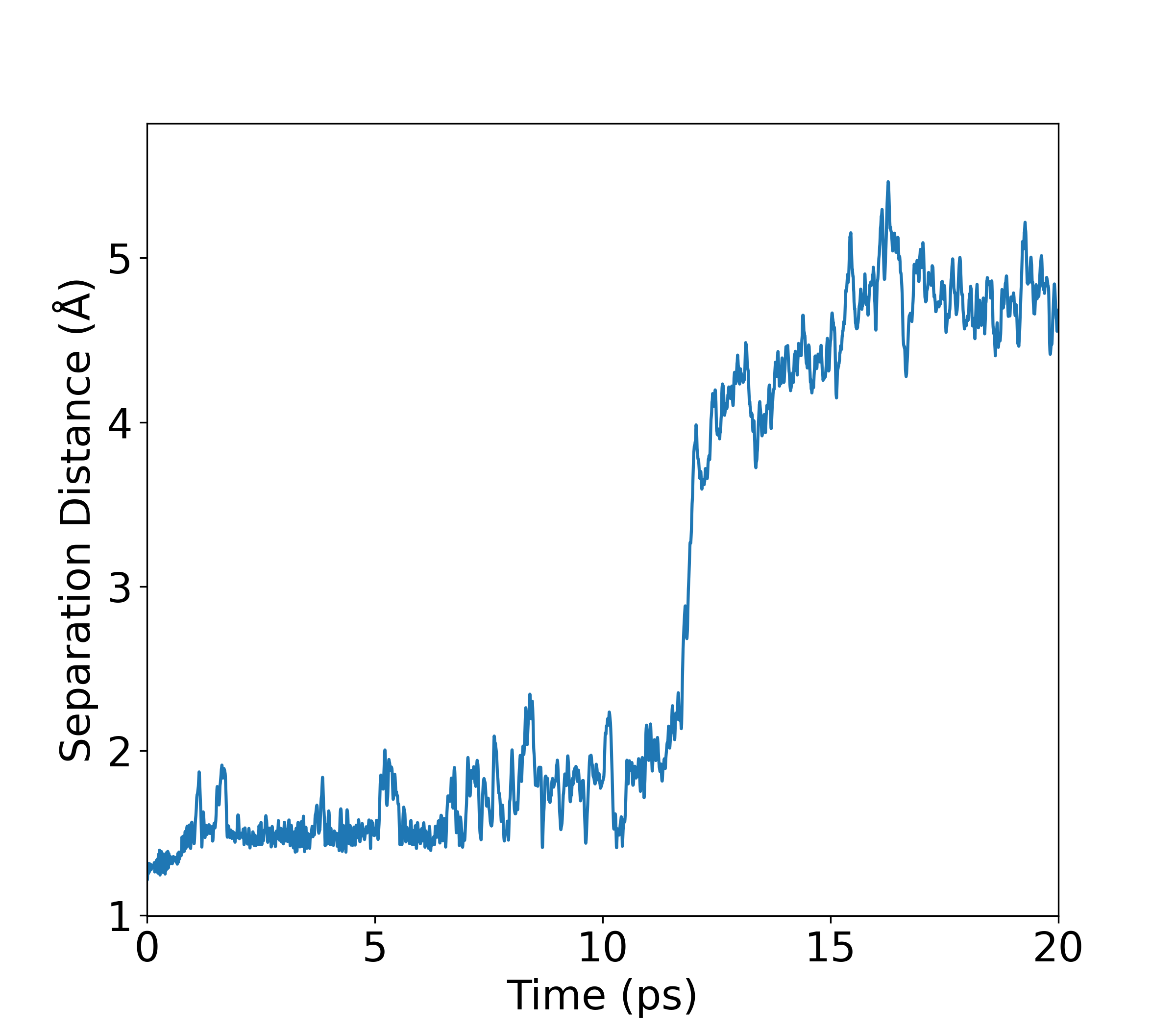}
    \caption{Distance between oxygen atoms in molecular oxygen as a function of time for 20.0 ps of QMD simulation with an applied bias (applied $\mu_e$ shift) of -0.90 eV. Between 12 and 13 ps, a sharp increase of O-O distance is clearly observed, demonstrating cleavage of the O-O bond.}
    \label{fig:oosep-appliedBias}
\end{figure}

In summary, the O$_2$ molecule fully dissociates via an inner-sphere mechanism during simulations using applied biases of -1.25 eV to -0.25 eV 
-- in each case, one oxygen atom stays adsorbed to the surface and the other forms an OH$^-$ anion. Also note that there is no evidence for a change in the inner-sphere mechanism or rate determining step (RDS) over this potential window. To provide an example of analyzing an inner-sphere reaction, Figure \ref{fig:mullMin09} shows the Mulliken charges for both molecular oxygen atoms, the NG sheet, and the oxygen atom of the water molecule that participates in the reaction for a simulation run with an applied bias of -0.9 eV. 
Two OH$^-$ anions are formed, shown clearly as the two oxygen atoms that take on a Mulliken charge of approximately -1.00 after $\sim$12 ps of simulation. Figure \ref{fig:oosep-appliedBias} shows the separation distance between the two oxygen atoms that are initially part of molecular oxygen, which also demonstrates molecular oxygen cleavage after $\sim$12 ps. Both of these analyses point to the partial reduction of O$_2$.

In the inner-sphere reaction pathway, adsorption is necessary for the ORR. In each simulation with an applied bias under 0.0, the O$_2$ molecule adsorbs to the NG surface (defined here as forming a bond of $\leq$ 1.4 Angstroms with a carbon of the NG sheet) prior to reduction. The time that O$_2$ stays adsorbed to the NG  sheet prior to reduction varies with applied bias, with a maximum observed adsorption time of 10.5 ps. The adsorbed oxygen atom remains bound to the sheet following reduction for the remainder of the simulation in all inner-sphere cases. Figure \ref{fig:adsTime} shows the adsorption times prior to oxygen reduction for each of the inner-sphere reactions.

We used several strategies to observe oxygen reduction within a time frame that is accessible by our simulation methods. (1) In the initial configuration the O$_2$ is close to the NG sheet. (2) In both the inner- and outer-sphere mechanisms, we are simulating conditions that most likely correspond to high overpotentials (Supporting Information section on ``Converting Applied Bias to Overpotential", Table \ref{chempot}, and Figure \ref{fig:trasattiScale}). (3) For the inner-sphere simulations, we use steered molecular dynamics (SMD) to draw O$_2$ to the sheet. Although these strategies will tend to accelerate diffusion and lower energy barriers, we do expect the time scales for the electron transfer steps in our simulations to be generally consistent with those in experimental systems. 

The final step of the inner-sphere ORR --- the reaction of the adsorbed oxygen with water to form two additional OH$^-$ anions --- is not seen in the present simulations, which we hypothesize is due to either the partial reduction (formation of O$^{\cdot-}_{\rm ads}$) or the length of the simulation. Additional simulations to determine the necessary conditions for the final step will be explored in future work. 
Even so, the observation of an inner-sphere mechanism in XL-BOMD simulations with applied bias is an exciting result which further demonstrates the utility and promise of this method in the electrochemistry field. 

\subsection{Dependence of Reaction Kinetics and Mechanism on Applied Bias}

Having observed both inner- and outer-sphere ORRs, we next investigated how varying applied bias could affect which mechanism is used and the relative reaction rates.
We performed 58 production simulations, each up to 20 ps in length, across a range of applied biases (Table \ref{allAppliedBiases}) and characterized changes in both mechanism and reaction rate.  This investigation was made possible by the computational efficiency of the XL-BOMD method, which enabled us to observe reactions in a sufficient number of simulations. Since applied bias is the actual variable we tune in our simulations, all of our simulations and analyses will be discussed and reported accordingly. We have estimated the conversion between applied bias and overpotential, which is described in the Supporting Information section on ``Converting Applied Bias to Overpotential", Table \ref{chempot}, and Figure \ref{fig:trasattiScale}, though further testing and validation is needed to fully verify the accuracy. Regardless of the exact conversion, simulating with relatively higher applied biases corresponds to higher overpotential conditions while simulating with relatively lower applied biases corresponds to lower overpotential conditions. 

The kinetics of electrocatalytic reactions are often described using the phenomenological Butler-Volmer equation \cite{Bockris00}. At small overpotentials, this is valid for outer-sphere electron transfer. However, at high overpotentials, or when inner-sphere electron transfer is involved, the Butler-Volmer equation is no longer accurate. This is problematic for electrocatalysis since the purpose of the electrocatalyst is to accelerate the bond making/breaking events that the Butler-Volmer equation fails to describe. 

Using N-doped graphene as an ideally polarizable electrode and XL-BOMD simulation as a means to probe the ORR provides a convenient means of studying the quantum mechanical aspects of ORR. By restricting the scope of our analysis to isolated, single-molecule ORR events, we are able to avoid the complexities typically encountered in dense electrocatalytic environments, for instance, coverage effects \cite{Nong20}. This idealization avoids convoluting variables such as adsorbate-adsorbate interactions, competitive multi-site binding, and coverage-dependent shifts in the local density of states.

\begin{figure}
    \centering
    \includegraphics[width=0.7\columnwidth]{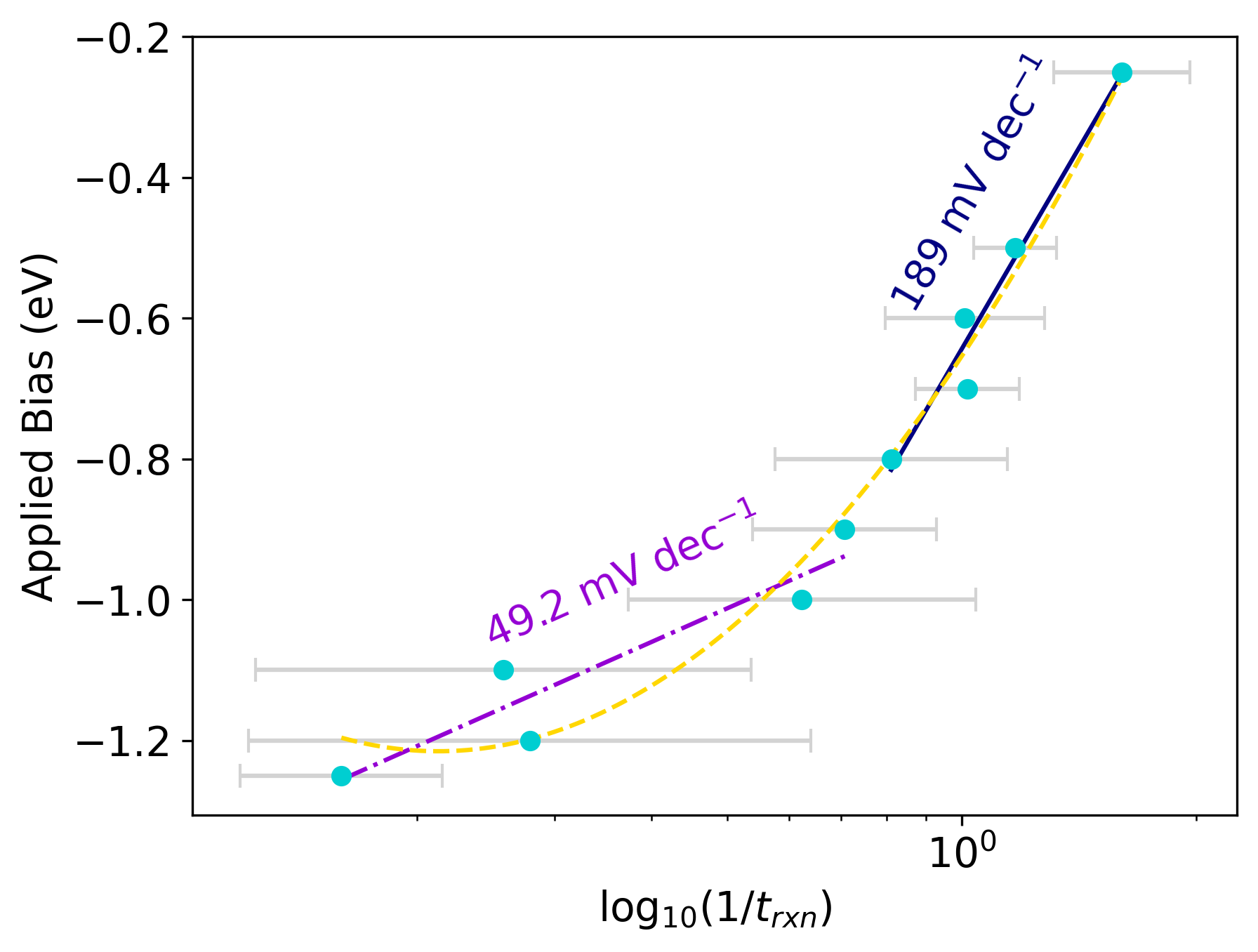}
    \caption{Plot of the log$_{10}$ of inverse reaction time (1/$t_{rxn}$) versus applied bias for the inner-sphere reactions. We here define reaction time as the time from the start of the simulation to the formation of all OH$^-$ ions. All points are plotted as mean values (turquoise dots) with error bars (gray lines) based on the simulation replicate results. Bi-linear Tafel slopes are fit to the data based on typical Tafel analysis with slopes of 49.2 mV dec$^{-1}$ (dot-dashed purple line) and 189 mV dec$^{-1}$ (solid navy line). An exponential fit is also included (dashed yellow line).}
    \label{fig:innersphere-rxnTimeVsBias}
\end{figure}

In the preceding, we showed that the inner-sphere regime dominates in simulations with lower applied biases, with the O$_2$ reacting after adsorbing onto the electrode. Inner-sphere reactions were observed at 10 applied biases ranging from -1.25 eV to -0.25 eV. 
Three to five replicate simulations were performed with each bias to determine variability in reaction time. We observed that reactions with lower applied biases took longer on average and also had greater variability in reaction time (Figure \ref{fig:innersphere-rxnTimeVsBias}). Here we define reaction time as the time from the start of the simulation to the formation of all OH$^-$ ions (Equations \ref{fullrxn-is} or \ref{fullrxn-is2}). We are using acceleration techniques, described above, however, we expect the relative reaction times for the tested systems to follow the same trends demonstrated here given that they are accelerated in the same way.
While we can fit this data with two Tafel slopes, which fall in the range observed for ORR on graphene electrodes \cite{Qu10, Dai15, Faisal17, Lu18}, this would imply a change in RDS/mechanism in a Butler-Volmer picture, and our simulations show no such changes. Moreover, our simulations reveal a continuously varying Tafel slope without a change in the underlying mechanism or RDS. Thus, Butler-Volmer kinetics do not describe the inner-sphere reaction.

\begin{figure}
    \centering
    \includegraphics[width=\columnwidth]{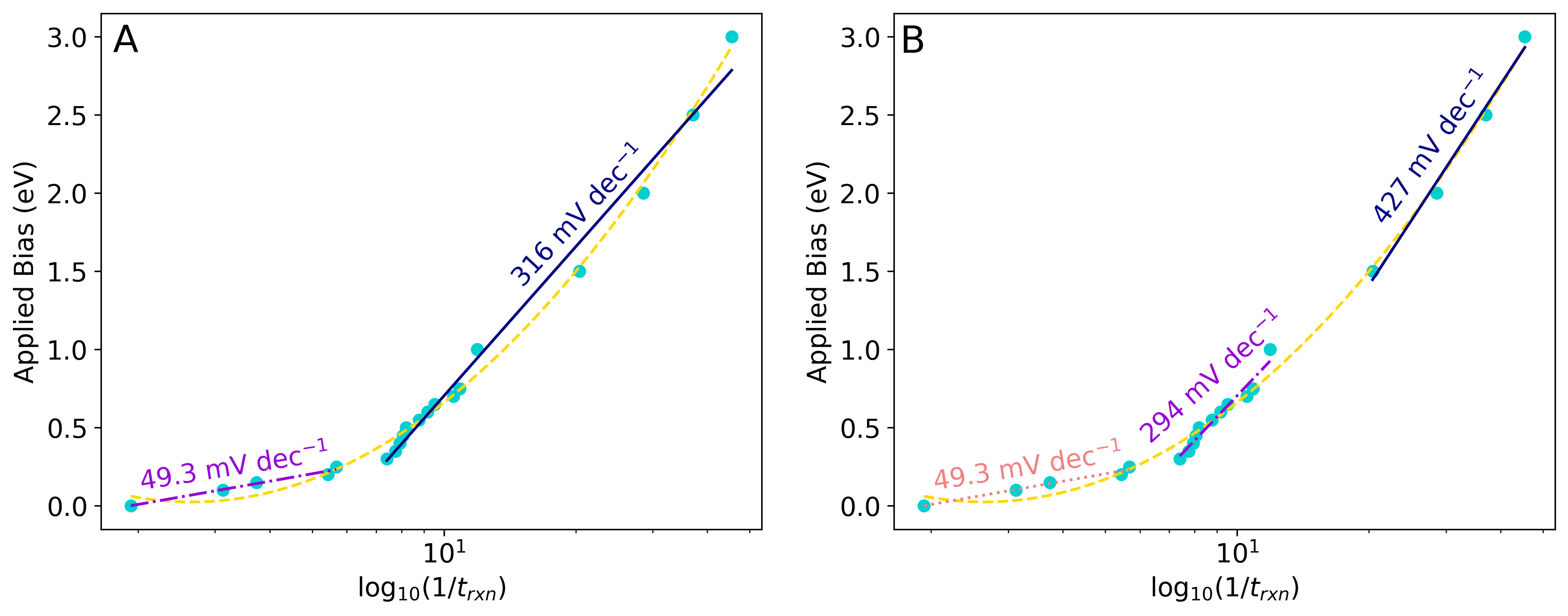}
    \caption{Plots of the log$_{10}$ of inverse reaction time (1/$t_{rxn}$) versus applied bias for the outer-sphere reactions (turquoise dots). We here define reaction time as the time from the start of the simulation to the formation of all OH$^-$ ions. (A) Bi-linear Tafel slopes are fit to the data based on typical Tafel analysis with slopes of 49.3 mV dec$^{-1}$ (dot-dashed purple line) and 316 mV dec$^{-1}$ (solid navy line). An exponential fit is also included (dashed yellow line). (B) Tri-linear Tafel slopes are fit to the data with slopes of 49.3 mV dec$^{-1}$ (dotted coral line), 294 mV dec$^{-1}$ (dot-dashed purple line), and 427 mV dec$^{-1}$ (solid navy line). An exponential fit is also included (dashed yellow line).}
    \label{fig:outersphere-rxnTimeVsBias}
\end{figure}

Using higher applied biases, we find that the reaction transitions to an outer-sphere electron transfer pathway, where the ORR takes place outside the inner Helmholtz plane. Outer-sphere reactions were observed at 20 applied biases ranging from 0 eV to 3.0 eV. 
Here, again, we define reaction time as the time from the start of the simulation to the formation of all OH$^-$ ions (Equations \ref{fullrxn} or \ref{fullrxn-hoo}).
As with the inner-sphere regime, we can fit the outer-sphere Tafel plot with two Tafel slopes that fall within the range observed for the ORR on doped carbon. No change in RDS/mechanism is observed between these two regions.
Alternatively, it is possible to fit three Tafel slopes to the outer-sphere data and to interpret them according to subtle differences in pathway seen during simulation (e.g., at the highest overpotentials, it appears that there is sufficient energy to ionize water prior to O$_2$ reduction). However, these differences are quite small and don't result in a marked mechanism change (e.g., from inner-sphere to outer-sphere) or a clear change in RDS. Additionally, none of the ranges span the preferred 1 decade of current that is typical for Tafel slope analysis \cite{Liu2014, Breslin2023}.
In fact, in the outer-sphere regime, the Tafel slope also appears to continuously bend with no clear change in mechanism (Figure \ref{fig:outersphere-rxnTimeVsBias}). This suggests the electrode, even while operating through an outer-sphere mechanism, fails to obey traditional Butler-Volmer kinetics.
This continuous bending is predicted for Tafel plots at high overpotentials, but is often difficult to observe experimentally given the speed of reactions at high overpotentials \cite{SchmicklerSantos2010}. Using XL-BOMD, we are able to observe such reactions and record their relative reaction times to provide a clearer picture of how applied bias might affect the ORR. 
%
%

\begin{figure}
    \centering
    \includegraphics[width=0.75\textwidth]{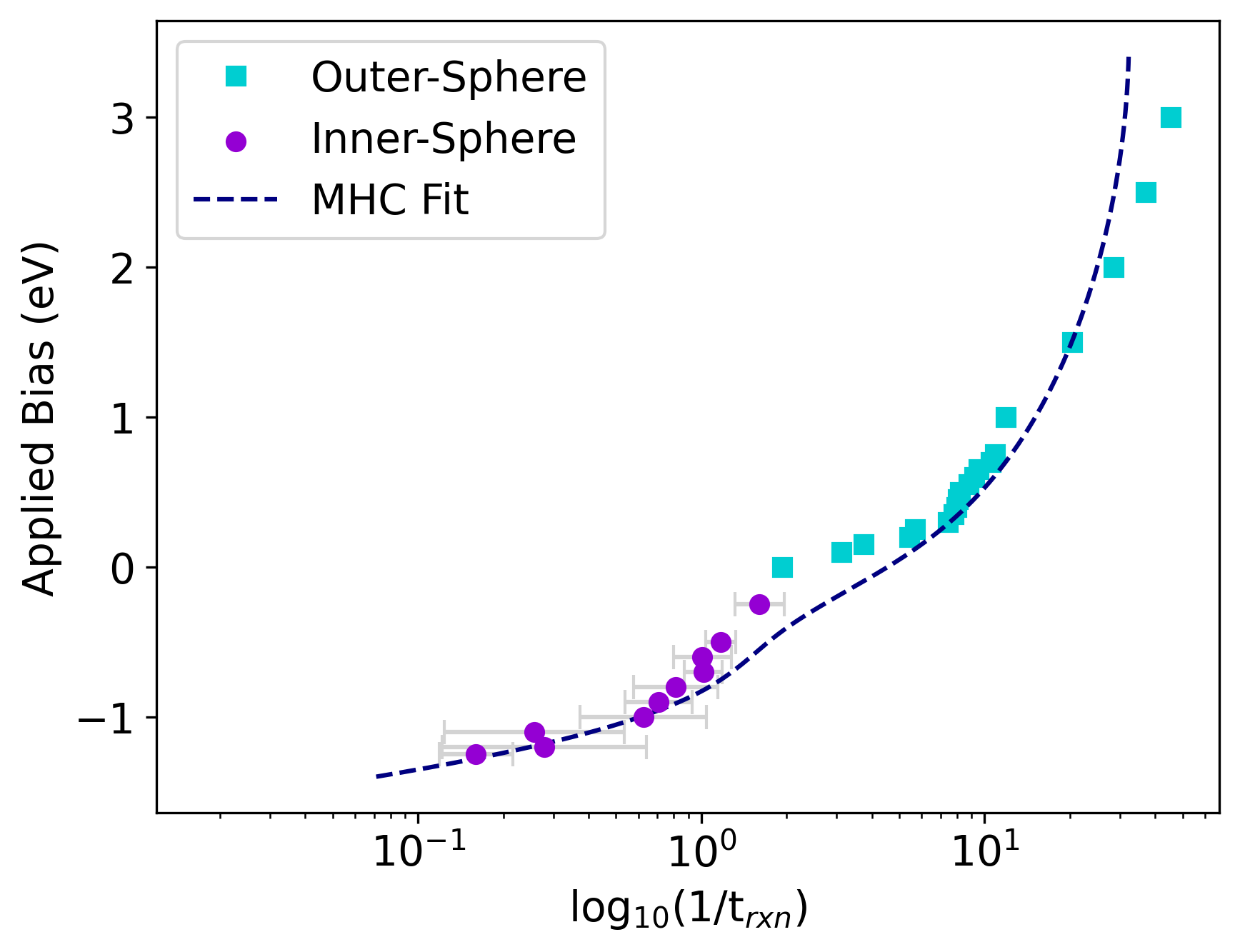}
    \caption{Plot of the log$_{10}$ of inverse reaction time (1/t$_{rxn}$) versus applied bias for the outer-sphere reactions (turquoise squares) and inner-sphere reactions (purple dots with gray error bars). We here define reaction time as the time from the start of the simulation to the formation of all OH$^-$ ions. A total fit using MHC theory is included (dashed navy line), which would typically be used to suggest that all reactions happen via an outer-sphere mechanism as the potential sweeps the O$_2$ energy through the Dirac cone. However, simulations clearly show the two different mechanisms.}
    \label{fig:mhcfit}
\end{figure}

Going beyond Butler-Volmer theory, it is common to use Marcus or Marcus-Hush-Chidsey (MHC) theory instead \cite{Chidsey1991, SchmicklerSantos2010, Henstridge12, Zeng14, Roman17, Zhu24} to explain the kinetics of reactions like the ORR. To use this theory, one must first assume that all of the reactions happen by an outer-sphere mechanism. We can fit our full range of simulation data points using MHC theory, and produce a convincing fit line (Figure \ref{fig:mhcfit}, fit parameters available in Supporting Information). 
Based on this common analysis, one would be lead to believe that all of our reactions happen via an outer-sphere mechanism and that the potential sweeps the O$_2$ energy through the Dirac cone. However, based on our simulations, we know that is not actually the case. This test provides a word of caution regarding the use of general theories without a way to validate the underlying mechanisms. It also points out how XL-BOMD can be used, as it is here, in conjunction with these well-known theories to develop a more detailed mechanistic understanding of the ORR. Using simulations, we are able to see reactions happen step-by-step, which provides a much clearer picture of what is happening during oxygen reduction on NG at the atomic level across a broad range of applied biases. 

\section{Conclusion}

 We have simulated the ORR on N-doped graphene across a broad range of applied biases, showing mechanistic transition between inner- and outer-sphere pathways and observing reaction kinetics. 
 In so doing, we have demonstrated that shadow extended Lagrangian Born-Oppenheimer molecular dynamics, implemented with DFTB, can simulate electrochemical processes at the solid–liquid interface of nitrogen-doped graphene with explicit inclusion of solvent. This capability goes beyond static DFT energetics by directly capturing time-resolved electron transfer events responsible for O$_2$ activation and dissociation. 
 
Using a novel electrochemical biasing scheme, we showed that reaction conditions can be systematically tuned within simulation. Higher applied biases promoted outer-sphere electron transfer mediated by the solvent, whereas lower applied biases favored inner-sphere pathways involving direct O$_2$ adsorption. This biasing approach establishes a generalizable route to explore how applied potential governs mechanistic regimes in electrocatalytic reactions.

In our kinetics analysis, we demonstrate that both Butler-Volmer theory and MHC theory are unable to capture the full mechanistic picture of the ORR on doped graphene. Our simulations clearly show a change in mechanism from inner-sphere to outer-sphere with changing applied bias that is not fully explained by either of these two theories. Using XL-BOMD, we are able to provide a detailed look at the changes that occur with applied bias and the step-by-step mechanisms of both inner- and outer-sphere ORR pathways. 

This work provides the first demonstration of XL-BOMD applied to heterogeneous electrocatalysis, simulating electrode and electrolyte together with potential biasing in a unified framework. Follow-up studies will extend this methodology to explore steady-state current conditions, explicit pH effects, and defective or layered catalyst architectures. Taken together, these results lay the foundation for XL-BOMD as a versatile tool for computational electrochemistry.

\begin{acknowledgement}

This work was supported by the LANL LDRD-ER program; the Department of Energy Offices of Basic Energy Sciences (Grant No. LANLE8AN); and the Exascale Computing Project (17-SC-20-SC), a collaborative effort of two U.S. Department of Energy organizations (Office of Science and the National Nuclear Security Administration) responsible for the planning and preparation of a capable exascale ecosystem, including software, applications, hardware, advanced system engineering, and early testbed platforms, in support of the nation's exascale computing imperative. Simulations were performed using Los Alamos National Laboratory Institutional Computing resources. 

This article has been approved for unlimited distribution with the following assigned LA-UR number: `LA-UR-25-20884'.

\end{acknowledgement}

\begin{suppinfo}

The following file is available free of charge

Supporting Information: PDF file including 
\begin{itemize}
  \item A comparison of adsorption energies calculated using DFTB and DFT
  \item A comparison of minimized water structures in vacuum using DFTB and DFT
  \item A simulation of molecular oxygen in water at the DFTB level of theory
  \item Molecular dynamics step timings for GPMDK
  \item A demonstration of the energy conservation of XL-BOMD simulations
  \item Component contributions to the oxygen species charges
  \item Information on control systems
  \item A discussion of an alternative reaction pathway seen in additional simulations
  \item Information on extended systems
  \item A discussion of the distribution of electron holes across the NG surface
  \item A full-length simulation with a $\mu_e$ shift of -2.0 eV 
  \item A comparison of adsorbed single oxygen species in DFTB simulations
  \item A figure showing the adsorption times prior to reaction for all inner-sphere simulations
  \item A discussion of the conversion of applied bias to overpotential
  \item A table showing all the applied biases used in simulations
  \item Parameters for the Marcus-Hush-Chidsey fitting
  
\end{itemize}

\end{suppinfo}

\bibliography{bibliography}

\newpage
\clearpage
\setcounter{page}{1}
\section*{Supporting Information}

\import{.}{suppInfo.tex}

\end{document}

%% file: suppInfo.tex
\newcommand{\beginsupplement}{%
        \setcounter{table}{0}
        \renewcommand{\thetable}{S\arabic{table}}%
        \setcounter{figure}{0}
        \renewcommand{\thefigure}{S\arabic{figure}}%
     }


\beginsupplement

\section*{Supporting Information for: Modeling Reactions on the Solid-Liquid Interface with Next Generation Extended Lagrangian Quantum-Based Molecular Dynamics}

Rae A. Corrigan Grove, Kevin G. Kleiner, Joshua Finkelstein, Ivana Matanovic, Michael E. Wall, Travis E. Jones, Anders M. N. Niklasson, and Christian F. A. Negre

This article has been approved for unlimited distribution with the LA-UR number: `LA-UR-25-20884'.









\newpage
\section*{Comparison to DFT}

To assess the suitability of our DFTB parameters for the current simulations, we compared the  adsorption energies for O$_2$, H$_2$O, OH$^{\cdot -}$, and O$^{\cdot -}$ with NG using both DFTB and DFT. All calculations were performed in vacuum.  For each species, we first performed energy minimization at the DFTB level of theory with a damped dynamics minimization method using a tolerance of about 10$^{-4}$ eV. DFT minimizations were performed using Quantum ESPRESSO (QE) \cite{Giannozzi09, Giannozzi17, Giannozzi20} using an L-BFGS minimization method. We used a plane wave basis set with a kinetic energy cutoff of 80 Ry and a charge density cutoff of 640 Ry with the PBE exchange-correlation functional using pseudopotentials from the SSSP database \cite{prandini2018precision}. All simulations were performed at the $\Gamma$ point to match the DFTB calculations. A comparison of adsorption energies calculated using DFTB and DFT is presented in Table \ref{table: adsEnergies}.

\begin{table}[H]
    \begin{center}
        \caption{Comparison of adsorption energies calculated using DFTB and DFT for key adsorbed species.}
        \begin{tabular}{ c c c }
         \hline
         Adsorbed Species & DFTB Energy (eV)& DFT Energy (eV)\\
         \hline
         OH$^{\cdot -}$ & -2.42 & -2.47\\
         O$^{\cdot -}$ & -3.92 & -5.03\\
         H$_2$O & -0.02 & -0.006\\
         O$_2$ & -0.48 & -0.43\\
         \hline
        \end{tabular}
        \label{table: adsEnergies}
    \end{center}
\end{table}
Table \ref{table: adsEnergies} shows that, although the \textit{lanl}31 parameter set is highly general and created to have transferability across different organic molecules,  the comparison of adsorption energies for important ORR related species is relatively good. However, a deviation exceeding 20 percent is observed in the case of oxygen (O$^{\cdot -}$). In the case of water, given that the values are lower than the threshold for chemical accuracy, a considerable relative deviation could be expected. Overall, this comparison indicates that using our DFTB parameters is reasonable for the present study.

\newpage
\section*{Water in Vacuum}
We compared minimized water structures in vacuum using DFTB and DFT levels of theory to assess the similarity of optimized molecular conformations. Results for bond lengths, angles, and hydrogen bond lengths are shown in Table \ref{table: vac-h2o}

\begin{table}[H]
    \begin{center}
        \caption{Comparison between bond lengths and angles for a single water molecule in vacuum and H-bond length for two water molecules in vacuum, minimized using DFT or DFTB.}
        \begin{tabular}{ c c c }
         \hline
         Property & DFT & DFTB\\
         \hline
         Bond Lengths (\AA) & 0.97 & 0.97\\
         Angle (degrees) & 104.27 & 101.70\\
         \hline
         H-Bond Length (\AA) & 1.92 & 2.02\\
         \hline
        \end{tabular}
        \label{table: vac-h2o}
    \end{center}
\end{table}

\newpage
\section*{Simulation of Molecular Oxygen in Water}
\begin{figure}[ht]
    \centering
    \includegraphics[width=0.75\textwidth]{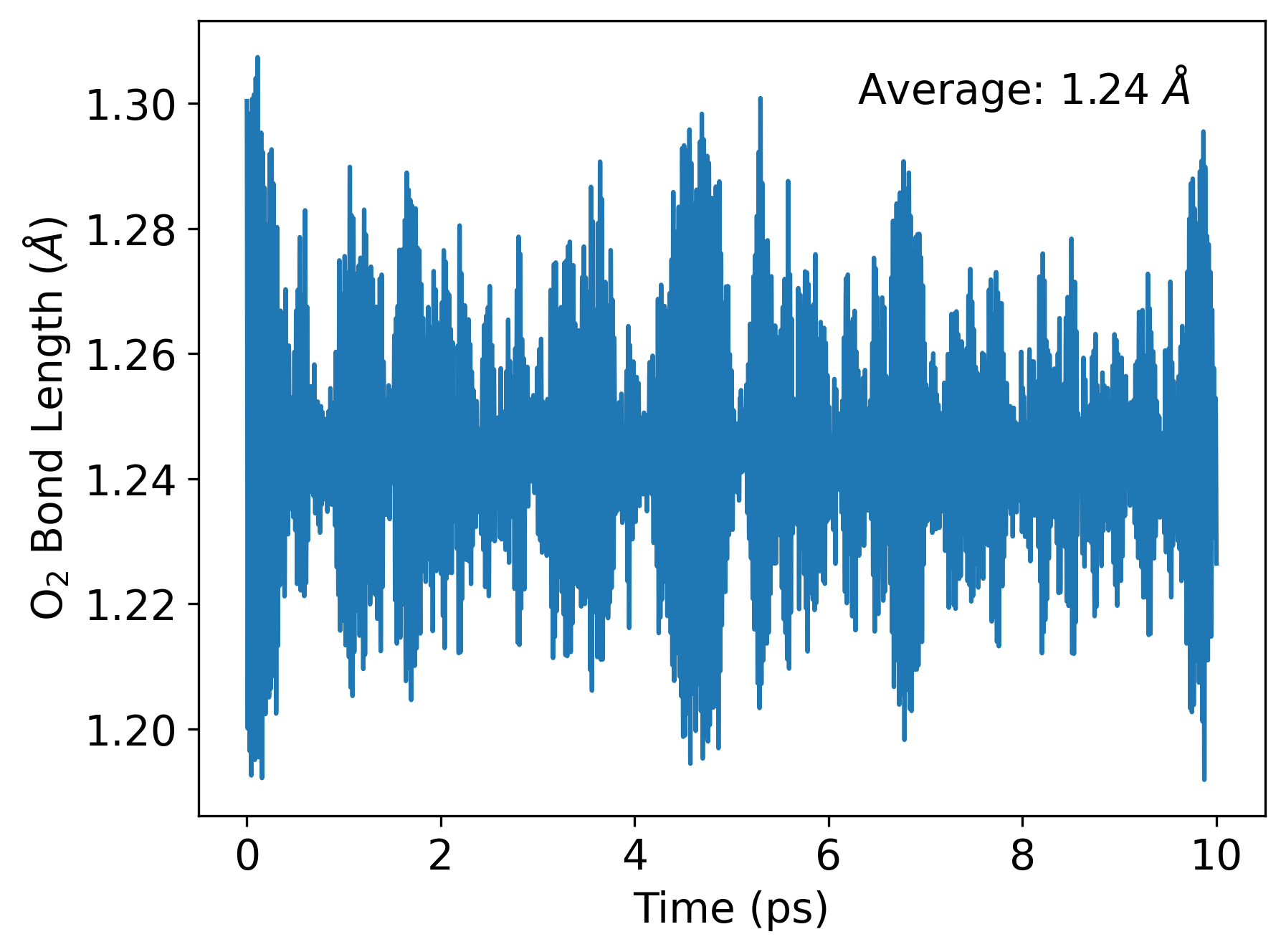}
    \caption{Separation distance between oxygen atoms in O$_2$ during a 10 ps simulation of O$_2$ in a box of water. }
    \label{fig:o2h2o-o2bondlength}
\end{figure}

It is important to confirm that molecular oxygen behaves as expected in aqueous solution. To test this, we performed a baseline simulation of a single molecular oxygen solvated in a box of 164 water molecules. We ran the simulation using XL-BOMD at the DFTB level of theory for 10 ps using a time step of 0.1 fs. We used a Langevin thermostat to maintain a temperature of approximately 300K. The average bond length of the O-O bond in molecular oxygen across simulation time was 1.24 \AA (Figure \ref{fig:o2h2o-o2bondlength}), which agrees with experimental reference \cite{Radjenovic19}. 

\newpage
\section*{GPMDK Timings}
\begin{figure}
    \centering
    \includegraphics[width=0.75\textwidth]{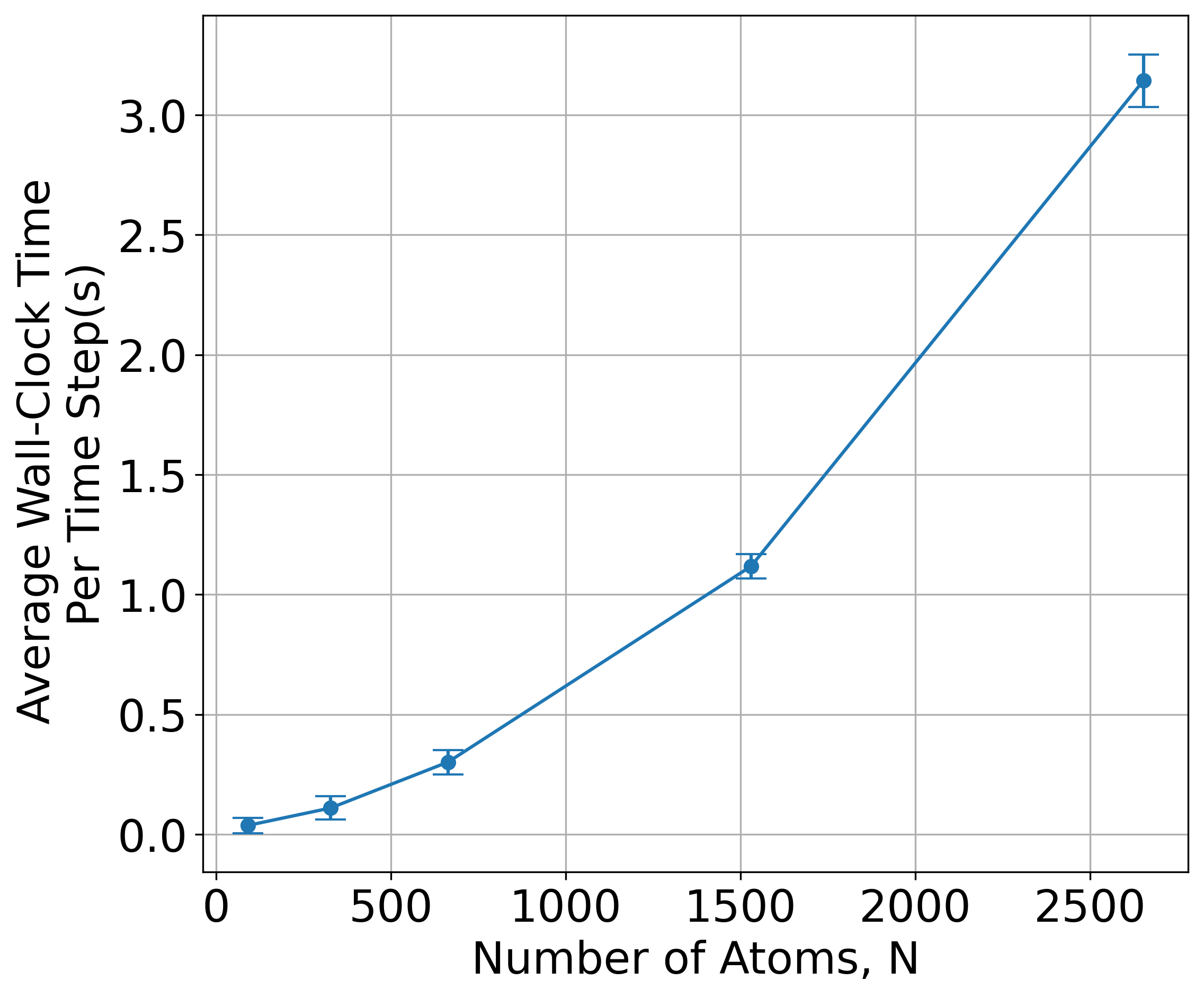}
    \caption{The average time for an MD step in GPMDK on a single CPU and associated standard deviations for five different water boxes (Table \ref{table: waterboxes}).}
    \label{fig:waterboxTimings}
\end{figure}

To demonstrate the timing scaling of of MD steps with simulation size using our code, GPMDK, we performed 100 MD steps using five different water boxes on a single CPU (Table \ref{table: waterboxes}). The average time for those 100 steps along with the standard deviations are shown in Figure \ref{fig:waterboxTimings}.

\begin{table}[H]
    \begin{center}
        \caption{Water boxes used for GPMDK MD timing tests.}
        \begin{tabular}{ c c }
         \hline
         Box Dimensions (\AA) & Number of Atoms\\
         \hline
         10 x 10 x 10 & 90\\
         15 x 15 x 15 & 327\\
         20 x 20 x 20 & 663\\
         25 x 25 x 25 & 1530\\
         30 x 30 x 30 & 2652 \\
         \hline
        \end{tabular}
        \label{table: waterboxes}
    \end{center}
\end{table}

\newpage
\section*{Energy Conservation in the \textit{NVE} Ensemble}
\begin{figure}
    \centering
    \includegraphics[width=0.75\textwidth]{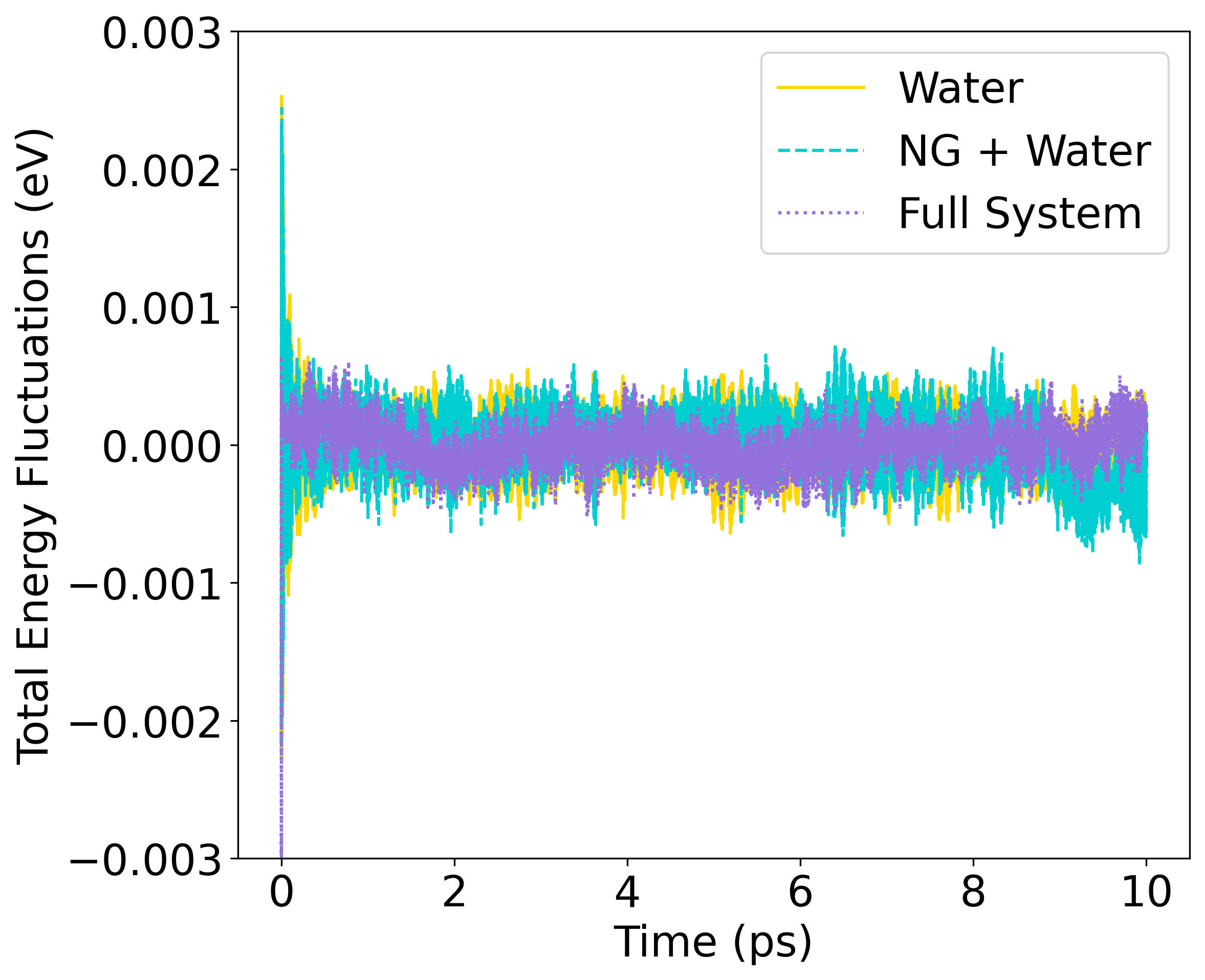}
    \caption{Total energy fluctuations from 10 ps \textit{NVE} simulations using XL-BOMD for three different systems: a box of water, the NG sheet with water, and the full system (NG, water, and O$_2$). Small fluctuations centered around zero demonstrate stability across simulation time. }
    \label{fig:nve}
\end{figure}

Figure \ref{fig:nve} demonstrates energy stability in XL-BOMD simulations of 10 ps with a box of water, water with NG, and the full reactive test system (NG, water, and O$_2$). From Figure \ref{fig:nve} we can see that there is virtually no energy drift for any of the three simulations. This result is crucial to demonstrate the validity of any atomistic simulation.

\newpage
\section*{Component Contributions to the Oxygen Charges}
The following is the system of equations used to determine component contributions, $C_x$, to the oxygen charges in Table \ref{table: contributions}:

\begin{equation}
    \begin{aligned}
        C_{-{\rm H}} + C_{e^-}= -1.01 \\
        C_{-{\rm R}} + C_{e^-} = -0.86\\
        C_{-{\rm OH}} + C_{e^-} = -0.65\\
        2C_{-{\rm H}} = -0.65\\
        C_{-{\rm H}} + C_\cdot = -0.59\\
        C_{-{\rm H}} + C_{-{\rm O}^-} = -0.46\\
        2C_{-{\rm H}} + C_{-{\rm H}^+} = -0.44\\
        C_{-{\rm OH}} + C_{-\rm{H}} = -0.32\\
        C_{-{\rm O}} + C_{-{\rm O}} = 0.00
    \end{aligned}
\end{equation}

The solution of this system of equations gives the components listed in Table \ref{table: contributions}.
\begin{table}
    \caption{Charge contributions of tested components bound to oxygen}
    \begin{tabular}{ c c }
     \hline
     Component &Mulliken Charge (\textit{e} units)\\
     \hline
     -H & -0.33\\
     e$^-$ & -0.67\\      
     -H$^+$ & 0.22\\
     -OH & 0.02\\
     -O$^-$ & -0.13\\
     -O & 0.00\\
     -R & -0.19\\
     $\cdot$ (Radical) & -0.26\\
     \hline
    \end{tabular}
    \label{table: contributions}
\end{table}

\newpage
\section*{Control Systems}
\begin{figure}
    \centering
    \includegraphics[width=\columnwidth]{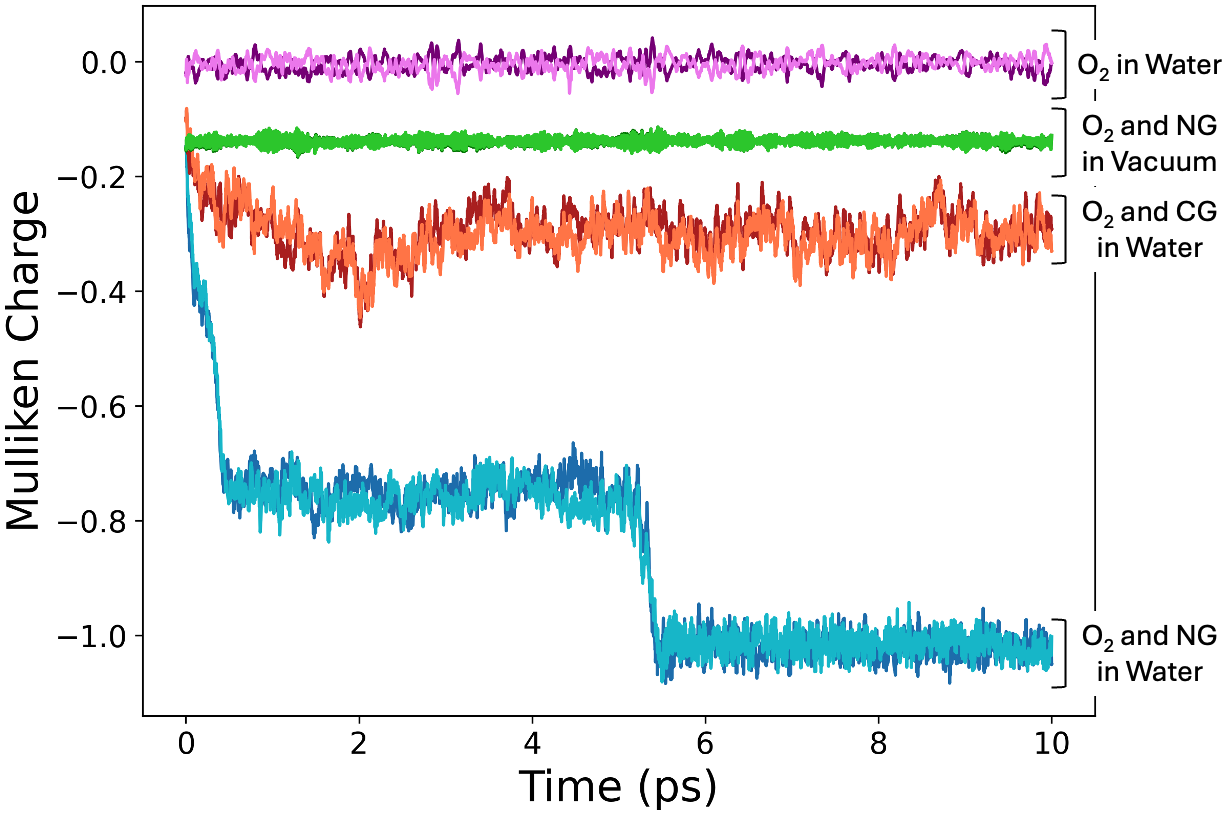}
    \caption{Mulliken charges in \textit{e} units for oxygen atoms that were initially part of the molecular oxygen across each of four simulations. In the main system, O$_2$ is reduced to form OH$^-$, while in the control systems, the initial O$_2$ molecules remain intact. Difference in Mulliken charges between the control systems can be attributed to the presence of differing additional species - i.e., water, nitrogen doped graphene (NG), and/or carbon only graphene sheet (CG)) in each system. }
    \label{fig:controlcomp}
\end{figure}

Three control systems were run to compare with the simulation results for the main system. The variant control systems are as follows: O$_2$ with N-doped graphene (NG) in vacuum, O$_2$ in water, and O$_2$ with carbon-only graphene (CG) in water. The starting coordinates for all three control systems were taken as pieces of the main system. The simulations were run for 10 ps using 0.1 fs time steps. A Langevin thermostat was used to maintain a simulation temperature of 300K and a $\beta$ value of 40 eV$^{-1}$ was used to ``smear'' the Fermi distribution and improve stability. These simulation conditions exactly match the simulation conditions of the main system to ensure the most direct comparison. We did not observe ORR in any of the control systems in the tested time range. This is in contrast to the observation of the ORR in the main system after less than 7 ps. Plots of the Mulliken charges of molecular oxygen atoms over simulation time for the control systems and the main system are shown in Figure \ref{fig:controlcomp}. It is clear that the molecular oxygen atoms in the control simulations do not undergo reduction to form HO$^-$ anions, since their Mulliken charges remain under 1.0 throughout the simulation time. In contrast, the oxygen atoms that initially form molecular oxygen in the main system are rapidly reduced to form OH$^-$ anions.

\newpage
\section*{Alternative Reaction Pathway}
\begin{figure}
    \centering
    \includegraphics[width=\columnwidth]{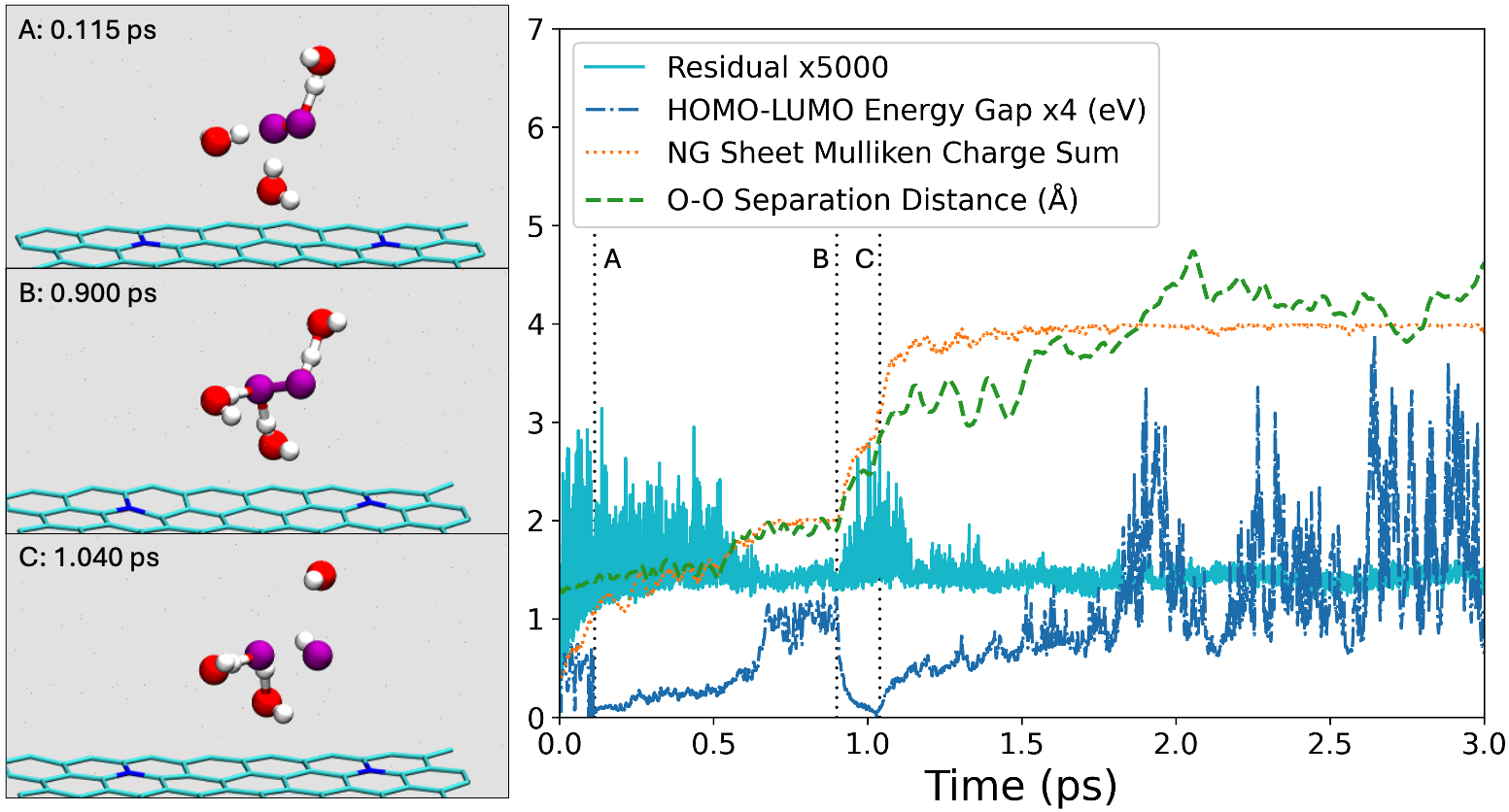}
    \caption{Overlay of four plots: residual ($q[n] - n$) multiplied by 5000 to show features (solid teal), HOMO-LUMO energy gap multiplied by 4 to show features (dot-dashed blue), separation distance between the two oxygen atoms that originally form molecular oxygen (dashed green), and the sum of the NG sheet atom Mulliken charges in \textit{e} units (dotted orange). Panels A, B, and C to the left correspond to gray dotted vertical lines A, B, and C on the plot.}
    \label{fig:overlay-hydroperoxo}
\end{figure}

The outer-sphere simulation described in the main text proceeded through a peroxo species, but another pathway of outer-sphere ORR is possible through a hydroperoxo species. We observed this pathway in another simulation of the ORR shown in Figure \ref{fig:overlay-hydroperoxo}. The residuals and HOMO-LUMO gap are also shown. We see HOMO-LUMO gap closure at points A and C in Figure \ref{fig:overlay-hydroperoxo} corresponding to the first and second injection of electrons from the NG sheet. Early in this simulation, we observe a short-lived transient species that has a bond length and Mulliken charge sum that closely resemble superoxide, O$_2^-$. This is similar to a version of the outer-sphere ORR reaction that has been observed on Pt as well \cite{Kumeda2023-ib}.

\newpage
\section*{Extended Systems}
To test the robustness of the XL-BOMD method in simulating ORR catalysis, two extended systems were constructed by replicating the entire system to obtain a larger NG-water interface with additional O$_2$ molecules. Both extended systems were simulated according to the same protocol using 0.1 fs time steps, a Langevin thermostat with a temperature of 300K, and a $\beta$ value of 40 eV$^{-1}$ to ``smear'' the Fermi distribution and improve stability. Rank-\textit{N} updates were performed with a rank of \textit{N}=4.
The two systems were created by replicating the original system along the $x$- or $x$- and $y$-axes. This replication extended the NG sheet and added additional O$_2$ molecules to the system. 

\begin{figure}
    \centering
    \includegraphics[width=0.75\textwidth]{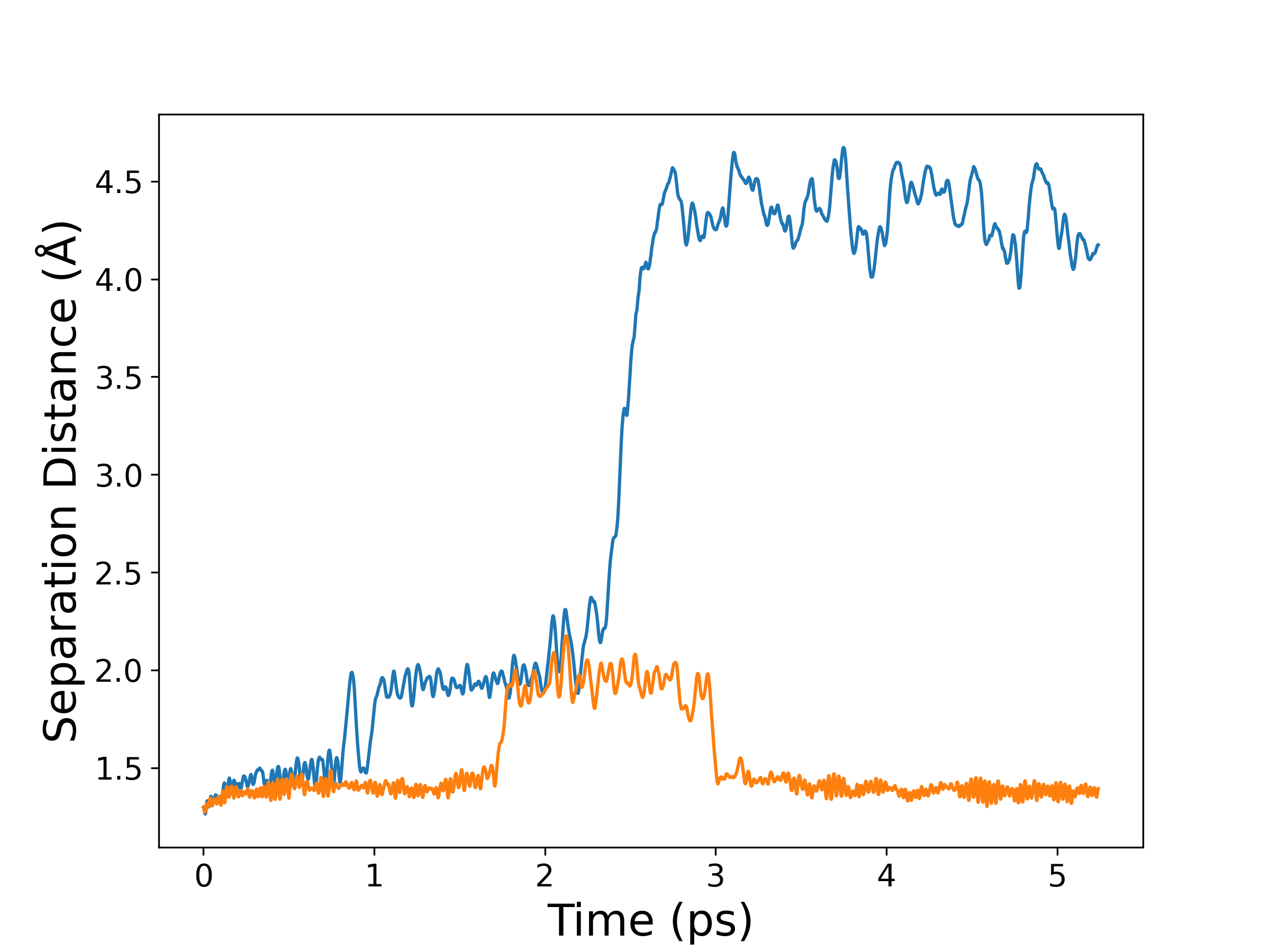}
    \caption{Separation distance between the atoms of each of the two initial O$_2$ molecules across simulation time. The noted increase in separation distance for one of the oxygen molecules after $\sim$2.5 ps demonstrates oxygen reduction.}
    \label{fig:oosepdouble}
\end{figure}

\begin{figure}
    \centering
    \includegraphics[width=0.75\textwidth]{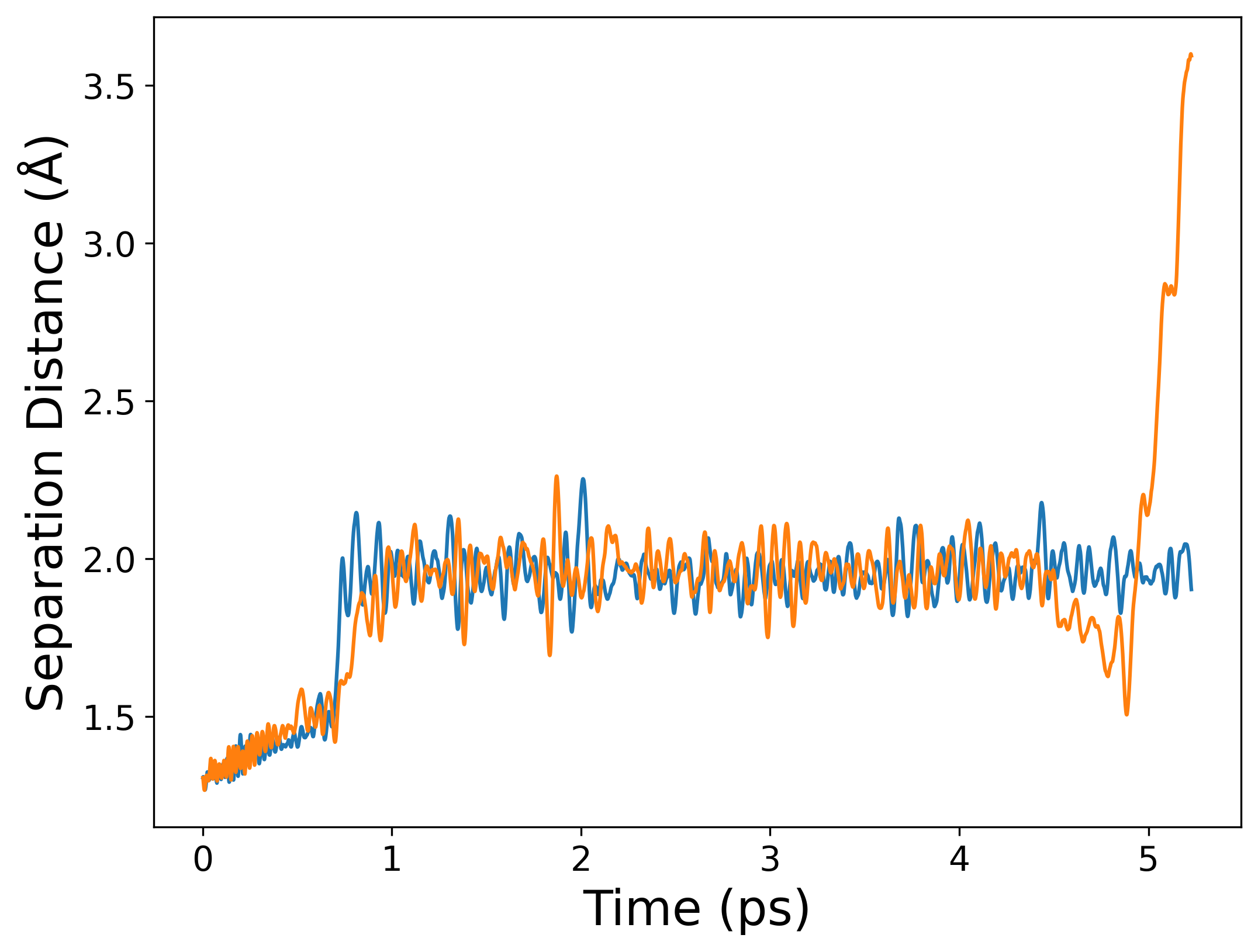}
    \caption{Separation distance between the atoms of each of the two initial O$_2$ molecules across simulation time. The noted increase in separation distance for one of the oxygen molecules after $\sim$5 ps demonstrates oxygen reduction.}
    \label{fig:oosep2x1plus1}
\end{figure}

The first system was created by adding a single additional replica, which doubled the size of the system. The ORR was observed for one of the O$_2$ molecules after $\sim$ 2.5 ps (Figure \ref{fig:oosepdouble}).
Additionally, this system was simulated with an applied bias of +1.0 V and the ORR was observed after $\sim$ 5 ps (Figure \ref{fig:oosep2x1plus1})

\begin{figure}
    \centering
    \includegraphics[width=0.75\textwidth]{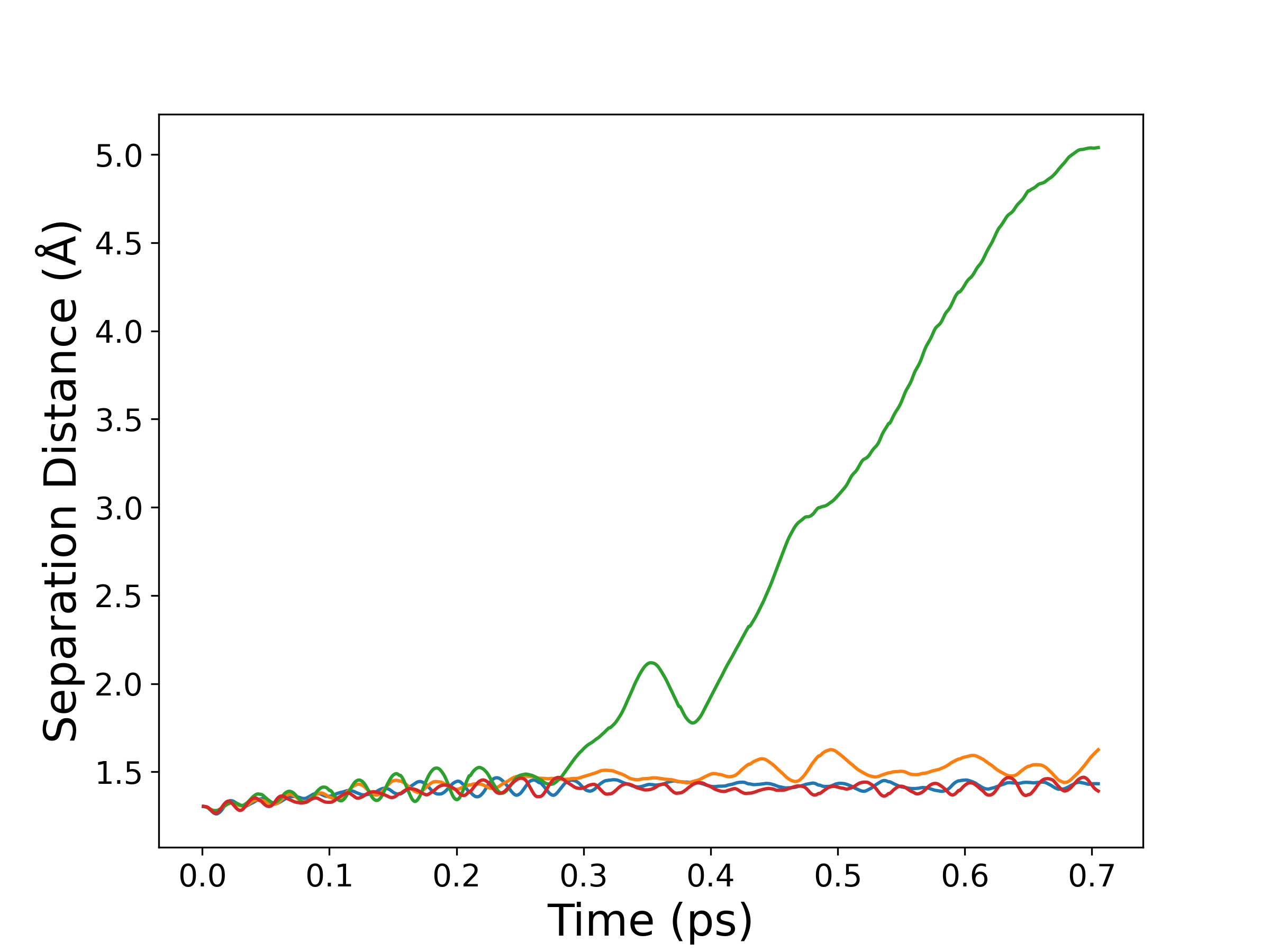}
    \caption{Separation distance between the atoms of each of the four initial O$_2$ molecules across simulation time. The noted increase in separation distance for one of the oxygen molecules between $\sim$0.4 and 0.7 ps demonstrates oxygen reduction.}
    \label{fig:oosepquad}
\end{figure}

The second of these systems was created by adding three additional replicas, quadrupling the system size. The ORR was observed for one O$_2$ molecules within 700 fs of simulation (Figure \ref{fig:oosepquad}). 
Additionally, this system was simulated with an applied bias of +1.0 V and the ORR was observed after $\sim$ 1 ps (Figure \ref{fig:oosep2x1plus1})
\begin{figure}
    \centering
    \includegraphics[width=0.75\textwidth]{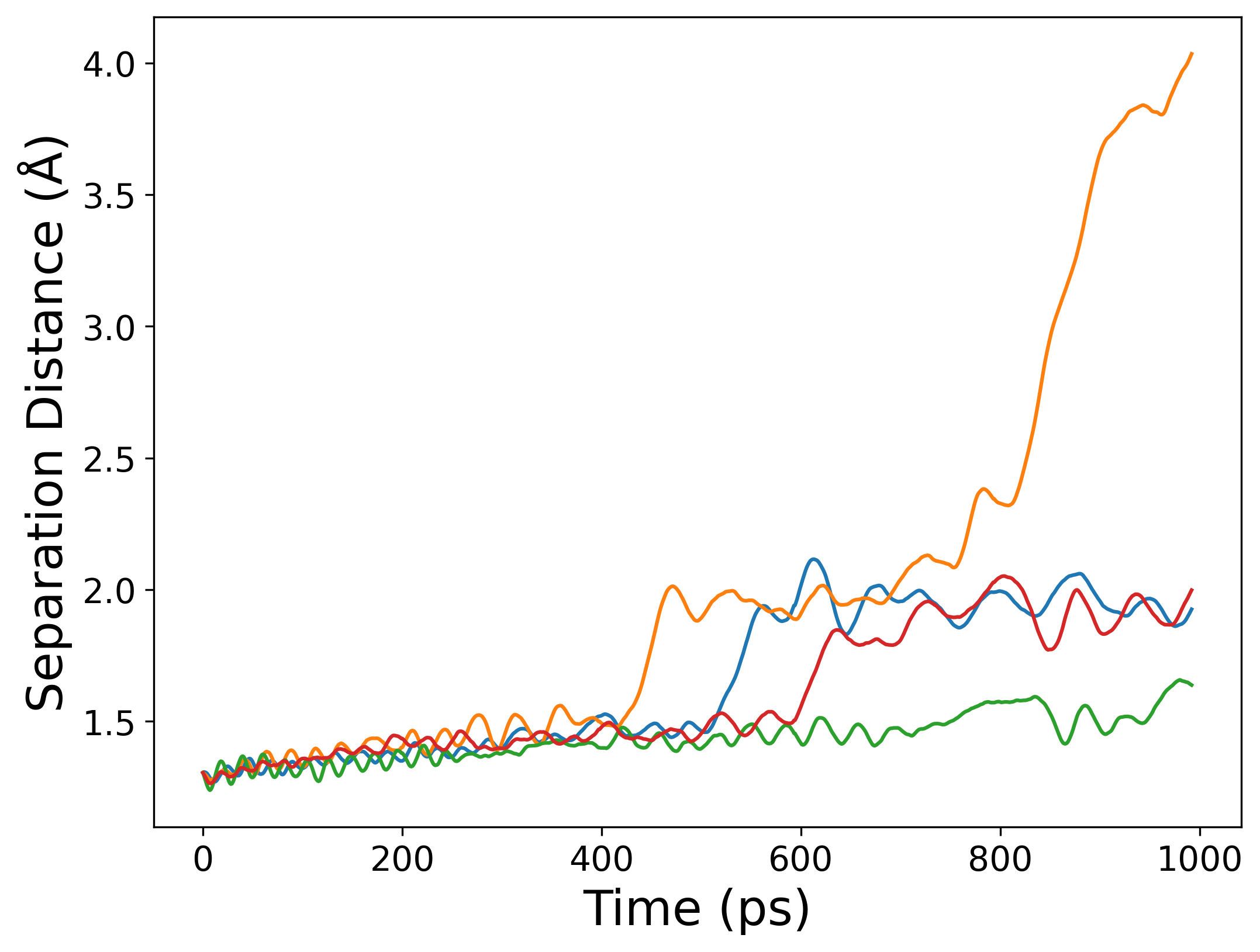}
    \caption{Separation distance between the atoms of each of the four initial O$_2$ molecules across simulation time. The noted increase in separation distance for one of the oxygen molecules after $\sim$1 ps demonstrates oxygen reduction.}
    \label{fig:oosep2x2plus1}
\end{figure}
In each extended system, we see the same outer-sphere mechanism as was observed in the zero and +1.0 V applied bias cases for the original test system. 

\newpage
\section*{Distribution of Electron Holes}
In Table \ref{table:mus} we show the distribution of holes between different carbon neighbors with respect to a central nitrogen. These populations are averages across all n$^{th}$-neighboring carbon atoms for all four nitrogen atoms present in the tested NG sheet over 0.5 ps of simulation time following the final injection of electrons. 

\begin{table}
    \caption{Mulliken charge averages in \textit{e} units for different carbon neighbors with respect to a central nitrogen during 0.5 ps of simulation time following the final injection of electrons showing the distribution of electron holes. First-neighbor carbon atoms have the largest positive Mulliken charge average, suggesting that electrons are primarily removed from those atoms. Smaller positive Mulliken charges on third- and fourth-neighbor carbons suggest that additional positive charge on the oxidized NG sheet can be de-localized to positions beyond the first-neighbor as well, though to a lesser degree.}
    \begin{tabular}{ c c }
     \hline
     Selected Atoms & \multicolumn{1}{p{3cm}}{\centering Average Mulliken \\ Charge}\\
     \hline
     Nitrogen Atoms&-0.093\\
     First-Neighbor Carbon Atoms & 0.202\\      
     Second-Neighbor Carbon Atoms& -0.019\\
     Third-Neighbor Carbon Atoms& 0.035\\
     Fourth-Neighbor Carbon Atoms& 0.015\\
     \hline
    \end{tabular}
    \label{table:mus}
\end{table}

\begin{figure}
        \centering
        \includegraphics[width=\columnwidth]{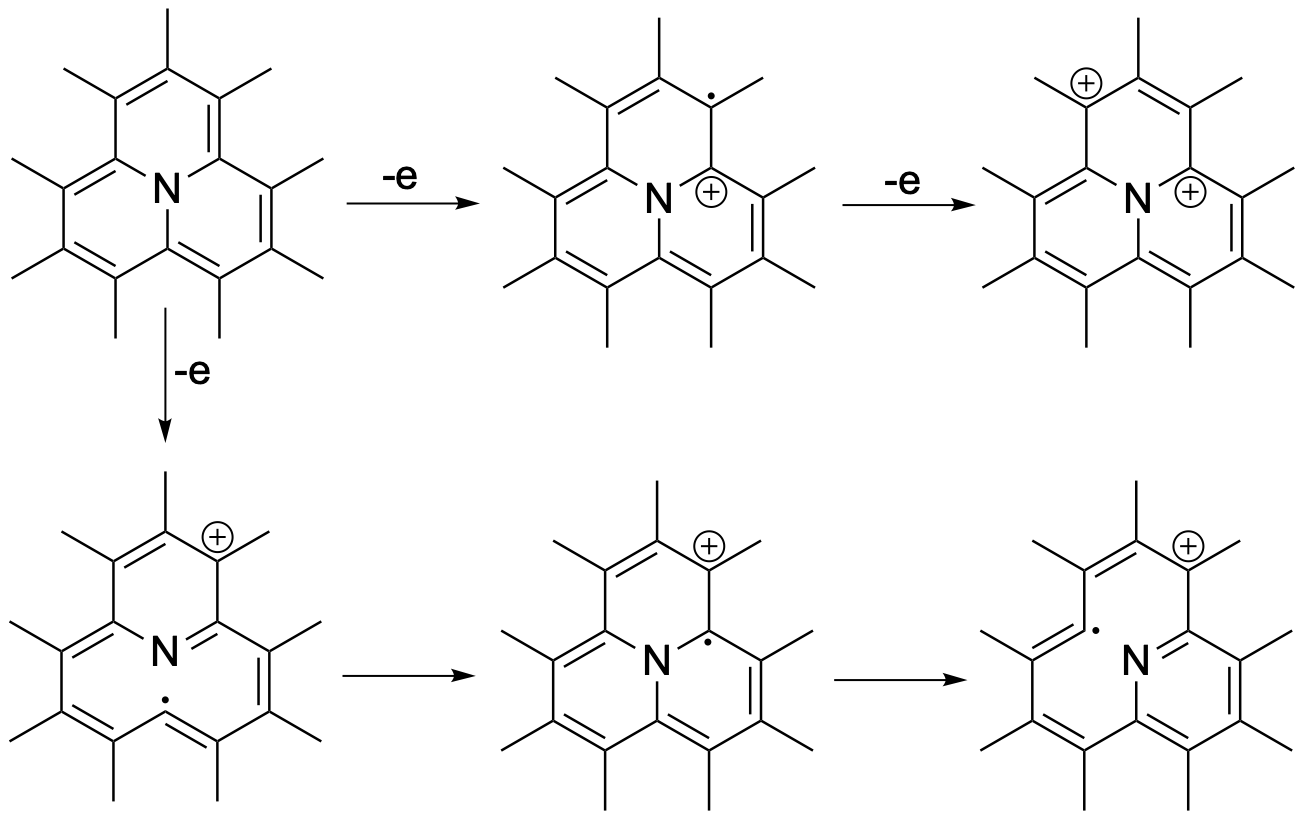}
        \caption{Possible resonance structures to explain the localization of the holes (positive charges) on the carbon atoms surrounding nitrogen doping atoms in the N-doped graphene sheet. These resonance structures show additional possibilities to those proposed in the main text, including how charge might be de-localized to have a positive charge on both even- and odd-neighboring carbons atoms.}
        \label{fig:mesomeric}
    \end{figure}

In Figure \ref{fig:mesomeric} we show different possible electron hole creation scenarios. Upon single-electron injection, a radical and a carbocation will be formed. If the carbocation is formed in the first neighbor, it can then be de-localized across all the other odd neighbors. To explain positive charge in even neighbors (i.e. the fourth neighbor carbons), one has to propose either a second carbocation formation or a structure as in the bottom left corner of Figure \ref{fig:mesomeric}. Note that if a carbocation were to be formed in the second neighbor, it would certainly lead to a more strained and unstable structure. The latter could explain why there is virtually no charge in the second neighbor, as evidenced in Table \ref{table:mus}.

\newpage
\section*{Complete Simulation with a $\mu_e$ Shift of -2.0 eV}
\begin{figure}
    \centering
    \includegraphics[width=0.75\textwidth]{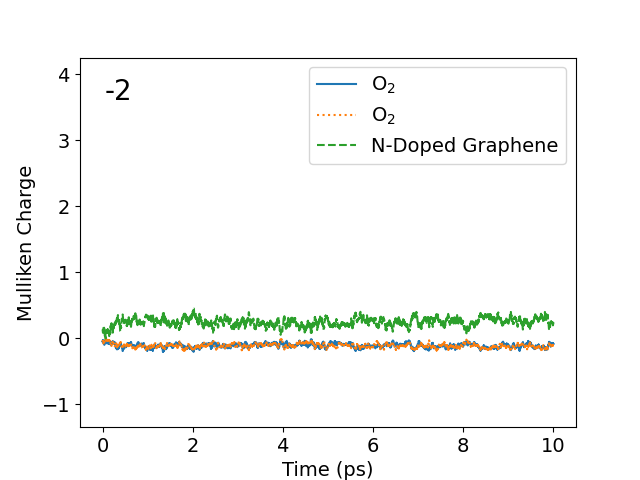}
    \caption{Mulliken charges in \textit{e} units across a 10 ps simulation time for both molecular oxygen atoms and the full NG sheet. For this simulation, we used an applied bias of -2.0 eV and did not observe the ORR in the 10 ps simulation time.}
    \label{fig:biasMinus2}
\end{figure}

Figure \ref{fig:biasMinus2} shows the Mulliken charges across the full 10 ps simulation time for the simulation of NG and O$_2$ in water with a $\mu_e$ shift of -2.0 eV, demonstrating that no ORR occurred across the tested simulation time. 

\newpage
\section*{Comparison of Adsorbed Single Oxygen Species}
\begin{figure}
    \centering
    \includegraphics[width=0.75\textwidth]{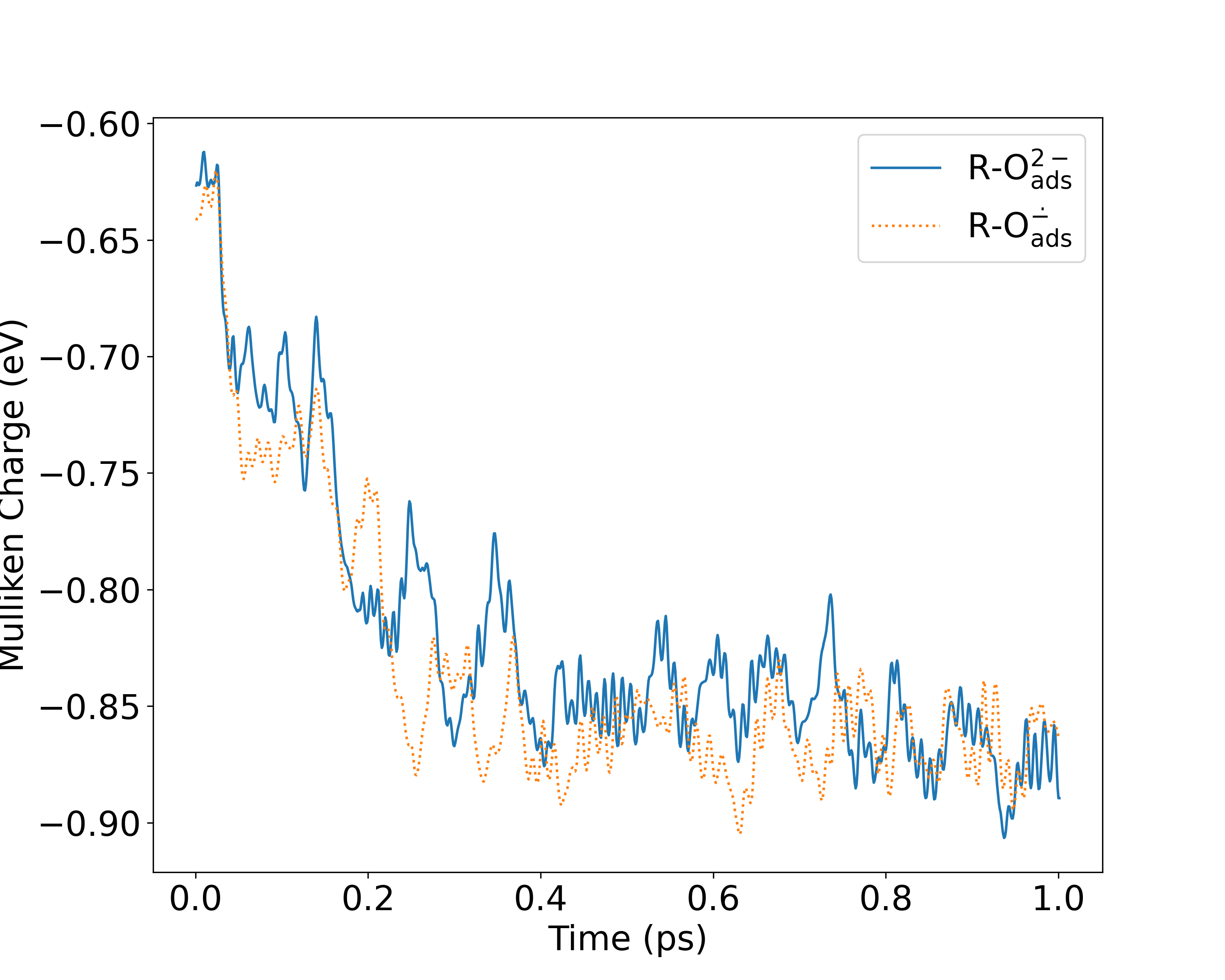}
    \caption{Mulliken charges in \textit{e} units for the two adsorbed oxygen species simulated in isolation across 1 ps of simulation time. In one simulation (solid blue line), an H$_3$O$^+$ was added to the water box to simulate O$^{2-}_{\rm ads}$. In the other simulation (dotted orange line), no ions or additional species beyond the sheet with the adsorbed oxygen and water were added in order to simulate O$^{\cdot-}_{\rm ads}$. Both species showed average Mulliken charges over the simulation time of $\sim$0.86 eV between 0.4 and 1.0 ps, following electronic rearrangement.}
    \label{fig:compOads}
\end{figure}

Figure \ref{fig:compOads} shows the Mulliken charges of the two different adsorbed single oxygen species simulated in isolation. Both species were simulated for 1 ps in a box of water using the same simulation conditions described for all other systems. In one simulation an H$_3$O$^+$ was added to the water box to simulate O$^{2-}_{\rm ads}$. In the other simulation, no ions or additional species beyond the sheet with the adsorbed oxygen and water were added in order to simulate O$^{\cdot-}_{\rm ads}$. Both species showed average Mulliken charges over the simulation time of $\sim$0.86 eV between 0.4 and 1.0 ps, following electronic rearrangement.

\newpage
\section*{Adsorption Times for Inner-Sphere Simulations}

\begin{figure}
    \centering
    \includegraphics[width=0.7\columnwidth]{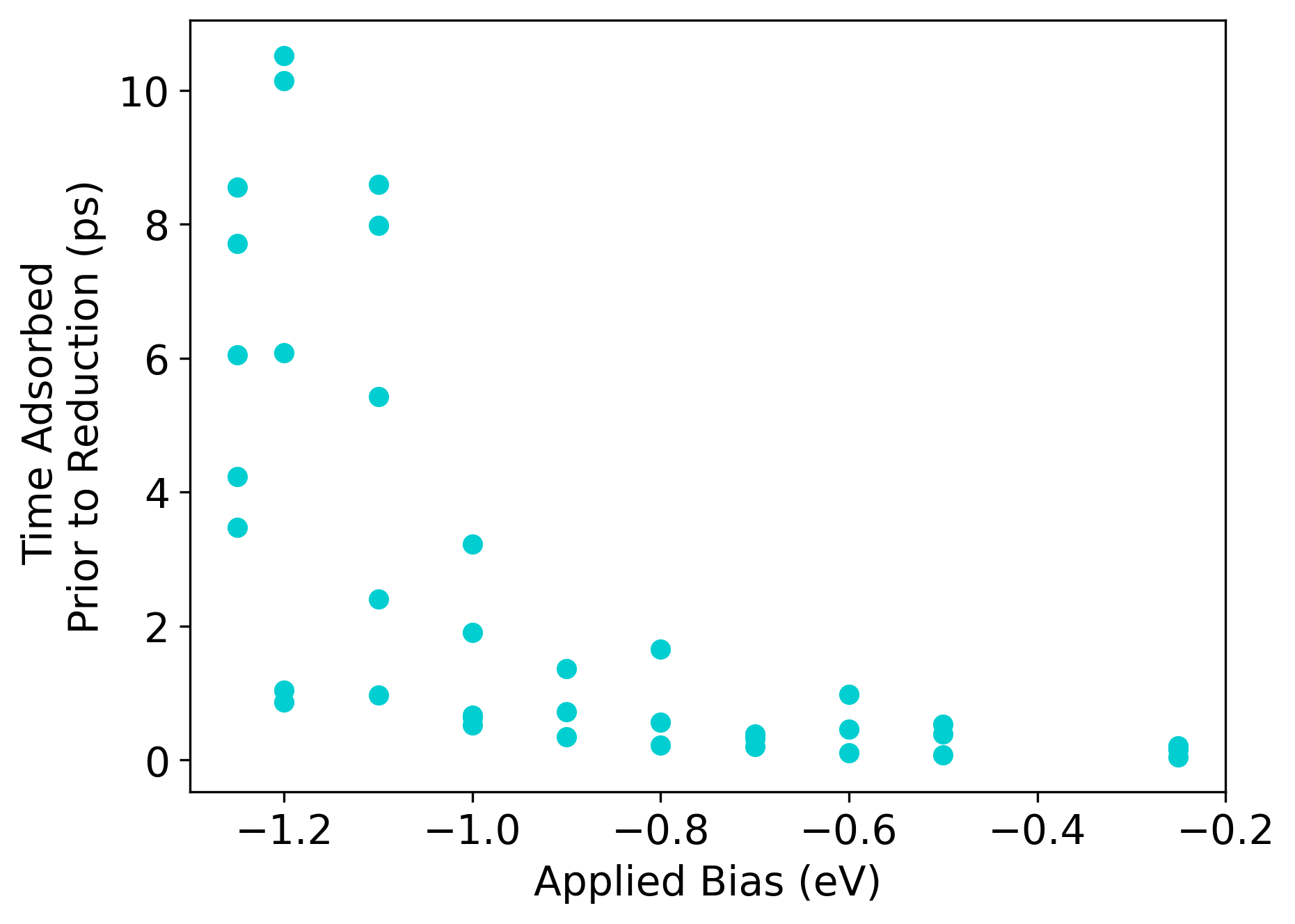}
    \caption{Applied bias vs adsorption time prior to reduction for all inner-sphere reaction simulations. Adsorption time prior to reduction tends to increase with lower applied bias.}
    \label{fig:adsTime}
\end{figure}

\newpage
\section*{Converting Applied Bias to Overpotential}

We use GPMDK to determine the Fermi energies, $\mu_e$, for various subsystems of the main and control systems (Table \ref{chempot}).

\begin{table}
    \caption{Values for the chemical potential of the electron, $\mu_e$ (eV), for the full simulation system (``All'') and various subsystems}
    \begin{tabular}{ c c }
     \hline
     Subsystem &$\mu_e$\\
     \hline
     All&-3.66\\
     N-Doped Graphene & -3.07\\      
     O$_2$& -6.47\\
     Water& -6.33\\
     Pure Carbon Graphene & -3.69\\
     \hline
    \end{tabular}
    \label{chempot}
\end{table}

\begin{figure}
    \centering
    \includegraphics[width=0.7\columnwidth]{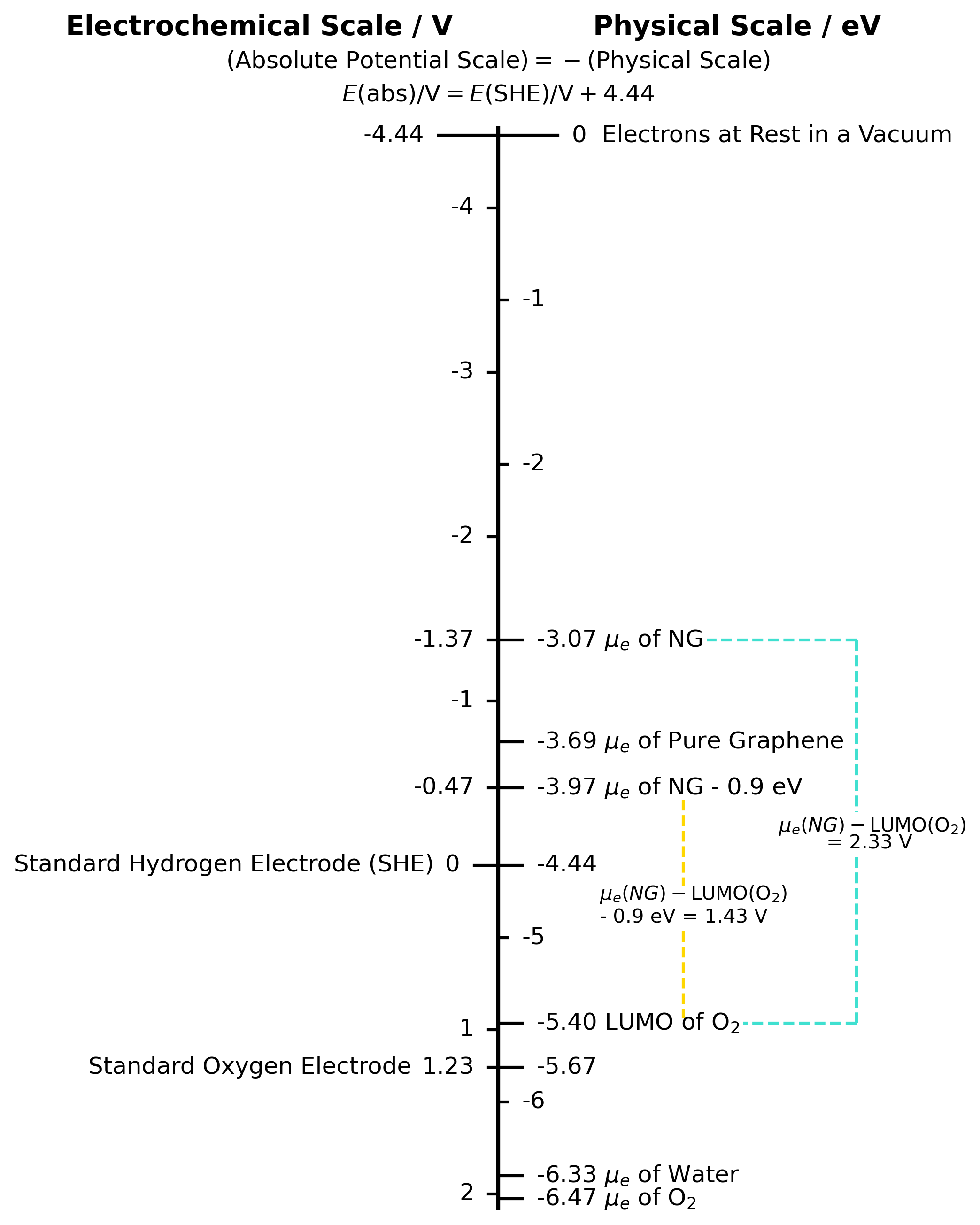}
    \caption{Placement of the Fermi energies, $\mu_e$, for the species listed in Table \ref{chempot} on the physical and electrochemical scales for comparison to SHE shifted values. Adapted from Trasatti \cite{Trasatti86}.}
    \label{fig:trasattiScale}
\end{figure}

By comparing these calculated Fermi energies, we can approximate the driving force of the reaction. At the beginning of the initial, unbiased simulations, the driving force for electron injection is given by the difference between $\mu_e$(NG) of the N-doped graphene sheet (given that $\mu_e$ is approximately equivalent to the HOMO of NG at the limit of a small gap) and the LUMO of O$_2$ (Figure \ref{fig:trasattiScale}); this difference is $\sim$ 2.33 eV. 
To expand on this a bit further, the reduction potential for the reversible hydrogen electrode $E(\mathrm{RHE})$ with respect to vacuum is at 4.44 eV \cite{Trasatti86} and ORR proceeds at $\leq$ 1.23 V with respect to the RHE. Combining these observations, we determine that ORR will proceed at $E(\mathrm{ORR})_{\mathrm{vacc}}$ = 5.67 eV. Given that $\mu_e$(NG) of -3.07 eV in Table \ref{chempot} corresponds to $E$(NG$^{+}$/NG)$_{vacc}$ = $\mu_e$(NG)/-e$_0$ = 3.07 eV, we can estimate that, in our initial unbiased simulations, an intrinsic potential difference of $E$(NG$^{+}$/NG) - $E$(ORR)$_{vacc}$ = -2.60 eV is in place. This intrinsic potential difference could be seen as, effectively, a highly reductive overpotential of 2.60 V for the ORR. Note that the driving force of 2.33 eV as determined from the electronic structure and the overpotential of 2.60 V are relatively close. 
A potential bias with a similar magnitude (-2.60 eV) should be applied to refill the positive charges accumulated on the NG sheet following the ORR to maintain a steady-state process. From the analysis above, we can infer that the observed outer-sphere mechanism is likely due to simulation conditions that translate to high overpotentials. Simulations using the method described in the "Applying an Electrochemical Bias" section to increase or reduce the bias on the NG sheet are discussed below. 

Applying a shift of -0.9 eV is equivalent to effectively reducing the $\mu_e(NG) - \mu_e(O_2)_{LUMO}$ difference from 2.33 eV to 1.43 eV. Using the $\mu_e$ values recorded in Table \ref{chempot}, an applied bias of -0.9 eV corresponds to an effective potential difference $E(\mathrm{NG}^{+}/\mathrm{NG})_{\mathrm{vacc}}$ - $E$(ORR)$_{\mathrm{vacc}}$ = -1.70 V (a reductive overpotential of 1.70 V, equivalent to running the ORR at -0.47 V vs RHE). 

\newpage
\section*{All Applied Biases Used for Production Simulations}

\begin{table}
    \caption{Applied biases used for inner- and outer-sphere simulations. All simulations were used to analyze the effects of applied bias on ORR mechanism and relative reaction rate. See Supporting Information section on "Converting Applied Bias to Overpotential" for an in depth discussion on how to convert applied biases into overpotentials.}
    \begin{tabular}{ c c }
     \hline
     \multicolumn{1}{p{3cm}}{\centering Inner-Sphere \\ Applied Biases (eV)} & \multicolumn{1}{p{3cm}}{\centering Outer-Sphere \\ Applied Biases (eV)}\\
     \hline
     -0.25& 0.00 \\
     -0.50 & 0.10 \\      
     -0.60 & 0.15 \\
     -0.70 & 0.20 \\
     -0.80 & 0.25 \\
     -0.90 & 0.30 \\
     -1.00 & 0.35 \\
     -1.10 & 0.40 \\
     -1.20 & 0.45 \\
     -1.25 & 0.50 \\
      & 0.55 \\
      & 0.60 \\
      & 0.65 \\
      & 0.70 \\
      & 0.75 \\
      & 1.00 \\
      & 1.50 \\
      & 2.00 \\
      & 2.50 \\ 
      & 3.00 \\
     \hline
    \end{tabular}
    \label{allAppliedBiases}
\end{table}

\newpage
\section*{Parameters for Marcus-Hush-Chidsey Fitting}

Outer-sphere reorganization energy, $\lambda_{OS} = 1.82$ eV

\noindent
Dirac cone shift = 0.4 eV 

\noindent
log(prefactor) of $\lambda_{OS}$ = 1.07


%% file: bibliography.bib
@misc{gpmdk,
  title       = "Graph-partitioning {MD} with Kernel ({GPMDK})",
  institution = "Github",
  howpublished = {\url{https://https://github.com/lanl/qmd-progress/tree/master/examples/gpmdk}},
  note = {Accessed: August 4, 2025},
   author = "Niklasson, A. M. N. and Negre C. F. A. and Cawkwell M. J. and Bock N. and Mniszewski S. M. and Wall M. E.",
  url	= {https://github.com/lanl/qmd-progress/tree/master/examples/gpmdk},
  language    = "en"
}

@misc{latte,
  title	= {LATTE},
  url		= {https://github.com/lanl/LATTE},
  howpublished =  {\url{https://github.com/lanl/LATTE}},
  author	= {Bock, N. and Cawkwell,  M. J.  and Coe, J. D.  and Krishnapriyan, A. and Kroonblawd, M. P.  and Lang, A. and Liu, C. and Martinez Saez, E. and Mniszewski, S. M. and Negre, C. F. A. and Niklasson, A. M. N.  and Sanville, E. and Wood, M. A. and Yang, P.},
  year	= {2008}
}

@article{Aradi07,
  author	= {Aradi, B. and Hourahine, B. and Frauenheim, Th.},
  title		= {DFTB+, a Sparse Matrix-Based Implementation of the DFTB Method},
  journal	= {J. Phys. Chem. A},
  volume	= {111},
  number	= {26},
  pages		= {5678-5684},
  year		= {2007},
  doi		= {10.1021/jp070186p},
  note		= {PMID: 17567110}
}

@article{Aradi15,
  author	= { Aradi, B. and Niklasson, A. M. N. and Frauenheim, Th.},
  title		= {Extended Lagrangian Density Functional Tight-Binding Molecular Dynamics for Molecules and Solids},
  journal	= {J. Chem. Theory Comput.},
  year		= {2015},
  volume	= {11},
  pages		= {3357-3363},
  number	= {7},
  doi		= {10.1021/acs.jctc.5b00324}
}

@book{Bockris00,
    author = {John O'M. Bockris and Amulya M. K. Reddy and Maria Gamboa-Aldeco},
    title = {Modern Electrochemistry 2A},
    publisher = {Springer},
    edition = {2nd},
    year = 2000
}

@article{Bonnet12,
  title = {First-Principles Molecular Dynamics at a Constant Electrode Potential},
  author = {Bonnet, Nic\'ephore and Morishita, Tetsuya and Sugino, Osamu and Otani, Minoru},
  journal = {Phys. Rev. Lett.},
  volume = {109},
  issue = {26},
  pages = {266101},
  numpages = {5},
  year = {2012},
  month = {Dec},
  publisher = {American Physical Society},
  doi = {10.1103/PhysRevLett.109.266101},
  url = {https://link.aps.org/doi/10.1103/PhysRevLett.109.266101}
}

@article{Boukhvalov2012-yw,
  title     = "Oxygen reduction reactions on pure and nitrogen-doped graphene: a first-principles modeling",
  author    = "Boukhvalov, D. W. and Son, Y-W.",
  journal   = "Nanoscale",
  publisher = "Royal Society of Chemistry (RSC)",
  volume    =  4,
  number    =  2,
  pages     = "417--420",
  month     =  jan,
  year      =  2012,
  language  = "en"
}

@article{Bredas14,
  author	= {Bredas, J-L.},
  title	= {Mind the gap!},
  journal	= {Mat. Horiz.},
  year	= {2014},
  volume	= {1},
  pages	= {17-19},
  number	= {},
  doi		= {10.1039/C3MH00098B}
}

@article{Breslin2023,
    author = {Alves, Daniele and Kasturi, P. Rupa and Collins, Gillian and Barwa, Tara N and Ramaraj, Sukanya and Karthik, Raj and Breslin, Carmel B.},
    title = {2D layered double hydroxides and transition metal dichalcogenides for applications in the electrochemical production of renewable hydrogen},
    journal = {Materials Advances},
    volume = {4},
    number = {24},
    pages = {6478-6497},
    year = {2023},
    month = {12},
    issn = {2633-5409},
    doi = {10.1039/d3ma00685a},
    url = {https://doi.org/10.1039/d3ma00685a},
    eprint = {https://pubs.rsc.org/ma/article-pdf/4/24/6478/9158851/d3ma00685a.pdf},
}

@article{CarParrinello85,
  title = {Unified Approach for Molecular Dynamics and Density-Functional Theory},
  author = {Car, R. and Parrinello, M.},
  journal = {Phys. Rev. Lett.},
  volume = {55},
  issue = {22},
  pages = {2471--2474},
  numpages = {0},
  year = {1985},
  publisher = {American Physical Society},
  doi = {10.1103/PhysRevLett.55.2471},
}

@article{Cawkwell2019,
    author = {Cawkwell, M. J. and Perriot, R.},
    title = "{Transferable density functional tight binding for carbon, hydrogen, nitrogen, and oxygen: Application to shock compression}",
    journal = {J. Chem. Phys.},
    volume = {150},
    number = {2},
    pages = {024107},
    year = {2019},
    month = {01},
    issn = {0021-9606},
    doi = {10.1063/1.5063385},
    url = {https://doi.org/10.1063/1.5063385},
    eprint = {https://pubs.aip.org/aip/jcp/article-pdf/doi/10.1063/1.5063385/14716411/024107\_1\_online.pdf},
}

@article{Chidsey1991,
    author = {C. E. D. Chidsey},
    title = {Free Energy and Temperature Dependence of Electron Transfer at the Metal-Electrolyte Interface},
    journal = {Science},
    volume = {251},
    pages = {919-922},
    year = 1991
}

@article{Choi14,
    author = {Choi, Chang Hyuck and Lim, Hyung-Kyu and Chung, Min Wook and Park, Jong Cheol and Shin, Hyeyoung and Kim, Hyungjun and Woo, Seong Ihl},
    title = {Long-Range Electron Transfer over Graphene-Based Catalyst for High-Performing Oxygen Reduction Reactions: Importance of Size, N-doping, and Metallic Impurities},
    journal = {Journal of the American Chemical Society},
    volume = {136},
    number = {25},
    pages = {9070-9077},
    year = {2014},
    doi = {10.1021/ja5033474},
    note ={PMID: 24905892},
}

@article{Clark2021-py,
  title    = "The Middle Science: Traversing Scale In Complex Many-Body Systems",
  author   = "Clark, A. E. and Adams, H. and Hernandez, R. and
              Krylov, A. I. and Niklasson, A. M. N. and Sarupria, S. and Wang, Y. and Wild, S. M. and Yang, Q.",
  journal  = "ACS Cent. Sci.",
  volume   =  7,
  number   =  8,
  pages    = "1271--1287",
  month    =  aug,
  year     =  2021,
  language = "en"
}

@article{CorriganGrove25,
    author = {Corrigan Grove, Rae A. and Moxley, Michael A. and Negre, Christian F. A. and Cawkwell, Marc J. and Niklasson, Anders M. N. and Mniszewski, Susan M. and Smith, Nathan and Prososki, Kevin and Wilson, Mark A. and Wall, Michael E.},
    title = {Combining Reactive Quantum-Mechanical Molecular-Dynamics Simulations with Mutagenesis, Crystallography, and Enzyme Kinetics to Reveal Plausible Steps of Isocyanide Hydratase Catalysis},
    journal = {Journal of Chemical Information and Modeling},
    volume = {65},
    number = {20},
    pages = {11079-11093},
    year = {2025},
    doi = {10.1021/acs.jcim.5c01152},
    note ={PMID: 41051317},
}

@article{Cui00,
  author	= {Cui, Q. and Elstner, M. and Kaxiras, E. and Frauenheim, Th. and Karplus, M.},
  title		= {A QM/MM Implementation of the Self-Consistent Charge Density Functional Tight Binding (SCC-DFTB) Method},
  journal	= {J. Phys. Chem. B},
  year		= {2000},
  volume	= {105},
  pages		= {569-585},
  number	= {2},
  doi		= {10.1021/jp0029109}
}

@article{Dai15,
    author = {Dai, Liming and Xue, Yuhua and Qu, Liangti and Choi, Hyun-Jung and Baek, Jong-Beom},
    title = {Metal-Free Catalysts for Oxygen Reduction Reaction},
    journal = {Chemical Reviews},
    volume = {115},
    number = {11},
    pages = {4823-4892},
    year = {2015},
    doi = {10.1021/cr5003563},
}

@article{Elstner98,
  author	= {Elstner, M. and Porezag, D. and Jungnickel, G. and Elsner, J. and Haugk, M. and Frauenheim, Th. and Suhai, S. and Seifert, G.},
  title	= {Self-consistent-charge density-functional tight-binding method for simulations of complex materials properties},
  journal	= {Phys. Rev. Lett.},
  year	= {1998},
  volume	= {58},
  pages	= {7260-7263},
  number	= {11},
  doi		= {10.1103/PhysRevB.58.7260}
}

@article{Elstner01,
  author	= {Elstner, M. and Jalkanen, K. J. and Knapp-Mohammady, M. and  Frauenheim, Th. and Suhai, S.},
  title		= {Energetics and structure of glycine and alanine based model peptides: Approximate SCC-DFTB, AM1 and PM3 methods in comparison with DFT, HF and MP2 calculations},
  journal	= {J. Phys. Chem. B},
  year		= {2001},
  volume	= {263},
  pages		= {203-219},
  number	= {2-3},
  doi		= {10.1016/S0301-0104(00)00375-X}
}

@article{Faisal17,
    author = {Shaikh Nayeem Faisal and Enamul Haque and Nikan Noorbehesht and Weimin Zhang and Andrew T. Harris and Tamara L. Church and Andrew I. Minett},
    title = {Pyridinic and graphitic nitrogen-rich graphene for high-performance supercapacitors and metal-free bifunctional electrocatalysts for ORR and OER},
    journal = {RSC Advances},
    issue = {29},
    doi = {10.1039/C7RA01355H},
    year = 2017
}

@article{JFinkelstein21,
  title={Bringing discrete-time Langevin splitting methods into agreement with thermodynamics},
  author={Finkelstein, J. and Cheng, C. and Fiorin, G. and Seibold, B. and Gr{\o}nbech-Jensen, N},
  journal={J. Chem. Phys.},
  volume={155},
  number={18},
  year={2021},
  publisher={AIP Publishing}
}

@article{Frauenheim00,
  author	= {Frauenheim, Th. and Seifert, G. and Elsterner, M.  and Hajnal, Z. and Jungnickel, G.  and Porezag, D.  and Suhai, S. and Scholz, R.},
  title	= {A Self‐Consistent Charge Density‐Functional Based
		  Tight‐Binding Method for Predictive Materials Simulations
		  in Physics, Chemistry and Biology},
  journal	= {Phys. Stat. Sol.},
  year		= {2000},
  volume	= {217},
  pages		= {41-43},
  number	= {1},
  doi		= {10.1002/(SICI)1521-3951(200001)217:1<41::AID-PSSB41>3.0.CO;2-V}
}

@article{Ganyecz2021-ne,
  title     = "Oxygen reduction reaction on {N}-doped graphene: Effect of positions and scaling relations of adsorption energies",
  author    = "Ganyecz, A. and Kállay, M.",
  journal   = "J. Phys. Chem. C Nanomater. Interfaces",
  publisher = "American Chemical Society (ACS)",
  volume    =  125,
  number    =  16,
  pages     = "8551--8561",
  month     =  apr,
  year      =  2021,
  language  = "en"
}

@article{Giannozzi09,
	Author = {Paolo Giannozzi and Stefano Baroni and Nicola Bonini and Matteo Calandra and Roberto Car and Carlo Cavazzoni and Davide Ceresoli and Guido L Chiarotti and Matteo Cococcioni and Ismaila Dabo and Andrea {Dal Corso} and Stefano de Gironcoli and Stefano Fabris and Guido Fratesi and Ralph Gebauer and Uwe Gerstmann and Christos Gougoussis and Anton Kokalj and Michele Lazzeri and Layla Martin-Samos and Nicola Marzari and Francesco Mauri and Riccardo Mazzarello and Stefano Paolini and Alfredo Pasquarello and Lorenzo Paulatto and Carlo Sbraccia and Sandro Scandolo and Gabriele Sclauzero and Ari P Seitsonen and Alexander Smogunov and Paolo Umari and Renata M Wentzcovitch},
	Journal = {Journal of Physics: Condensed Matter},
	Number = {39},
	Pages = {395502 (19pp)},
	Title = {QUANTUM ESPRESSO: a modular and open-source software project for quantum simulations of materials},
	Url = {http://www.quantum-espresso.org},
	Volume = {21},
	Year = {2009}
}

@article{Giannozzi17,
  author={P Giannozzi and O Andreussi and T Brumme and O Bunau and M Buongiorno Nardelli and M Calandra and R Car and C Cavazzoni and D Ceresoli and M Cococcioni and N Colonna and I Carnimeo and A Dal Corso and S de Gironcoli and P Delugas and R A DiStasio Jr and A Ferretti and A Floris and G Fratesi and G Fugallo and R Gebauer and U Gerstmann and F Giustino and T Gorni and J Jia and M Kawamura and H-Y Ko and A Kokalj and E Küçükbenli and M Lazzeri and M Marsili
  and N Marzari and F Mauri and N L Nguyen and H-V Nguyen and A Otero-de-la-Roza and L Paulatto and S Poncé and D Rocca and R Sabatini and B Santra and M Schlipf and A P Seitsonen and A Smogunov and I Timrov and T Thonhauser and P Umari and N Vast and X Wu and S Baroni},
  title={Advanced capabilities for materials modelling with QUANTUM ESPRESSO},
  journal={Journal of Physics: Condensed Matter},
  volume={29},
  number={46},
  pages={465901},
  url={http://stacks.iop.org/0953-8984/29/i=46/a=465901},
  year={2017},
}

@article{Giannozzi20,
 author = {Giannozzi,Paolo  and Baseggio,Oscar  and Bonfà,Pietro  and Brunato,Davide  and Car,Roberto  and Carnimeo,Ivan and Cavazzoni,Carlo  and de Gironcoli,Stefano  and Delugas,Pietro  and Ferrari Ruffino,Fabrizio  and Ferretti,Andrea and Marzari,Nicola  and Timrov,Iurii  and Urru,Andrea  and Baroni,Stefano },
 title = {Quantum ESPRESSO toward the exascale},
 journal = {The Journal of Chemical Physics},
 volume = {152},
 number = {15},
 pages = {154105},
 year = {2020},
 doi = {10.1063/5.0005082},
eprint = {https://doi.org/10.1063/5.0005082}
}

@article{Goyal14,
    author = {Goyal, Puja and Qian, Hu-Jun and Irle, Stephan and Lu, Xiya and Roston, Daniel and Mori, Toshifumi and Elstner, Marcus and Cui, Qiang},
    title = {Molecular Simulation of Water and Hydration Effects in Different Environments: Challenges and Developments for DFTB Based Models},
    journal = {The Journal of Physical Chemistry B},
    volume = {118},
    number = {38},
    pages = {11007-11027},
    year = {2014},
    doi = {10.1021/jp503372v},
    note ={PMID: 25166899},
}

@article{Grob23,
    author = {Gro\ss{}, Axel},
    title = {Challenges for ab initio molecular dynamics simulations of electrochemical interfaces},
    journal = {Current Opinion in Electrochemistry},
    volume = {40},
    year = 2023
}

@article{NGronbechJensen20,
  title={Complete set of stochastic Verlet-type thermostats for correct Langevin simulations},
  author={Gr{\o}nbech-Jensen, N.},
  journal={Mol. Phys.},
  volume={118},
  number={8},
  pages={e1662506},
  year={2020},
  publisher={Taylor \& Francis}
}

@article{Haunschild2019-rj,
  title     = "A comprehensive analysis of the history of {DFT} based on the bibliometric method {RPYS}",
  author    = "Haunschild, R. and Barth, A. and French, B.",
  journal   = "J. Cheminform.",
  publisher = "Springer Science and Business Media LLC",
  volume    =  11,
  number    =  1,
  pages     =  72,
  month     =  nov,
  year      =  2019,
  language  = "en"
}

@article{Henstridge12,
    title = {Marcus–Hush–Chidsey theory of electron transfer applied to voltammetry: A review},
    journal = {Electrochimica Acta},
    volume = {84},
    pages = {12-20},
    year = {2012},
    note = {ELECTROCHEMICAL SCIENCE AND TECHNOLOGYState of the Art and Future PerspectivesOn the occasion of the International Year of Chemistry (2011)},
    issn = {0013-4686},
    doi = {https://doi.org/10.1016/j.electacta.2011.10.026},
    author = {Martin C. Henstridge and Eduardo Laborda and Neil V. Rees and Richard G. Compton},
}

@article{HohenbergKohn64,
  author	= {Hohenberg, P. and Kohn, W.},
  title		= {Inhomogeneous Electron Gas},
  journal	= {Phys. Rev.},
  year		= {1964},
  volume	= {136},
  pages		= {B864-B870},
  number	= {3B},
  doi		= {10.1103/PhysRev.136.B864}
}

@article{Hormann19,
    author = {Hörmann, Nicolas G. and Andreussi, Oliviero and Marzari, Nicola},
    title = {Grand canonical simulations of electrochemical interfaces in implicit solvation models},
    journal = {The Journal of Chemical Physics},
    volume = {150},
    number = {4},
    pages = {041730},
    year = {2019},
    month = {01},
    issn = {0021-9606},
    doi = {10.1063/1.5054580},
    url = {https://doi.org/10.1063/1.5054580},
    eprint = {https://pubs.aip.org/aip/jcp/article-pdf/doi/10.1063/1.5054580/15556657/041730_1_online.pdf},
}

@article{Hourahine20,
    author = {Hourahine, B. and Aradi, B. and Blum, V. and Bonafé, F. and Buccheri, A. and Camacho, C. and Cevallos, C. and Deshaye, M. Y. and Dumitrică, T. and Dominguez, A. and Ehlert, S. and Elstner, M. and van der Heide, T. and Hermann, J. and Irle, S. and Kranz, J. J. and Köhler, C. and Kowalczyk, T. and Kubař, T. and Lee, I. S. and Lutsker, V. and Maurer, R. J. and Min, S. K. and Mitchell, I. and Negre, C. and Niehaus, T. A. and Niklasson, A. M. N. and Page, A. J. and Pecchia, A. and Penazzi, G. and Persson, M. P. and Řezáč, J. and Sánchez, C. G. and Sternberg, M. and Stöhr, M. and Stuckenberg, F. and Tkatchenko, A. and Yu, V. W.-z. and Frauenheim, T.},
    title = "{DFTB+, a software package for efficient approximate density functional theory based atomistic simulations}",
    journal = {J. Chem. Phys.},
    volume = {152},
    number = {12},
    pages = {124101},
    year = {2020},
    month = {03},
    issn = {0021-9606},
    doi = {10.1063/1.5143190},
    url = {https://doi.org/10.1063/1.5143190},
    eprint = {https://pubs.aip.org/aip/jcp/article-pdf/doi/10.1063/1.5143190/16711771/124101\_1\_online.pdf},
}

@article{Humphrey96,
    author={William Humphrey and Andrew Dalke and Klaus Schulten},
    title={{VMD} -- {V}isual {M}olecular {D}ynamics},
    journal={Journal of Molecular Graphics},
    year=1996,
    volume=14,
    pages={33-38},
    note={},
    tbstatus={Published.},
    techrep={},
    tbreference={222}
}

@book{Hutter_Book, 
    author={Marx, D. and Hutter, J.},
    place={Cambridge}, 
    title={Ab Initio Molecular Dynamics: Basic Theory and Advanced Methods}, 
    publisher={Cambridge University Press}, 
    year={2009},
    DOI={10.1017/CBO9780511609633}, 
}

@article{NGJensen13,
  title={A simple and effective Verlet-type algorithm for simulating Langevin dynamics},
  author={Gr{\o}nbech-Jensen, N. and Farago, O.},
  journal={Mol. Phys.},
  volume={111},
  number={8},
  pages={983--991},
  year={2013},
  publisher={Taylor \& Francis}
}

@article{ROJones89,
  title = {The density functional formalism, its applications and prospects},
  author = {Jones, R. O. and Gunnarsson, O.},
  journal = {Rev. Mod. Phys.},
  volume = {61},
  issue = {3},
  pages = {689--746},
  numpages = {0},
  year = {1989},
  month = {Jul},
  publisher = {American Physical Society},
  doi = {10.1103/RevModPhys.61.689},
  url = {https://link.aps.org/doi/10.1103/RevModPhys.61.689}
}

@article{KohnSham65,
  title = {Self-Consistent Equations Including Exchange and Correlation Effects},
  author = {Kohn, W. and Sham, L. J.},
  journal = {Phys. Rev.},
  volume = {140},
  issue = {4A},
  pages = {A1133--A1138},
  numpages = {0},
  year = {1965},
  month = {Nov},
  publisher = {American Physical Society},
  doi = {10.1103/PhysRev.140.A1133},
  url = {https://link.aps.org/doi/10.1103/PhysRev.140.A1133}
}

@article{Krishnapriyan17,
  author	= {Krishnapriyan, A. and Yang, P. and Niklasson, A. M. N. and Cawkwell, M. J.},
  title		= {Numerical Optimization of Density Functional Tight Binding Models: Application to Molecules Containing Carbon, Hydrogen, Nitrogen, and Oxygen},
  journal	= {J. Chem. Theory Comput.},
  year		= {2017},
  volume	= {13},
  pages		= {6191-6200},
  number	= {12},
  doi		= {10.1021/acs.jctc.7b00762}
}

@article{Kronberg21,
    author = {Kronberg, Rasmus and Laasonen, Kari},
    title = {Reconciling the Experimental and Computational Hydrogen Evolution Activities of Pt(111) through DFT-Based Constrained MD Simulations},
    journal = {ACS Catalysis},
    volume = {11},
    number = {13},
    pages = {8062-8078},
    year = {2021},
    doi = {10.1021/acscatal.1c00538},
}

@article{Kudin02,
  author	= {Kudin, K. N. and Scuseria, G. E.},
  title		= {A black-box self-consistent field convergence algorithm:
		  One step closer},
  journal	= {Phys. Rev. Lett.},
  year		= {2002},
  volume	= {116},
  pages		= {8255-8256},
  number	= {19},
  doi		= {10.1063/1.1470195}
}

@article{Kumeda2023-ib,
  title     = "Cations determine the mechanism and selectivity of alkaline
               oxygen reduction reaction on Pt(111)",
  author    = "Kumeda, Tomoaki and Laverdure, Laura and Honkala, Karoliina and
               Melander, Marko M and Sakaushi, Ken",
  journal   = "Angew. Chem. Int. Ed Engl.",
  publisher = "Wiley",
  volume    =  62,
  number    =  51,
  pages     = "e202312841",
  month     =  nov,
  year      =  2023,
  language  = "en"
}

@article{Liao22,
    author = {Liao, Xiaobin and Lu, Ruihu and Xia, Lixue and Liu, Qian and Wang, Huan and Zhao, Kristin and Wang, Zhaoyang and Zhao, Yan},
    title = {Density Functional Theory for Electrocatalysis},
    journal = {ENERGY \& ENVIRONMENTAL MATERIALS},
    volume = {5},
    number = {1},
    pages = {157-185},
    doi = {https://doi.org/10.1002/eem2.12204},
    url = {https://onlinelibrary.wiley.com/doi/abs/10.1002/eem2.12204},
    eprint = {https://onlinelibrary.wiley.com/doi/pdf/10.1002/eem2.12204},
    year = {2022}
}

@article{Liu2014,
    author = {Fang, Ya-Hui and Liu, Zhi-Pan},
    title = {Tafel Kinetics of Electrocatalytic Reactions: From Experiment to First-Principles},
    journal = {ACS Catalysis},
    volume = {4},
    number = {12},
    pages = {4364-4376},
    year = {2014},
    month = {10},
    issn = {2155-5435},
    doi = {10.1021/cs501312v},
    url = {https://doi.org/10.1021/cs501312v},
    eprint = {https://pubs.acs.org/accacs/article-pdf/4/12/4364/16692416/cs501312v.pdf},
}

@article{Low24,
    author = {Low, J. L. and Roth, C. and Paulus, B.},
    title = {Exploring the Inner- and Outer-Sphere Mechanistic Pathways of ORR on M-N-Cs with Pyrrolic MN4 Motifs},
    journal = {J. Phys. Chem. C.},
    volume = {128},
    pages = {5075-5083},
    year = {2024}
}

@article{Lozovoi01,
    author = {Lozovoi, A. Y. and Alavi, A. and Kohanoff, J. and Lynden-Bell, R. M.},
    title = {Ab initio simulation of charged slabs at constant chemical potential},
    journal = {The Journal of Chemical Physics},
    volume = {115},
    number = {4},
    pages = {1661-1669},
    year = {2001},
    month = {07},
    issn = {0021-9606},
    doi = {10.1063/1.1379327},
    url = {https://doi.org/10.1063/1.1379327},
    eprint = {https://pubs.aip.org/aip/jcp/article-pdf/115/4/1661/19076484/1661_1_online.pdf},
}

@article{Lu18,
    author = {Xiangyu Lu and Dan Wang and Liping Ge and Lihui Xiao and Haiyan Zhang and Lilai Liu and Jinqiu Zhang and Maozhong An and Peixia Yang},
    title = {Enriched graphitic N in nitrogen-doped graphene as a superior metal-free electrocatalyst for the oxygen reduction reaction},
    journal = {New Journal of Chemistry},
    doi = {10.1039/C8NJ04857F},
    issue = {24},
    year = 2018
}

@article{Ma2022-xo,
  title     = "{N}-doped graphene for electrocatalytic O$_2$ and CO$_2$ reduction",
  author    = "Ma, R. and Wang, K. and Li, C. and Wang, C.
               and Habibi-Yangjeh, A. and Shan, G.",
  journal   = "Nanoscale Adv.",
  publisher = "Royal Society of Chemistry (RSC)",
  volume    =  4,
  number    =  20,
  pages     = "4197--4209",
  month     =  oct,
  year      =  2022,
  language  = "en"
}

@article{Man2020-py,
  title     = "First principle studies of oxygen reduction reaction on {N} doped graphene: Impact of {N} concentration, position and co-adsorbate effect",
  author    = "Man, I-C. and Trancă, I. and Soriga, S-G.",
  journal   = "Appl. Surf. Sci.",
  publisher = "Elsevier BV",
  volume    =  510,
  number    =  145470,
  pages     =  145470,
  month     =  apr,
  year      =  2020,
  language  = "en"
}

@article{Markovic01,
    author = {Marković, N. M. and Schmidt, T. J. and Stamenković, V. and Ross, P. N.},
    title = {Oxygen Reduction Reaction on Pt and Pt Bimetallic Surfaces: A Selective Review},
    journal = {Fuel Cells},
    volume = {1},
    number = {2},
    pages = {105-116},
    doi = {https://doi.org/10.1002/1615-6854(200107)1:2<105::AID-FUCE105>3.0.CO;2-9},
    url = {https://onlinelibrary.wiley.com/doi/abs/10.1002/1615-6854%28200107%291%3A2%3C105%3A%3AAID-FUCE105%3E3.0.CO%3B2-9},
    eprint = {https://onlinelibrary.wiley.com/doi/pdf/10.1002/1615-6854%28200107%291%3A2%3C105%3A%3AAID-FUCE105%3E3.0.CO%3B2-9},
    year = {2001}
}

@article{Martinez2015-vm,
  title     = "Thermostating extended Lagrangian Born-Oppenheimer molecular dynamics",
  author    = "Martínez, E. and Cawkwell, M. J. and Voter, A. F. and
               Niklasson, A. M. N.",
  journal   = "J. Chem. Phys.",
  publisher = "AIP Publishing",
  volume    =  142,
  number    =  15,
  pages     =  154120,
  month     =  {Apr},
  year      =  2015,
}

@article{Matanovic18,
  author	= {Matanovic, I. and Artyushkova, K. and Atanassov, P.},
  title		= {Understanding PGM-free catalysts by linking density functional theory calculations and structural analysis: Perspectives and challenges},
  journal	= {Curr. Opin. Electrochem.},
  year	= {2018},
  volume	= {9},
  pages	= {137-144},
  number	= {11169},
  doi		= {10.1016/j.coelec.2018.03.009}
}

@article{Melander19,
    author = {Melander, Marko M. and Kuisma, Mikael J. and Christensen, Thorbjørn Erik Køppen and Honkala, Karoliina},
    title = {Grand-canonical approach to density functional theory of electrocatalytic systems: Thermodynamics of solid-liquid interfaces at constant ion and electrode potentials},
    journal = {The Journal of Chemical Physics},
    volume = {150},
    number = {4},
    pages = {041706},
    year = {2018},
    month = {11},
    issn = {0021-9606},
    doi = {10.1063/1.5047829},
    url = {https://doi.org/10.1063/1.5047829},
    eprint = {https://pubs.aip.org/aip/jcp/article-pdf/doi/10.1063/1.5047829/15553742/041706_1_online.pdf},
}

@article{Montemore17,
    author = {Montemore, M.
    M. and van Spronsen, M. A. and Madix, R. J. and Friend, C. M.},
    title = {O$_2$ Activation by Metal Surfaces: Implications for Bonding and Reactivity on Heterogeneous Catalysts},
    journal = {Chem. Rev.},
    volume = {118},
    number = {5},
    pages = {2816-2862},
    year = {2018},
    doi = {10.1021/acs.chemrev.7b00217},
}

@article{Morozan11,
  author	= {Morozan, A. and Jousselme, B. and Palacin, S. },
  title		= {Low-platinum and platinum-free catalysts for the oxygen reduction reaction at fuel cell cathodes},
  journal	= {Energy Environ. Sci.},
  year		= {2011},
  volume	= {4},
  pages		= {1238-1254},
  number	= {},
  doi		= {10.1039/C0EE00601G}
}

@article{Mosyak96,
    author = {Mosyak, A. and Nitzan, A. and Kosloff, R.},
    title = {Numerical simulations of electron tunneling in water},
    journal = {J. Chem. Phys.},
    volume = {104},
    number = {4},
    pages = {1549-1559},
    year = {1996},
    doi = {10.1063/1.470743},
}

@article{Negre11,
  author	= {Negre, C. F. A.  and Jara, G. E. and  Vera, D. M. A. and Pierini, A. B. and S{\'{a}}nchez, C. G. },
  title		= {Detailed analysis of water structure in a solvent mediated electron tunneling mechanism},
  journal	= {J. Phys.: Condens. Matter},
  pages		= {245305},
  year		= {2011},
  volume	= {23},
  number	= {24},
  doi		= {10.1088/0953-8984/23/24/245305}
}

@article{Negre23,
    author = {Negre, C. F. A. and Wall, M. E. and Niklasson, A. M. N.},
    title = "{Graph-based quantum response theory and shadow Born–Oppenheimer molecular dynamics}",
    journal = {J. Chem. Phys.},
    volume = {158},
    number = {7},
    pages = {074108},
    year = {2023},
    month = {02},
    issn = {0021-9606},
    doi = {10.1063/5.0137119},
    url = {https://doi.org/10.1063/5.0137119}
}

@article{Niklasson08,
  author	= {Niklasson, A. M. N. },
  title		= {Extended Born-Oppenheimer Molecular Dynamics},
  journal	= {Phys. Rev. Lett.},
  year		= {2008},
  volume	= {100},
  pages		= {123004},
  number	= {12},
  doi		= {110.1103/PhysRevLett.100.123004}
}

@article{Niklasson09,
  author	= {Niklasson, A. M. N. and Steneteg, P. and Odell, A. and Bock, N. and Challacombe, M. and Tymczak, C. J. and Holmstr\"{o}m, E. and Zheng, G. and Weber, V.},
  title		= {Extended Lagrangian Born-Oppenheimer molecular dynamics with dissipation},
  journal	= {J. Chem. Phys.},
  year		= {2009},
  volume	= {130},
  pages		= {214109},
  number	= {21},
  doi		= {10.1063/1.3148075}
}

@article{Niklasson14,
  author	= {Niklasson, A. M. N. and Cawkwell, M. J.},
  title		= {Generalized extended Lagrangian Born-Oppenheimer molecular dynamics},
  journal	= {J. Chem. Phys.},
  year		= {2014},
  volume	= {141},
  pages		= {1-5},
  number	= {164123},
  doi		= {10.1063/1.4898803}
}

@article{Niklasson15,
  author	= {Niklasson, A. M. N.  and Cawkwell, M. J. and 
		  Rubensson, E. H. and Rudberg, E.},
  title		= {Canonical density matrix perturbation theory},
  journal	= {Phys. Rev. E.},
  year		= {2015},
  volume	= {92},
  pages		= {1-2},
  number	= {063301},
  doi		= {10.1103/PhysRevE.92.063301}
}

@article{Niklasson17,
  author	= {Niklasson, A. M. N. },
  title	= {Next generation extended Lagrangian first principles molecular dynamics},
  journal	= {J. Chem. Phys.},
  year	= {2017},
  volume	= {147},
  pages	= {1-5},
  number	= {054103},
  doi		= {10.1063/1.4985893}
}

@article{ANiklasson20,
    author = {Niklasson, Anders M. N.},
    title = {Extended Lagrangian Born–Oppenheimer molecular dynamics using a Krylov subspace approximation},
    journal = {The Journal of Chemical Physics},
    volume = {152},
    number = {10},
    pages = {104103},
    year = {2020},
    month = {03},
    issn = {0021-9606},
    doi = {10.1063/1.5143270},
    url = {https://doi.org/10.1063/1.5143270},
    eprint = {https://pubs.aip.org/aip/jcp/article-pdf/doi/10.1063/1.5143270/15572649/104103_1_online.pdf},
}

@article{Niklasson21,
    author = {Niklasson, A. M.},
    title = {Extended Lagrangian Born-Oppenheimer molecular dynamics: from density functional theory to charge relaxation models},
    journal = {Eur. Phys. J. B},
    volume = {94},
    issue = {164},
    year = {2021},
    doi = {10.1140/epjb/s10051-021-00151-6}
}

@article{ANiklasson23,
    author = {Niklasson, A. M. N. and Negre, C. F. A.},
    title = "{Shadow energy functionals and potentials in Born–Oppenheimer molecular dynamics}",
    journal = {J. Chem. Phys.},
    volume = {158},
    number = {15},
    pages = {154105},
    year = {2023},
    month = {04},
    issn = {0021-9606},
    doi = {10.1063/5.0146431},
    url = {https://doi.org/10.1063/5.0146431},
    eprint = {https://pubs.aip.org/aip/jcp/article-pdf/doi/10.1063/5.0146431/19665239/154105\_1\_5.0146431.pdf},
}

@article{Nong20,
    author = {Hong Nhan Nong and Lorenz J. Falling and Arno Bergmann and Malte Klingenhof and Hoang Phi Tran and Camillo Sp{\\"o}ri and Rik Mom and Janis Timoshenko and Guido Zichittella and Axel Knop-Gericke and Simone Piccinin and Javier P{\'{e}}rez-Ram{\'{i}}rez and Beatriz Roldan Cuenya and Robert Schl{\\"o}gl and Peter Strasser and Detre Teschner and Travis E. Jones},
    title = {Key role of chemistry versus bias in electrocatalytic oxygen evolution},
    journal = {Nature},
    pages = {408-413},
    doi = {10.1038/s41586-020-2908-2},
    year = 2020
}

@article{Norskov2004-qu,
  title     = "Origin of the overpotential for oxygen reduction at a fuel-cell cathode",
  author    = "Nørskov, J. K. and Rossmeisl, J. and Logadottir, A. and Lindqvist, L. and Kitchin, J. R. and Bligaard, T. and Jónsson, H.",
  journal   = "J. Phys. Chem. B",
  publisher = "American Chemical Society (ACS)",
  volume    =  108,
  number    =  46,
  pages     = "17886--17892",
  month     =  nov,
  year      =  2004,
  language  = "en"
}

@article{Okamoto2009-tg,
  title     = "First-principles molecular dynamics simulation of O$_2$ reduction on nitrogen-doped carbon",
  author    = "Okamoto, Y.",
  journal   = "Appl. Surf. Sci.",
  publisher = "Elsevier BV",
  volume    =  256,
  number    =  1,
  pages     = "335--341",
  month     =  oct,
  year      =  2009,
  language  = "en"
}

@book{RParr89,
	author = {R. G. Parr and W. Yang},
	title = {Density-functional theory of atoms and molecules},
	publisher = {Oxford University Press},
	year = 1989,
	address = {Oxford}
}

@article{Perriot2018-cg,
  title     = "Density functional tight binding calculations for the simulation of shocked nitromethane with {LATTE}-{LAMMPS}",
  author    = "Perriot, R. and Negre, C. F. A. and McGrane, S. D. and Cawkwell, M. J.",
  journal   = "AIP Conf. Proc.",
  publisher = "American Institute of Physics",
  volume    =  1979,
  number    =  1,
  pages     =  050014,
  month     =  jul,
  year      =  2018
}

@article{prandini2018precision,
   title={Precision and efficiency in solid-state pseudopotential calculations},
   author={Prandini, Gianluca and Marrazzo, Antimo and Castelli, Ivano E and Mounet, Nicolas and Marzari, Nicola},
   journal={npj Computational Materials},
   volume={4},
   number={1},
   pages={72},
   year={2018},
   issn = {2057-3960},
   url = {https://www.nature.com/articles/s41524-018-0127-2},
   doi = {10.1038/s41524-018-0127-2},
   note = {\href{http://materialscloud.org/sssp}{http://materialscloud.org/sssp}},
   publisher={Nature Publishing Group UK London}
}

@article{Qu10,
    author = {Liangti Qu and Jong-Beom Baek and Liming Dai},
    title = {Nitrogen-Doped Graphene as Efficient Metal-Free Electrocatalyst for Oxygen Reduction in Fuel Cells},
    journal = {ACS Nano},
    volume = {4},
    number = {3},
    pages = {1321-1326},
    doi = {10.1021/nn901850u},
    year = 2010
}

@article{Rabuck99,
  author	= {Rabuck, A. D. and Scuseria, G. E. },
  title	= {Improving self-consistent field convergence by varying occupation numbers},
  journal	= {J. Chem. Phys.},
  year		= {1999},
  volume	= {110},
  pages		= {695-696},
  number	= {2},
  doi		= {10.1063/1.478177}
}

@article{Radjenovic19,
    author ="Radjenovic, Petar M. and Hardwick, Laurence J.",
    title  ="Evaluating chemical bonding in dioxides for the development of metal–oxygen batteries: vibrational spectroscopic trends of dioxygenyls{,} dioxygen{,} superoxides and peroxides",
    journal  ="Phys. Chem. Chem. Phys.",
    year  ="2019",
    volume  ="21",
    issue  ="3",
    pages  ="1552-1563",
    publisher  ="The Royal Society of Chemistry",
    doi  ="10.1039/C8CP04652B",
    url  ="http://dx.doi.org/10.1039/C8CP04652B",
}

@article{Ramaswamy01,
  author	= {Ramaswamy, N. and Mukerjee, S.},
  title	= {Influence of Inner- and Outer-Sphere Electron Transfer Mechanisms during Electrocatalysis of Oxygen Reduction in Alkaline Media},
  journal	= {J. Phys. Chem. C.},
  year		= {2001},
  volume	= {115},
  pages		= {18015-18026},
  number	= {36},
  doi		= {10.1021/jp204680p}
}

@article{Reyimjan06,
  author	= {Reyimjan, A. S. and Alfred, B. A. and Nalini, P. S. and Swaminatha, P. K. and Branko, N. P.},
  title	= {O$_2$ Reduction on Graphite and Nitrogen-Doped Graphite: Experiment and Theory},
  journal	= {J. Phys. Chem. B.},
  year		= {2006},
  volume	= {110},
  pages		= {1787-1793},
  number	= {4},
  doi		= {10.1021/jp055150g}
}

@article{Roman17,
    author = {Román, Alex M. and Dudoff, Jessica and Baz, Adam and Holewinski, Adam},
    title = {Identifying “Optimal” Electrocatalysts: Impact of Operating Potential and Charge Transfer Model},
    journal = {ACS Catalysis},
    volume = {7},
    number = {12},
    pages = {8641-8652},
    year = {2017},
    doi = {10.1021/acscatal.7b03235},
}

@book{SchmicklerSantos2010,
    author = {W. Schmickler and E. Santos},
    title = {Interfacial Electrochemistry},
    publisher = {Springer-Verlag Berlin, Heidelberg},
    doi = {10.1007/978-3-642-04937-8},
    edition = {2},
    year = {2010}
}

@article{Schwarz2020,
    title = {The electrochemical interface in first-principles calculations},
    journal = {Surface Science Reports},
    volume = {75},
    number = {2},
    pages = {100492},
    year = {2020},
    issn = {0167-5729},
    doi = {https://doi.org/10.1016/j.surfrep.2020.100492},
    url = {https://www.sciencedirect.com/science/article/pii/S0167572920300133},
    author = {Kathleen Schwarz and Ravishankar Sundararaman},
}

@article{Singh2024-ms,
  title     = "Surface studies of beta-1,3,5,7-tetranitro-1,3,5,7-tetrazoctane and pentaerythritol tetranitrate from density functional tight-binding calculations and implications on crystal shape",
  author    = "Singh, H. and Negre, C. F. A. and Redondo, A. and
               Perriot, R.",
  journal   = "Cryst. Growth Des.",
  publisher = "American Chemical Society (ACS)",
  volume    =  24,
  number    =  9,
  pages     = "3681--3690",
  month     =  may,
  year      =  2024,
  language  = "en"
}

@article{Sivak,
    author = {Sivak, D. A. and Chodera, J. D. and Crooks, G. E.},
    title = {Time Step Rescaling Recovers Continuous-Time Dynamical Properties for Discrete-Time Langevin Integration of Nonequilibrium Systems},
    journal = {J. Phys. Chem. B},
    volume = {118},
    number = {24},
    pages = {6466-6474},
    year = {2014},
    doi = {10.1021/jp411770f},
    note ={PMID: 24555448},
    URL = { https://doi.org/10.1021/jp411770f},
    eprint = {https://doi.org/10.1021/jp411770f}
}

@Inbook{Song08,
    author="Song, C. and Zhang, J.",
    editor="Zhang, Jiujun",
    title="Electrocatalytic Oxygen Reduction Reaction",
    bookTitle="PEM Fuel Cell Electrocatalysts and Catalyst Layers: Fundamentals and Applications",
    year="2008",
    publisher="Springer London",
    address="London",
    pages="89--134",
    isbn="978-1-84800-936-3",
    doi="10.1007/978-1-84800-936-3_2",
}

@article{Taylor06,
  title = {First principles reaction modeling of the electrochemical interface: Consideration and calculation of a tunable surface potential from atomic and electronic structure},
  author = {Taylor, Christopher D. and Wasileski, Sally A. and Filhol, Jean-Sebastien and Neurock, Matthew},
  journal = {Phys. Rev. B},
  volume = {73},
  issue = {16},
  pages = {165402},
  numpages = {16},
  year = {2006},
  month = {Apr},
  publisher = {American Physical Society},
  doi = {10.1103/PhysRevB.73.165402},
  url = {https://link.aps.org/doi/10.1103/PhysRevB.73.165402}
}

@article{Tian25,
    author = {Tian, Yumiao and Hou, Pengfei and Zhou, Yuanyuan and Li, Quan},
    title = {Alkali Metal Cations Induce Pseudo-outer-sphere Oxygen Reduction Reaction Mechanism in Electrolyte-Catalyst Synergy},
    journal = {Journal of the American Chemical Society},
    volume = {147},
    number = {38},
    pages = {35118-35127},
    year = {2025},
    doi = {10.1021/jacs.5c12947},
    note ={PMID: 40947988},
}

@article{Trasatti86,
    author = {Trasatti, S.},
    title = {The absolute electrode potential: an explanatory note},
    journal = {PAC},
    volume = {58},
    pages = {955-969},
    year = {1986}
}

@article{Vegge18,
    title = {DFT study of stabilization effects on N-doped graphene for ORR catalysis},
    journal = {Catalysis Today},
    volume = {312},
    pages = {118-125},
    year = {2018},
    note = {Computational Catalysis at NAM25},
    issn = {0920-5861},
    doi = {https://doi.org/10.1016/j.cattod.2018.02.015},
    url = {https://www.sciencedirect.com/science/article/pii/S092058611830066X},
    author = {Mateusz Reda and Heine Anton Hansen and Tejs Vegge},
}

@article{Yang19,
    author = {Yang, L. and Shui, J. and Du, L. and Shao, Y. and Liu, J. and Dai, L. and Hu, Z.},
    title = {Carbon-Based Metal-Free ORR Electrocatalysts for Fuel Cells: Past, Present, and Future},
    journal = {Adv. Mater.},
    volume = {31},
    number = {13},
    pages = {1804799},
    doi = {https://doi.org/10.1002/adma.201804799},
    url = {https://onlinelibrary.wiley.com/doi/abs/10.1002/adma.201804799},
    eprint = {https://onlinelibrary.wiley.com/doi/pdf/10.1002/adma.201804799},
    year = {2019}
}

@article{Yeh11,
author = {Yeh, Kuan-Yu and Janik, Michael J.},
title = {Density functional theory-based electrochemical models for the oxygen reduction reaction: Comparison of modeling approaches for electric field and solvent effects},
journal = {Journal of Computational Chemistry},
volume = {32},
number = {16},
pages = {3399-3408},
doi = {https://doi.org/10.1002/jcc.21919},
url = {https://onlinelibrary.wiley.com/doi/abs/10.1002/jcc.21919},
eprint = {https://onlinelibrary.wiley.com/doi/pdf/10.1002/jcc.21919},
year = {2011}
}

@article{Yu2011-ua,
  title     = "Oxygen reduction reaction mechanism on nitrogen-doped graphene: A density functional theory study",
  author    = "Yu, L. and Pan, X. and Cao, X. and Hu, P. and Bao, X.",
  journal   = "J. Catal.",
  publisher = "Elsevier BV",
  volume    =  282,
  number    =  1,
  pages     = "183--190",
  month     =  aug,
  year      =  2011,
  language  = "en"
}

@article{Zeng14,
    author = {Yi Zeng and Raymond B. Smith and Peng Bai and Martin Z. Bazant},
    title = {Simple formula for Marcus-Hush-Chidsey kinetics},
    journal = {Journal of Electroanalytical Chemistry},
    year = {2014}
}

@article{Zhang2011-cw,
  title     = "Mechanisms of oxygen reduction reaction on nitrogen-doped graphene for fuel cells",
  author    = "Zhang, L. and Xia, Z.",
  journal   = "J. Phys. Chem. C Nanomater. Interfaces",
  publisher = "American Chemical Society (ACS)",
  volume    =  115,
  number    =  22,
  pages     = "11170--11176",
  month     =  jun,
  year      =  2011,
  language  = "en"
}

@article{Zhao22,
    author = {Zhao, Hongyan and Cao, Hao and Zhang, Zisheng and Wang, Yang-Gang},
    title = {Modeling the Potential-Dependent Kinetics of CO2 Electroreduction on Single-Nickel Atom Catalysts with Explicit Solvation},
    journal = {ACS Catalysis},
    volume = {12},
    number = {18},
    pages = {11380-11390},
    year = {2022},
    doi = {10.1021/acscatal.2c02383},
}

@article{Zheng11,
  author	= {Zheng, G. and Niklasson, A. M. N. and Karplus, M.},
  title		= {Lagrangian formulation with dissipation of Born-Oppenheimer molecular dynamics using the density-functional tight-binding method},
  journal	= {J. Chem. Phys.},
  year		= {2011},
  volume	= {135},
  pages		= {044122},
  number	= {4},
  doi		= {10.1063/1.3605303}
}

@article{Zhu24,
    author = {Zhu, Xinwei and Huang, Jun and Eikerling, Michael},
    title = {Hierarchical Modeling of the Local Reaction Environment in Electrocatalysis},
    journal = {Accounts of Chemical Research},
    volume = {57},
    number = {15},
    pages = {2080-2092},
    year = {2024},
    doi = {10.1021/acs.accounts.4c00234}
}
